\documentclass[useAMS,usenatbib,usegraphicx]{mn2e}

\usepackage{subfigure}
\usepackage{url}
\usepackage{times}

\makeatletter

\renewcommand{\@thesubfigure}{(\alph{subfigure})\hskip\subfiglabelskip}
\renewcommand{\@@thesubfigure}{(\alph{subfigure})}

\newcommand\ion[2]{#1$\;${\small\rmfamily\@Roman{#2}}\relax}%

\newcommand\aap{{A\&A}}%
\newcommand\araa{{ARA\&A}}%
\newcommand\aaps{{A\&AS}}%
\newcommand\mnras{{MNRAS}}%
\newcommand\apj{{ApJ}}%
\newcommand\apjs{{ApJS}}%
\newcommand\apjl{{ApJ}}%
\newcommand\aj{{AJ}}%
\newcommand\pasp{{PASP}}%
\newcommand\memsai{{Mem.~Soc.~Astron.~Italiana}}%
\newcommand\procspie{{Proc.~SPIE}}%

\makeatother

\title[Spectrum Syntheses of High Resolution ILS of Galactic
GCs]{Spectrum Syntheses of High Resolution Integrated Light Spectra of
Galactic Globular Clusters\footnote{Based on observations obtained
with the Hobby-Eberly Telescope, which is a joint project of the
University of Texas at Austin, the Pennsylvania State University,
Stanford University, Ludwig-Maximilians-Universit\"{a}t M\"{u}nchen,
and Georg-August-Universität G\"{o}ttingen.}}

\author[C.M. Sakari et al.]{Charli M. Sakari$^{1,5}$\thanks{E-mail:
sakaricm@uvic.ca}, Matthew Shetrone$^{2}$, Kim Venn$^{1}$, Andrew
McWilliam$^{3}$,
\newauthor and Aaron Dotter$^{4}$\\
$^{1}$Department of Physics and Astronomy, University of Victoria,
Victoria, BC V8W 3P2, Canada\\
$^{2}$McDonald Observatory, University of Texas at Austin, HC75 Box
1337-MCD, Fort Davis, TX 79734, USA\\
$^{3}$The Observatories of the Carnegie Institute of Washington, 813
Santa Barbara Street, Pasadena, CA 91101-1292, USA\\
$^{4}$Research School of Astronomy and Astrophysics, The Australian
National University, Weston, ACT 2611, Australia\\
$^{5}$Vanier Canada Graduate Scholar}
\begin{document}

\maketitle

\label{firstpage}

\begin{abstract}
Spectrum syntheses for three elements (Mg, Na, and Eu) in
high-resolution integrated light spectra of the Galactic globular
clusters 47~Tuc, M3, M13, NGC~7006, and M15 are presented, along with
calibration syntheses of the Solar and Arcturus spectra.  Iron
abundances in the target clusters are also derived from integrated
light equivalent width analyses.  Line profiles in the spectra of
these five globular clusters are well fit after careful consideration
of the atomic and molecular spectral features, providing levels of
precision that are better than equivalent width analyses of the same
integrated light spectra, and that are comparable to the precision in
individual stellar analyses. The integrated light abundances from the
5528 and 5711 \AA \hspace{0.025in} \ion{Mg}{1} lines, the 6154 and
6160 \AA  \hspace{0.025in} \ion{Na}{1} lines, and the 6645 \AA
\hspace{0.025in} \ion{Eu}{2} line fall within the observed
\textit{ranges} from individual stars; however, these integrated light
abundances do not always agree with the \textit{average} literature
abundances.  Tests with the second parameter clusters M3, M13, and
NGC~7006 show that assuming an incorrect horizontal branch morphology
is likely to have only a small ($\la 0.06$ dex) effect on these Mg,
Na, and Eu abundances. These tests therefore show that integrated
light spectrum syntheses can be applied to unresolved globular
clusters over a wide range of metallicities and horizontal branch
morphologies.  Such high precision in integrated light spectrum
syntheses is valuable for interpreting the chemical abundances of
globular cluster systems around other galaxies.
\end{abstract}


\section{Introduction}\label{sec:Intro}
Detailed chemical abundances from high-resolution spectroscopy of
individual stars in globular clusters (GCs) associated with the Milky
Way and its nearby satellite galaxies have provided valuable
information about stellar evolution and nucleosynthesis
(e.g., \citealt{Gratton2004}), the chemical evolution of low- and
high-mass galaxies (e.g., \citealt{Tolstoy2009}), and the assembly
history of the Milky Way
(e.g., \citealt{FreemanBlandHawthorn2002}). In
particular, chemical tagging has enabled the identification of
potentially accreted stellar streams and GCs
(e.g., \citealt{Cohen2004},  \citealt{Sbordone2005},
\citealt{Chou2010}, \citealt{Sakari2011}), which shows that
hierarchical merging has played an important role in the formation of
the Milky Way Galaxy.  Observations of extragalactic GC systems
suggest that most galaxies have experienced complicated assembly
histories that may include a significant component of accretion from
dwarf galaxies (see \citealt{HarrisSaasFee} and
\citealt{BrodieStrader2006} for reviews on the observed properties of
GC systems).  Certain models suggest that most GCs form in small
dark matter halos at high redshift, and are accreted by larger
galaxies at later times \citep{Bekki2008,MuratovGnedin2010}.  Detailed
chemical abundances of  extragalactic globular cluster systems,
particularly those associated with galaxy types (e.g., massive
ellipticals vs. spirals) and environments (e.g., located within galaxy
clusters) that are not found in the Local Group, can provide
information about the assembly histories of their host galaxies.

Individual stars are too faint for detailed spectral analyses beyond
the Local Group.  Distant extragalactic GCs, however, appear as
bright, point-like sources, and can be observed at much greater
distances than individual stars.  Therefore, an alternative way to
determine the flux-weighted average chemistry of a GC is from an
integrated light spectrum (ILS) of the entire population.  Information
on the chemical abundance ratios for a system of GCs can then be used
to study the chemical evolution and early assembly history of the host
galaxy.  Integrated light spectroscopy has historically been done at
low- to medium-resolution.  Lower-resolution ILS analysis methods have
been tested and calibrated on Galactic GCs within the last few decades
(e.g., \citealt{Schiavon2002,LeeWorthey2005})---such methods have
proven capable of determining the ages, metallicities, and
$\alpha$-abundances of Galactic GCs.  However, detailed abundances (of,
e.g., iron-peak or neutron capture elements) require moving to higher
resolution.  By providing additional elements and higher precision,
high resolution observations provide information about, for example,
the contributions from different types of supernovae or AGB stars.

The high resolution integrated light abundance analysis program
ILABUNDS (presented in \citealt{McWB}, hereafter MB08), has been
tested on several Galactic GCs.  The majority of ILABUNDS analyses
thus far have been equivalent width analyses, which are sufficient for
elements with multiple, fairly strong lines since with a large number
of lines the error in the mean abundance will be less sensitive to
individual measurement errors.  Equivalent width techniques have
successfully reproduced the literature abundances (from individual
stars) of $\alpha$, iron peak, and neutron capture elements in several
Galactic GCs (MB08, \citealt{Cameron2009}) and produced reasonable
abundances for inner halo M31 GCs, Large Magellanic Cloud GCs, and GCs
associated with other Local Group dwarf galaxies
\citep{Colucci2009,Colucci2011,Colucci2011b,Colucci2012}.  However,
many elements do not have enough (or any) strong lines available for
an \textit{equivalent width} analysis.  The features in ILS are also
often weak and/or blended, which makes an \textit{equivalent width}
analysis particularly difficult.  There are many elements with
detectable spectral lines in high-resolution spectra, e.g., Cu, Zn,
Ba, and Eu, which are valuable for galaxy and GC formation and
chemical evolution theories (e.g., \citealt{Matteucci1993},
\citealt{Mishenina2002}, \citealt{Travaglio2004}) but which are
unsuitable for an \textit{equivalent width} analysis.

With individual stellar analyses, the preferred method for determining
abundances of such elements is to \textit{synthesize} the entire
spectral region around the line of interest.  This technique is
different from an equivalent width analysis, because the line profile,
width, \textit{and} depth are be fit simultaneously, while the effects
of nearby lines are taken into account. When a line is too
weak or a spectrum is too noisy, upper limits can be obtained with
spectrum synthesis, which can still provide valuable
constraints. \citet{Colucci2009,Colucci2012} have synthesized
high-resolution ILS to obtain abundances of Na, Mg, Al, Sr, Ba, La,
and Eu, among other elements, with random $1\sigma$ errors that are
typically on the order of $0.15-0.3$ dex.  These errors are quite
large given the spectral resolution; lower uncertainties would be more
useful for chemical comparisons between clusters or galaxies, or for
comparisons with chemical evolution models.  Tests on Galactic GCs can
better determine the accuracy (through comparisons with literature
abundances) and precision (through stringent tests on high S/N
spectra) of integrated light spectrum syntheses.

In order to understand and quantify the accuracy and precision of an
IL spectrum synthesis method, this paper presents tests on the five
Galactic GCs 47~Tuc, M3, M13, NGC~7006, and M15 in four different 10
\AA \hspace{0.025in} regions around the 5528 and 5711 \AA
\hspace{0.025in} \ion{Mg}{1} lines, the 6154 and 6160 \AA
\hspace{0.025in} \ion{Na}{1} lines, and the 6645 \AA 
\hspace{0.025in} \ion{Eu}{2} line.  These elements have been selected
because they are important for galaxy/GC formation theories.  In
addition, the lines are easily detectable in 
the majority of the spectra, providing abundances rather than upper
limits. Other than the \ion{Eu}{2} line, these lines and features also
avoid the complications of significant hyperfine structure or isotopic
corrections.

The target clusters were selected to cover a range of metallicities
and horizontal branch morphologies, as discussed in Section
\ref{subsec:Targets}.  Sections \ref{subsec:Observations} and
\ref{subsec:DataReduction} describe the observations of these nearby
targets, and the subsequent data reductions, while Section
\ref{sec:MB08} describes the abundance analysis method used to infer
the chemical abundances.  The syntheses are finally presented in
Section \ref{sec:Syntheses}, along with detailed discussions of the
errors.  The important findings of this paper are discussed in Section
\ref{sec:Discussion}.  Finally, the results are summarized in Section
\ref{sec:Summary}.

\section{Observations and Data Reduction}\label{sec:Observations}

\subsection{Target Selection}\label{subsec:Targets}
The targets were selected to span a range of metallicities (from
$[\rm{Fe/H}] = -0.7$ to $-2.4$) and HB morphologies (from red to very
blue), though all appear to be standard Galactic GCs (i.e. old and
$\alpha$-enhanced).  In particular, M3/M13/NGC~7006 form a ``second
parameter'' triad, i.e. the three GCs have similar ages and
metallicities, yet different HB morphologies. Basic information about
the target clusters is provided in Table \ref{table:Targets}.
Positions and integrated $V$ magnitudes are from \citet{Harris}; the
[Fe/H], [$\alpha$/Fe], and age estimates come from isochrone fits by
\citet{Dotter2010,Dotter2011}; and the horizontal branch (HB) index,
$(B~-~R)~/~(B~+~V~+~R)$, (where $B$, $R$, and $V$ are the number of
stars blueward, redward, and inside of the instability strip) comes
from \citet{MackeyVanDenBergh}.  In order to minimize the
uncertainties that occur when modeling the stellar populations,
resolved photometry was utilized to estimate atmospheric parameters
for the GC stars (see Section \ref{subsubsec:Photometry}).

\begin{table*}
\centering
\begin{minipage}{140mm}
\caption{The target clusters.\label{table:Targets}}
  \begin{tabular}{@{}lccccccc@{}}
  \hline
Cluster & RA (J2000) & Dec (J2000) & $V_{\rm{int}}$ & [Fe/H] &
[$\alpha$/Fe] & Age & HB index\\
 \hline
47~Tuc (NGC 104)& $00^{\rm{h}}24^{\rm{m}}05^{\rm{s}}.67$ & $-72^{\circ}04\arcmin52\arcsec.6$ & 3.95 & -0.70 & 0.2 & $12.75\pm0.50$ & -0.99\\
M3  (NGC 5272)  & $13^{\rm{h}}42^{\rm{m}}11^{\rm{s}}.62$ & $+28^{\circ}22\arcmin38\arcsec.2$ & 6.19 & -1.60 & 0.2 & $12.50\pm0.50$ & \phantom{-}0.08 \\
M13 (NGC 6205)  & $16^{\rm{h}}41^{\rm{m}}41^{\rm{s}}.24$ & $+36^{\circ}27\arcmin35\arcsec.5$ & 5.78 & -1.60 & 0.2 & $13.00\pm0.50$ & \phantom{-}0.97 \\
NGC 7006        & $21^{\rm{h}}01^{\rm{m}}29^{\rm{s}}.38$ & $+16^{\circ}11\arcmin14\arcsec.4$ &10.56 & -1.50 & 0.2 & $12.25\pm0.75$ & -0.28 \\
M15 (NGC 7078)  & $21^{\rm{h}}29^{\rm{m}}58^{\rm{s}}.33$ & $+12^{\circ}10\arcmin01\arcsec.2$ & 6.20 & -2.40 & 0.2 & $13.25\pm1.00$ & \phantom{-}0.67 \\
\hline
\end{tabular}
\end{minipage}\\
\medskip
\raggedright {\bf References: } Positions and integrated magnitudes are from
\citet{Harris}.  The [Fe/H], [$\alpha$/Fe], and age estimates are from
isochrone fitting \citep{Dotter2010,Dotter2011}.  The HB index,
$(B~-~R)~/~(B~+~V~+~R)$, comes from \citet{MackeyVanDenBergh}.
\end{table*}

\subsection{Observations}\label{subsec:Observations}
The 47~Tuc spectrum was kindly provided by R. Bernstein \&
A. McWilliam.  It is the same spectrum analyzed in MB08; details on
the observations and data reduction can be found there.  Additional
normalizations with low-order polynomials were performed and the
apertures were combined, as described below.

The other spectra were observed with the High Resolution Spectrograph
(HRS; \citealt{HRSref}) on the Hobby-Eberly Telescope (HET;
\citealt{HETref,HETQueueref}) at McDonald Observatory in Fort Davis,
TX in 2011 and 2012.  A slit width of 1$\arcsec$ was used, leading to
an instrumental spectral resolution of $R\approx 30,000$. Given the
velocity dispersions of the targets (see below), a higher spectral
resolution is unnecessary.  The 600 gr/mm cross disperser was used,
with a central wavelength of 6302.9 \AA.  The spectral coverage is
therefore $\sim 5320-6290$ \AA \hspace{0.025in} on the blue chip and
$\sim 6360-7340$ \AA \hspace{0.025in} on the red chip.  This
wavelength range was chosen for future unresolved targets in order to
minimize the effects of improperly modeled HBs, since blue HB stars
should contribute less to the integrated light at red wavelengths.

\begin{figure*}
\begin{center}
\centering
\subfigure[M3]{\includegraphics[scale=0.35]{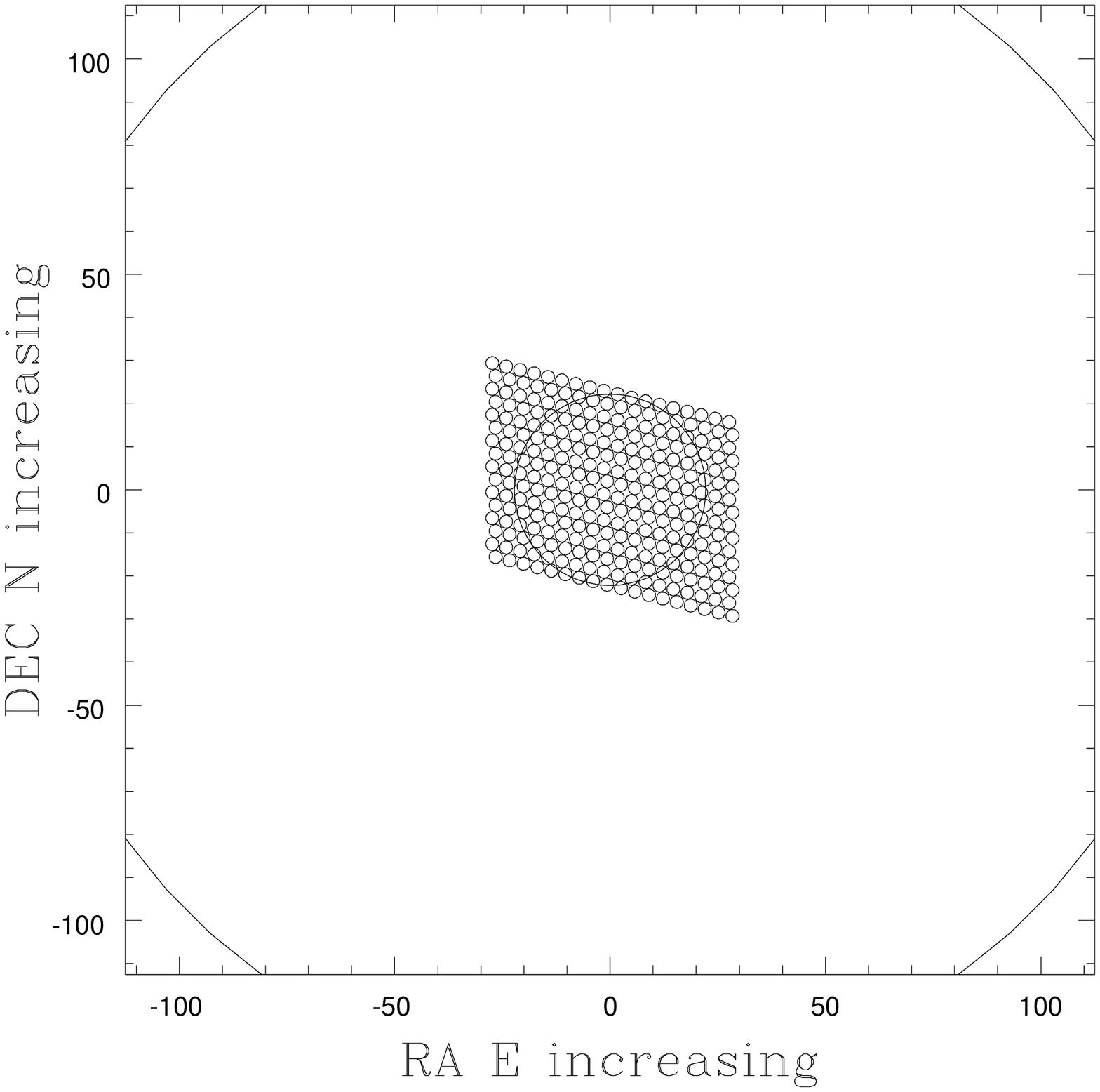}\label{fig:M3Pointings}}
\subfigure[M13]{\includegraphics[scale=0.35]{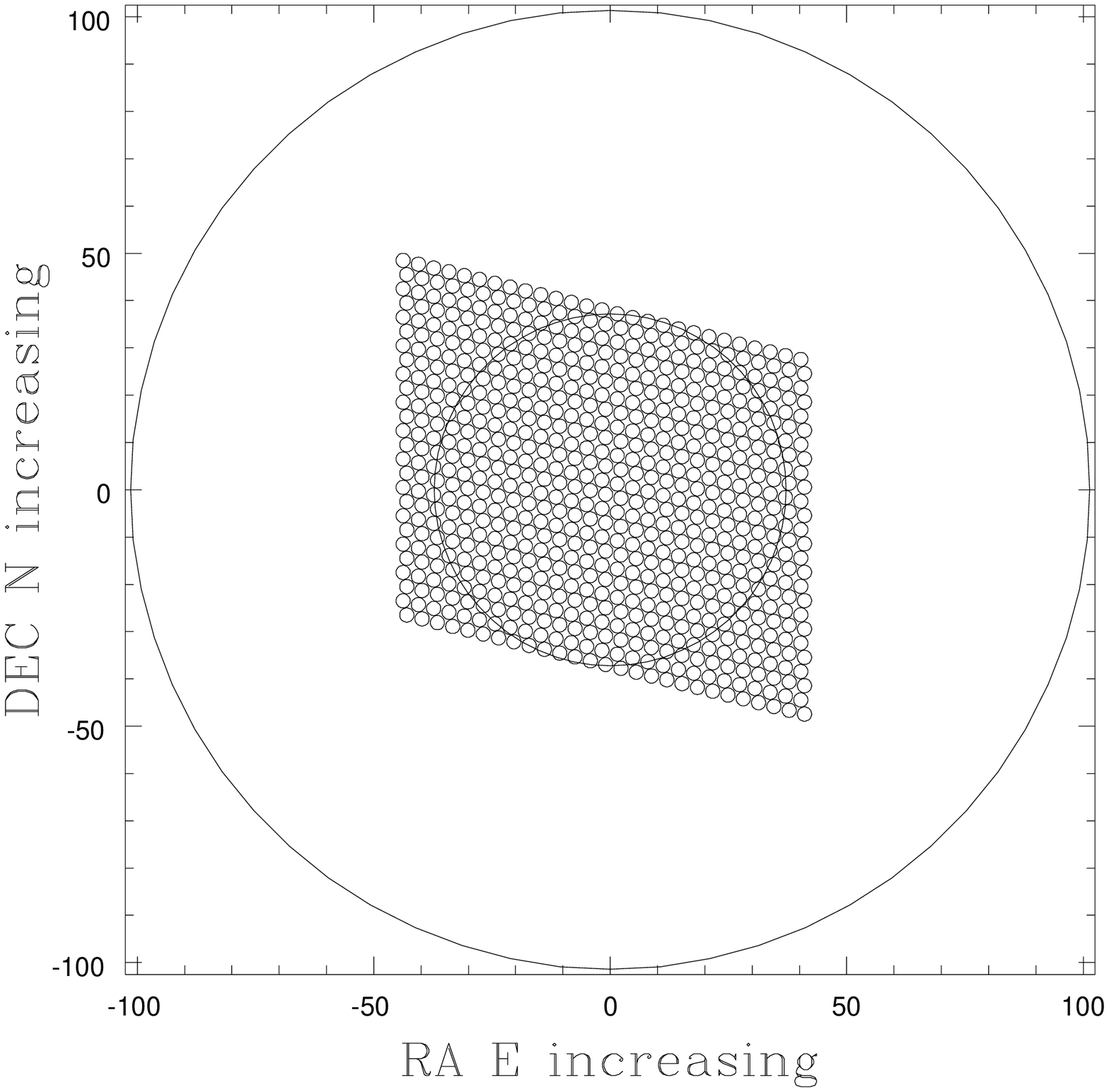}\label{fig:M13Pointings}}
\subfigure[NGC7006]{\includegraphics[scale=0.35]{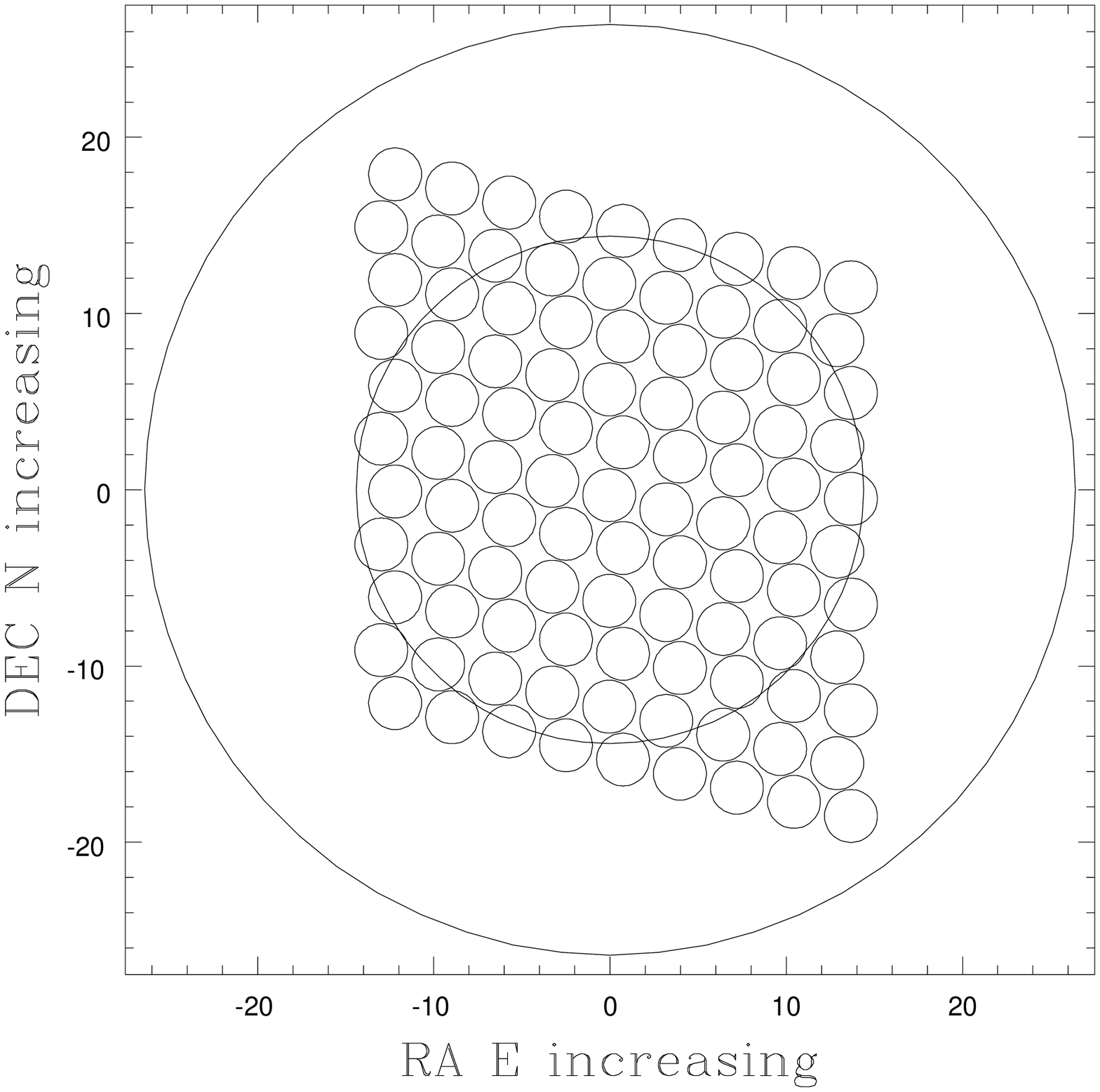}\label{fig:NGC7006Pointings}}
\subfigure[M15]{\includegraphics[scale=0.35]{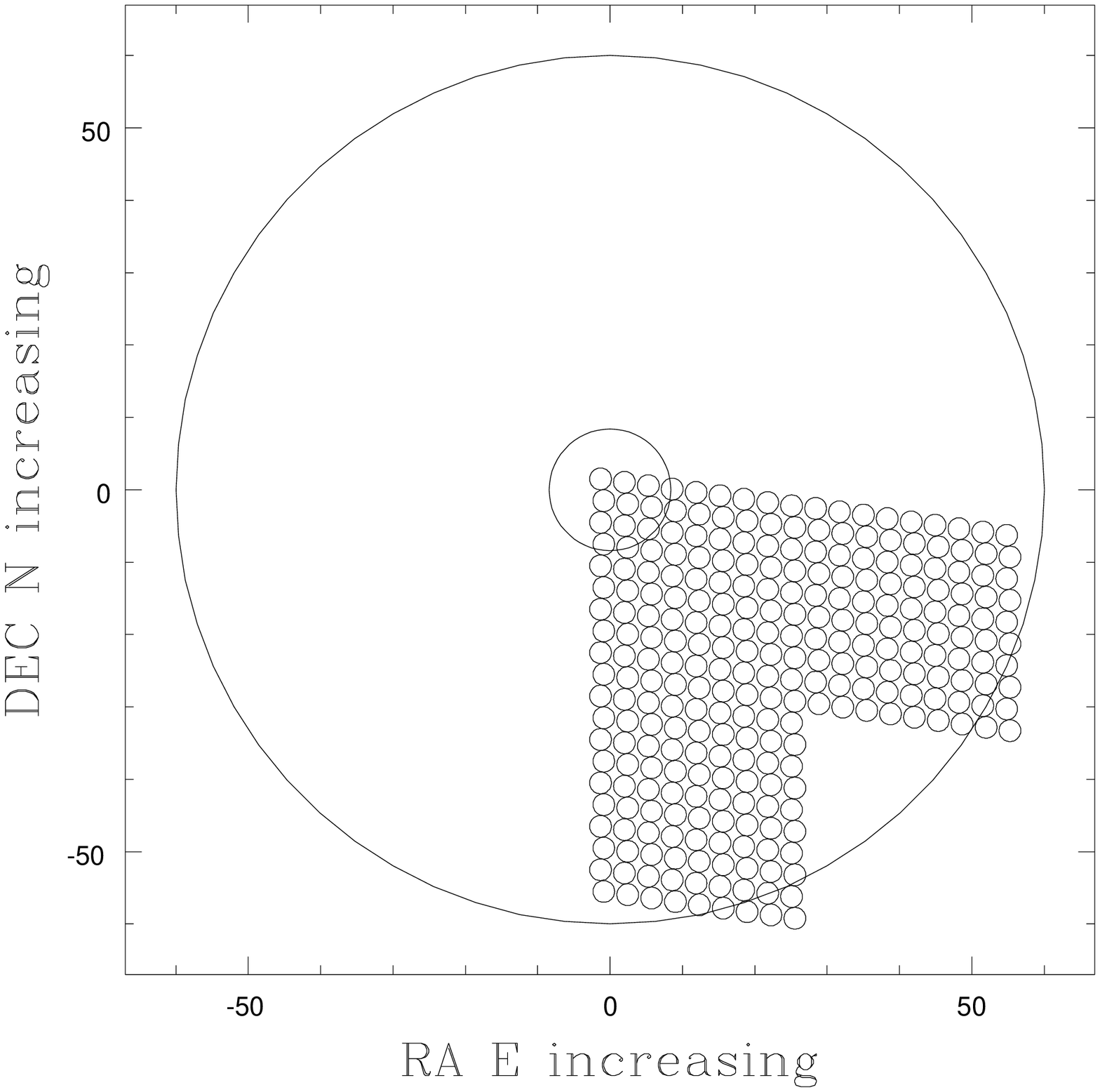}\label{fig:M15Pointings}}
\caption{Fiber pointings for the HET observations of the target
clusters.  The small circles show the positions of the 3$\arcsec$
fibers (both sky and object), while the larger circles show the core
and half-light radii (from \citealt{Harris}).  The centers of the
clusters are shown at arbitrary positions.  Each observation scans the
three fibers across the cluster; for an individual GC, each pointing
lasts for the same amount of time, and the pointings are not
overlapped.\label{fig:Pointings}}
\end{center}
\end{figure*}

ILS observations of distant, point-like targets are relatively simple.
Nearby GCs, however, are much more difficult to observe for integrated
light studies, given their large sizes on the sky and the fact that
their stars \textit{are} resolved.  It is essential to observe these
difficult targets, however, because the ILS methods can be calibrated
through comparisons with abundances from individual stars.  To
overcome these observational difficulties, the ILS of M3, M13,
NGC~7006, and M15 were obtained by scanning the HRS fibers across the
cluster cores.  A number of specific pointings on the cluster were
selected, and the fibers were moved to each position.  The large
3$\arcsec$ fiber was used in order to maximize spatial coverage on the
clusters.  HRS provides two additional 3$\arcsec$ fibers located
10$\arcsec$ from the center of the object fiber.  In typical
observations these extra fibers are intended to serve as sky fibers;
however, because of the spatial extent of the Galactic GCs, these sky
fibers served as additional object fibers during the
observations. Separate sky observations with all three fibers were
therefore taken after each GC observation. These pointing patterns are
shown in Figure \ref{fig:Pointings}, along with the clusters' core and
half-light radii.

For M3, M13, and NGC~7006, the entire core of each cluster was
observed, though the clusters were too large to observe the entire
area within the half-light radius in a reasonable amount of time.
Note that M15 was mapped differently than NGC 7006, M3, or M13; its
wedge-shaped map represents a slice of the cluster out to the
half-light radius, assuming the cluster is spherically symmetric.  Any
differences in the spectra as a result of the different mappings
will be negligible because the input photometry have been selected to
reflect the spatial coverage of the ILS (see Section
\ref{subsubsec:Photometry}).

The total exposure times and S/N ratios in blue (5500 \AA) and red
(7000 \AA) regions are shown in Table
\ref{table:Observations}. Generally, the exposure times were
calculated to allow an observation of a single HB star to reach S/N =
70, though NGC 7006 did not receive sufficient time to meet this
goal.

\begin{table*}
\centering
\begin{minipage}{165mm}
\caption{GC observations.\label{table:Observations}}
  \begin{tabular}{@{}llccccccc@{}}
  \hline
Cluster & Observation & Exposure & S/N$^{a}$ & S/N$^{a}$ & $v_{\rm{helio, obs}}$ &
$v_{\rm{helio, lit}}$ & $\sigma_{\rm{obs}}$ & $\sigma_{\rm{lit}} $\\
 & Dates & Time (s) & (5500 \AA) & (7000 \AA)& (km/s) & (km/s) & (km/s) & (km/s)\\
  \hline
47~Tuc$^{b}$&2000 Jul 18, 19& 11030 & 120  & 180  & -             & -      & $11.50\pm 0.30^{c}$ & 11.0\\
 & & & & & & & & \\
M3       & 2012 Mar 25, & 9940  & 180 & 230 & $-146.0 \pm 1.1$ & -147.6 & $5.66\pm 0.15$ & 5.5 \\
 &  \phantom{20}Apr 16, 17, 18                     &       &     &     &
&        &                &  \\
 & & & & & & & & \\
M13      & 2012 Apr 17, 20, 22         & 11569 & 130 & 250 & $-247.5 \pm 1.3$ & -244.2 & $7.23\pm 0.33$ & 7.1 \\
 & & & & & & & & \\
NGC 7006  & 2011 Sep 24,                & 8903  & 65  & 130  & $-380.4 \pm 0.7$ & -384.1 & $4.49\pm 0.60$ & -\\
 & 2012 May 29, Jun 19                &       &     &     &
&        &                &  \\
 & & & & & & & & \\
M15      & 2011 Sep 27                 & 3280  & 95  & 220  & $-106.6 \pm 0.2$ & -107.0 & $12.54\pm0.60$ & 13.5\\
\hline
\end{tabular}
\end{minipage}\\
\medskip
\raggedright {\bf References: } Literature values are from the Harris
Catalog \citep{Harris}.\\
$^{a}$S/N ratios (per pixel) are measured in IRAF.\\
$^{b}$47~Tuc was observed with the Las Campanas 2.5 m Du Pont
Telescope by R. Bernstein \& A. McWilliam; see MB08 for more details.\\
$^{c}$This velocity dispersion has been determined in the same way as
the other GCs, for consistency.\\
\end{table*}

\subsection{Data Reduction}\label{subsec:DataReduction}
The data reduction was performed in the Image Reduction and Analysis
Facility program (IRAF)\footnote{IRAF is distributed by the National
Optical Astronomy Observatory, which is operated by the Association of
Universities for Research in Astronomy, Inc., under cooperative
agreement with the National Science Foundation.} according to the
standard HRS data reduction methods, with several exceptions.  A
separate bias frame removal was not completed, as the process can add
noise to the spectra.\footnote{In addition, the scattered light
subtraction removes some of the bias level; see the online HRS Data
Reduction Tips at \url{http://hydra.as.utexas.edu/?a=help&h=29#HRS}}
To remove cosmic rays the aperture extraction was performed
with variance weighting, where IRAF considers the gain and read noise
of the CCD and identifies and removes any pixels that deviate
significantly from the noise.  This variance weighting procedure
occasionally affects the shape of the continuum; thus, the shape of
the non-weighted extraction was maintained.

Separate sky observations were obtained after all ILS observations.
Again, all three fibers (object and sky) were used in each exposure;
the spectra from the three fibers were averaged together to remove
noise and cosmic rays. Because the sky spectra were still fairly
noisy, the spectra were replaced with continuum fits and the emission
lines were added back in (using the sky line identifications from the
UVES Quality Control sky spectrum
website\footnote{\url{http://www.eso.org/observing/dfo/quality/UVES/pipeline/sky_spectrum.html}}).
The IRAF task \textit{skytweak} determined the necessary scaling
factor between a target object spectrum and its sky spectrum, and the
sky spectrum was multiplied by that factor before it was subtracted
from the object spectrum. Telluric standards were also observed during
each night, allowing separate removal of the telluric absorption
lines.

Because ILS may have many undetectable, blended, weak features that
can obscure/distort the continuum level, the spectra must be carefully
normalized.  To remove the blaze function of the spectrograph and the
blackbody function of the cluster, an extremely metal-poor (EMP) giant
was observed with the same instrumental setup.  An EMP star should
have very few spectral lines, and finding the continuum should
therefore be quite simple.  Furthermore, an EMP giant should have a
similar effective temperature as the average temperature of the
cluster.  Information on the observed EMP star, CS29502-092, is given
in Table \ref{table:EMP}.  The EMP star's continuum was fit with the
IRAF task \textit{continuum}, and the target spectra were divided by
these continuum fits.  Additional low-order polynomial fits were
required to fully normalize the target spectra.

\begin{table*}
\centering
\begin{minipage}{170mm}
\caption{Properties of the observed EMP star.\label{table:EMP}} 
  \begin{tabular}{@{}lccccccccc@{}}
  \hline
Star & RA & Dec & V & $T_{\rm{eff}}$ & Observation & Exposure & S/N$^{a}$ & S/N$^{a}$ & $v_{\rm{helio}}^{b}$ \\
 & (J2000) & (J2000) & & (K) & Dates & Times (s) & (5500 \AA) & (7000
\AA)& (km/s)\\
\hline
CS29502-092 & $22^{\rm{h}}22^{\rm{m}}36^{\rm{s}}.0$ & $-01^{\circ}38\arcmin27\arcsec.5$ & 11.87 & $5001$ & 2011 Nov 16  & 690  & 167 & 341 & $-67.0 \pm 0.4$ \\
\hline
\end{tabular}
\end{minipage}\\
\medskip
\raggedright{\bf References: } The position and magnitude are from the
SIMBAD database. The stellar temperature is an average from the
Stellar Abundances for Galactic Archaeology (SAGA) database
\citep{SAGA}.\\ $^{a}$S/N ratios (per pixel) are measured in IRAF.\\
$^{b}$The radial velocity was determined in the same way
as the GC targets.\\
\end{table*}

The normalized GC spectra were then cross-correlated with a reference
spectrum (using the IRAF task \textit{fxcor}) to determine the radial
velocity and velocity dispersion.  Arcturus was used as the template
spectrum for the cross-correlation; this very high resolution
($R=150,000$) spectrum was observed with the Fourier Transform
Spectrometer (FTS) on the McMath Telescope \citep{Hinkle2003}, and was
obtained from the Arcturus
Atlas.\footnote{\url{ftp://ftp.noao.edu/catalogs/arcturusatlas/}} The
IRAF task \textit{fxcor} correlates the target spectra with the
Arcturus template spectrum, and identifies the values with the highest
correlation; the peak of the correlation occurs at the observed radial
velocity of the cluster, and the width determines the velocity
dispersion (as discussed below). The heliocentric velocities were
measured from each observation, and were averaged together to produce
the final, averaged heliocentric velocities (shown in Table
\ref{table:Observations}).

\begin{figure*}
\begin{center}
\centering
\includegraphics[scale=0.8]{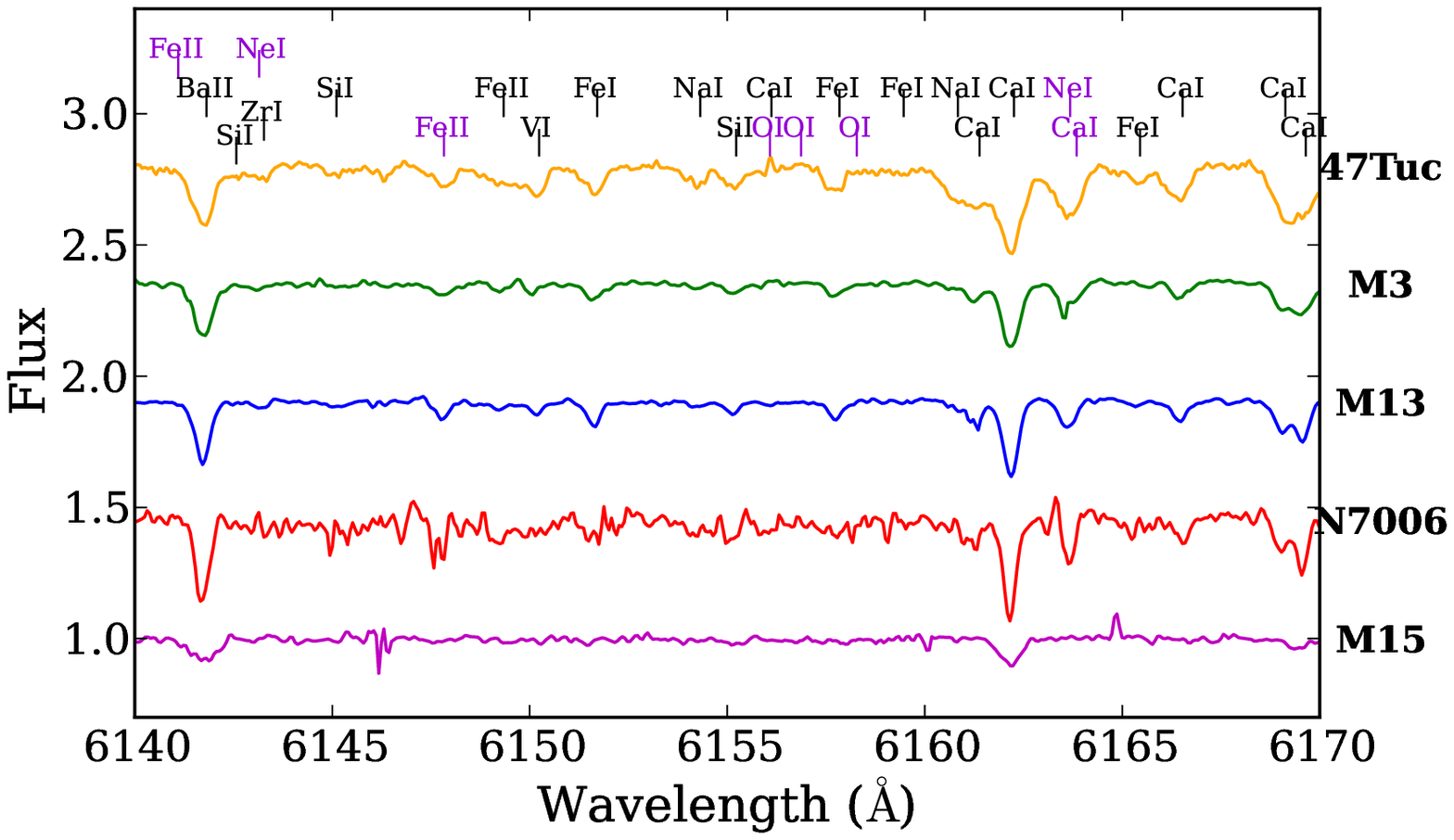}
\caption{The ILS of the five Galactic GCs studied in this work.  The
47~Tuc spectrum is from MB08, while the other four are HET spectra.
Lines of interest are noted; the lines identified in black are lines
that are typically used in standard RGB stellar analyses, while those
in purple are used with hotter stars.  The differences in line
strengths are due to population, composition, and velocity dispersion
differences between the five GCs.\label{fig:Spectra}}
\end{center}
\end{figure*}

Each of the individual reduced and normalized spectra were shifted to
the rest frame, using the radial velocity from \textit{fxcor}.  The
rest frame spectra were then combined with average sigma-clipping
rejection routines.  The individual observations were weighted by flux
during the averaging.  The blue ends of the individual apertures often
suffered from lower S/N, especially at blue wavelengths.  These noisy
regions were removed from the spectrum before the individual
apertures were combined.  Examples of the final spectra are shown in
Figure \ref{fig:Spectra}, along with notable spectral features.

After the individual observations were combined, the final spectra
were again cross-correlated with Arcturus to determine the cluster
velocity dispersions.  The width of the cross-correlation peak in
\textit{fxcor} depends on the width of the spectral lines.  In
individual stars the line width is often dominated by the instrumental
broadening, though rotation, microturbulence, etc. can also affect the
line width.  In an ILS, the velocity dispersion of the target is often
the dominant source of broadening. The full-width at half maximum
(FWHM) of the cross-correlation peak of an ILS is therefore related to
the velocity dispersion, $\sigma$, of the cluster, as described by
\citet{Alpaslan2009}.  Since the instrumental broadening is still
significant in the target GCs,\footnote{Recall that $R = 30,000$,
which means that the FWHM from instrumental broadening alone should be
10 km/s.  This instrumental broadening is likely to vary across the
CCD---however, these variations are likely to be insignificant in ILS
given the width of the lines.} the observed velocity dispersion is
found by subtracting (in quadrature) the instrumental broadening.  The
broadening of the target spectrum is taken into account in the
calibration curve, though there may be additional sources of
broadening that are unaccounted for.  Thus, the derived velocity
dispersions may be upper limits.  This is not crucial for this
analysis, since only the total width of the lines is important.  These
velocity dispersions for the target GCs are reported in Table
\ref{table:Observations}, and are in good agreement with the
literature values.  Line blends in the ILS may lead to overestimates
of the velocity dispersion; however, this does not seem to be
significant for these GCs.

\subsection{Calibration Spectra}\label{subsec:CalibrationSpectra}
As shown in Section \ref{sec:Syntheses}, the success of spectrum
synthesis relies upon a complete and accurate input line list.  To
test the accuracy of these input line lists, spectrum syntheses of the
Solar and Arcturus spectra have been performed. The Solar spectrum ($R
= 300,000$; \citealt{Kurucz2005}) comes from the Kurucz 2005 solar
flux atlas.\footnote{\url{http://kurucz.harvard.edu/sun.html}}  Solar
atmospheric parameters of $T_{\rm{eff}} = 5777$ K, $\log g = 4.44$
dex, $\chi = 0.85$ km s$^{-1}$, and $[\rm{M/H}] = 0.0$ were adopted
\citep{Yong}.

The Arcturus spectrum is the same one used for the cross-correlation
with the target ILS (see Section \ref{subsec:DataReduction}).  Arcturus
atmospheric parameters of $T_{\rm{eff}} = 4300$ K, $\log g = 1.50$
dex, $\chi = 1.56$~km~s$^{-1}$, and $[\rm{M/H}] = -0.6$ \citep{Yong}
were adopted.  This Arcturus temperature and surface gravity are in
excellent agreement with \citet{Fulbright2006} and
\citet{RamirezAllendePrieto2011}, though the microturbulence values
and metallicities differ slightly between the three studies. Both
\citet{Fulbright2006} and \citet{RamirezAllendePrieto2011} find higher
metallicities and microturbulence values, with
$\xi~=~1.67$~km~s$^{-1}$ and $[\rm{Fe/H}]~=~-0.50$, and
$\xi~=~1.74$~km~s$^{-1}$ and $[\rm{Fe/H}]~=~-0.52$, respectively.  The
Yong et al. microturbulence and metallicity are adopted here, because
their values agree best with the \ion{Fe}{1} and \ion{Fe}{2}
abundances derived in this work (Section
\ref{subsubsec:Metallicities}).

\section{Detailed Abundances with ILABUNDS}\label{sec:MB08}
The chemical abundances in the target GCs were determined with the
program ILABUNDS (MB08).  ILABUNDS is a modification of the 1997
version of the local thermodynamic equilibrium (LTE) line analysis and
spectrum synthesis code
MOOG\footnote{\url{http://www.as.utexas.edu/~chris/moog.html}}
\citep{Sneden}.  The equivalent width version of ILABUNDS was
described and presented in MB08; here, a spectrum synthesis version of
ILABUNDS (recently developed by A. McWilliam) is employed.

For spectrum syntheses of individual stars, MOOG requires an input
model atmosphere and line list.  ILABUNDS operates similarly, except
that a model atmosphere must be provided for each star in the cluster,
and the line list must be fairly complete.  The input model
atmospheres for ILABUNDS are discussed in Section
\ref{subsec:ModelAtms}, while the input line list is discussed in
Section \ref{subsec:LineList}.  With these inputs, the spectrum
synthesis code calculates a synthetic ILS for each population box
(Section \ref{subsubsec:CMDBoxes}) and combines the spectra together,
weighted by flux.  The final synthetic spectrum is broadened by the
instrumental broadening and velocity dispersion (see Section
\ref{subsec:DataReduction}), and is compared to the observed
spectrum.

\subsection{Model Atmospheres}\label{subsec:ModelAtms}
For each star observed in the ILS the atmospheric parameters
(effective temperature, $T_{\rm{eff}}$; surface gravity, $\log g$;
microturbulence, $\xi$; and metallicity, [Fe/H]) must be known in
order to determine a cluster's integrated light chemical abundances.
The stellar temperature, gravity, and microturbulence values are
estimated from observed photometry (Section
\ref{subsubsec:Photometry}), which is grouped into color-magnitude
diagram (CMD) boxes to simplify the process (Section
\ref{subsubsec:CMDBoxes}).  The metallicities are found through a
standard ILS equivalent width analysis (Section
\ref{subsubsec:Metallicities}).  Once the atmospheric parameters of a
CMD box are known, a corresponding model atmosphere (from the Kurucz
database;\footnote{\url{http://kurucz.harvard.edu/grids.html}}
\citealt{KuruczModelAtmRef}) is then assigned.

\subsubsection{Input Photometry}\label{subsubsec:Photometry}
For unresolved targets the atmospheric parameters of the stars in a
cluster must be modeled with isochrones, and observational diagnostics
must be developed in order to ensure that the population is correctly
modeled (see, e.g., \citealt{Colucci2009,Colucci2011}).  With
nearby targets, the uncertainties from modeling the stellar
populations can be removed with the use of resolved photometry,
which provides the color and magnitude of each star in the observed
region.  These observable quantities are then converted to physical
quantities via empirical relations.

For 47~Tuc, the  Hubble Space Telescope (HST) $B$, $V$ data from
R. Schiavon \citep{Guhathakurta1992,Howell2000} were used to estimate
the stellar parameters---more information on this data can be found in
MB08.  For M3, M13, NGC~7006, and M15, the $V$, $I$ HST photometry
from the ACS Galactic Globular Cluster Survey 
(e.g. \citealt{Sarajedini2007,Anderson2008,Dotter2011}) provided the
estimates for the stellar atmospheric parameters.  The $(m-M)_V$ and
$E(B-V)$ values from \citet{Harris}, the $E(V-I)$ extinction
corrections of \citet{McCall2004}, and the $(V-I)$ color-temperature 
conversions from \citet{RamirezMelendez2005} then provided the
temperatures and surface gravities of the stars, as described in MB08.
For stars that fell outside the \citet{RamirezMelendez2005}
calibration space, the atmospheric parameters of the Kurucz models
that best matched the observed values were adopted.  An empirical
relationship between the surface gravity and the microturbulence (also
described in MB08) provided values for $\xi$.

The input photometry does not \textit{exactly} match the populations
observed in the ILS.  Although the photometry were selected to only
include stars within the maximum radii covered by the HET, the
non-circular coverage of the ILS introduces some population
differences (particularly for M15; see Figure \ref{fig:Pointings}).
Field star contamination may also affect the input photometry and the
observed spectrum.  The presence of even one additional bright star
can affect the observed ILS and the subsequent analysis (for example,
the bright blue star in NGC~7006's CMD).  The effects of stochastic
sampling of the brightest stars in unresolved targets have been
investigated by \citet{Colucci2011}, who used Monte Carlo sampling to
populate their stellar populations.  They found, for an old,
metal-poor GC, that the maximum difference in [Fe/H] is only $\sim
0.08$ dex when randomly repopulating the entire GC.  Resolved Galactic
GCs with HST photometry provide a unique opportunity to test these
results directly. For example, ILS of different scanned areas of the
clusters can be compared (e.g., different wedges of M15), or test
stars can be artificially added to the input photometry.  These latter
tests will be discussed in a subsequent paper (Sakari et al., in
prep.), though preliminary syntheses of the lines presented in this
paper suggest that field stars are unlikely to have a significant
effect on the synthesized abundances.  In particular, if the bright
blue star in NGC~7006's CMD were a Solar metallicity field star with
slow rotation, several sharp features would be present in the observed
ILS; however, these sharp features are not observed, and they would
have a negligible effect on the final abundances.

\subsubsection{CMD Boxes}\label{subsubsec:CMDBoxes}
The process of assigning model atmospheres is simplified by binning
the stars together into boxes, based on their position in the CMD.
Each CMD box is then assigned a model atmosphere with the average
color, magnitude, temperature, etc. of all the stars in the box; the
contributions to the total ILS from each box are then weighted by the
number of stars in each box.

\begin{figure*}
\begin{center}
\centering
\subfigure[M3]{\includegraphics[scale=0.6]{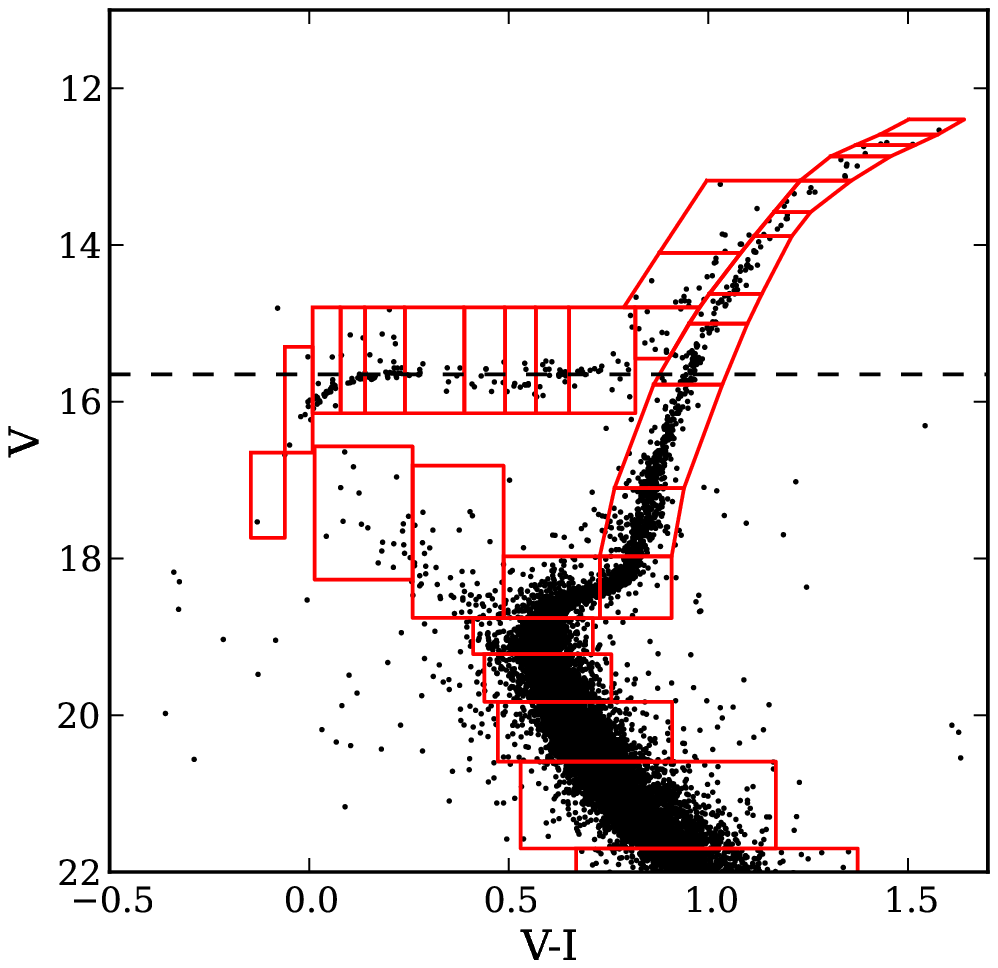}\label{fig:M3CMD}}
\subfigure[M13]{\includegraphics[scale=0.6]{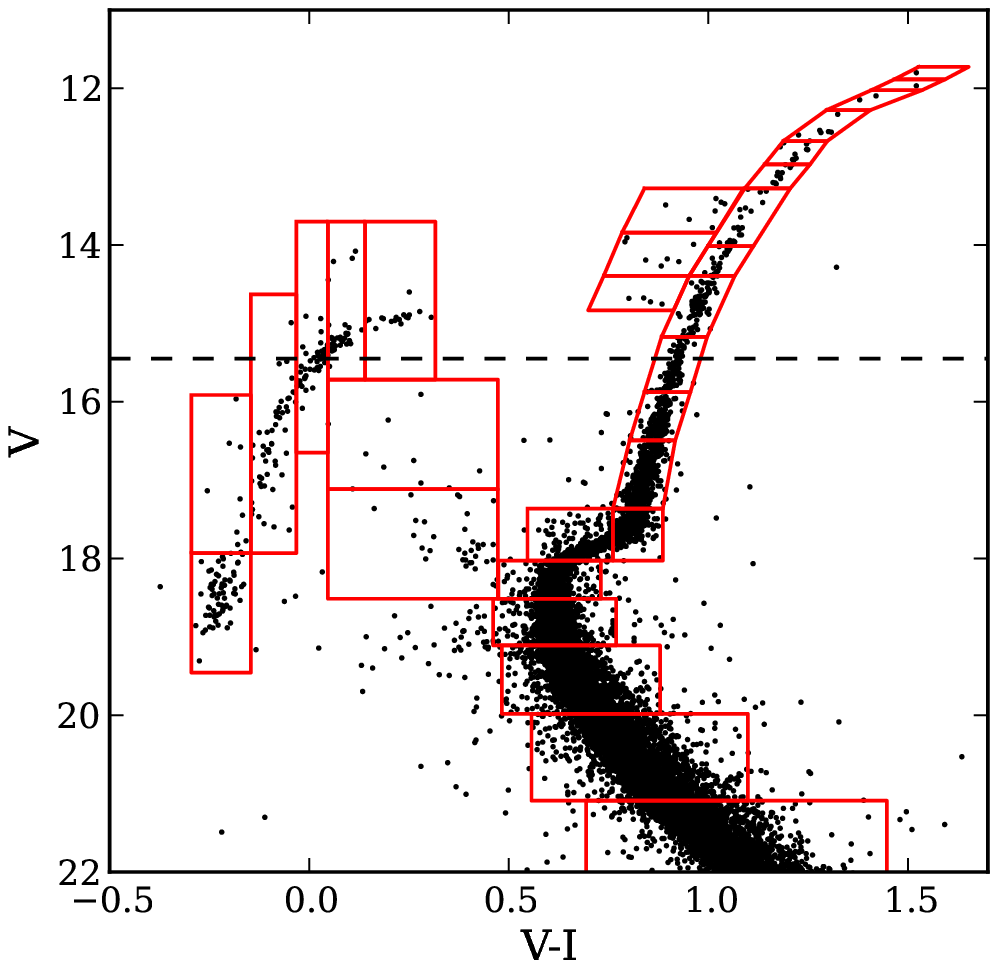}\label{fig:M13CMD}}
\subfigure[NGC7006]{\includegraphics[scale=0.6]{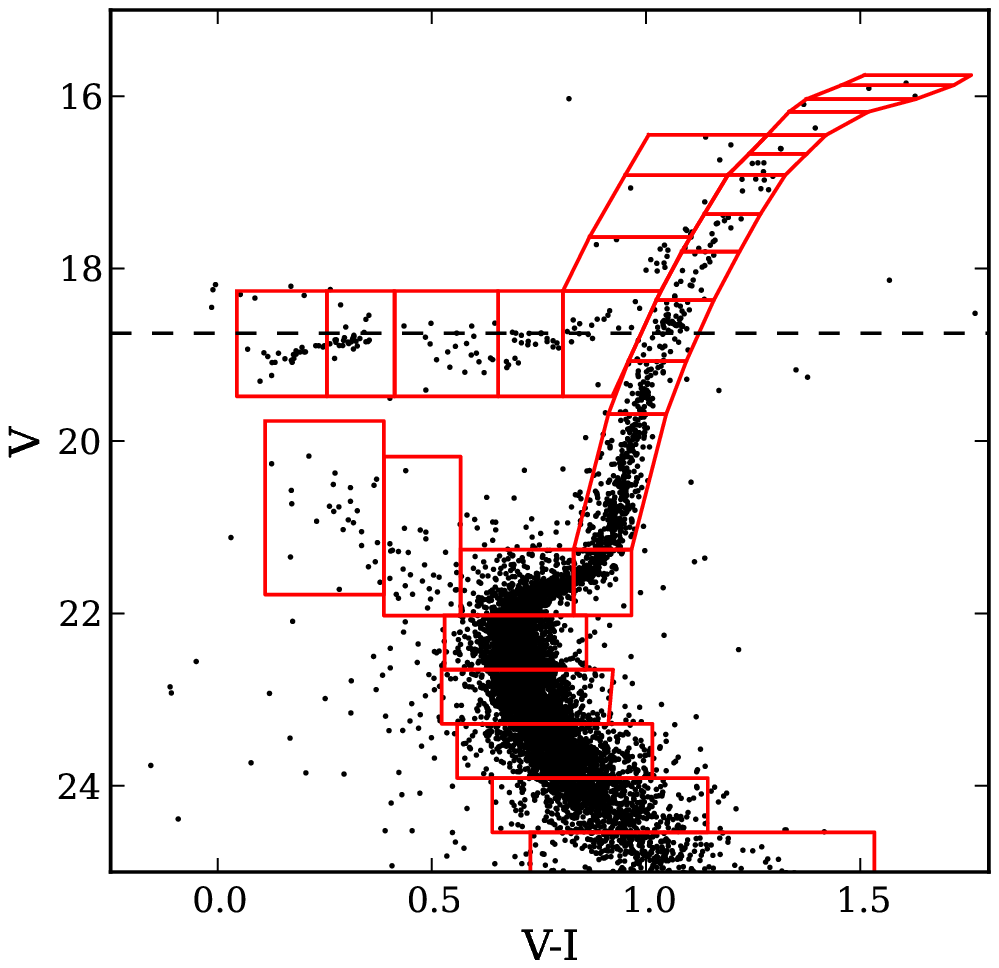}\label{fig:NGC7006CMD}}
\subfigure[M15]{\includegraphics[scale=0.6]{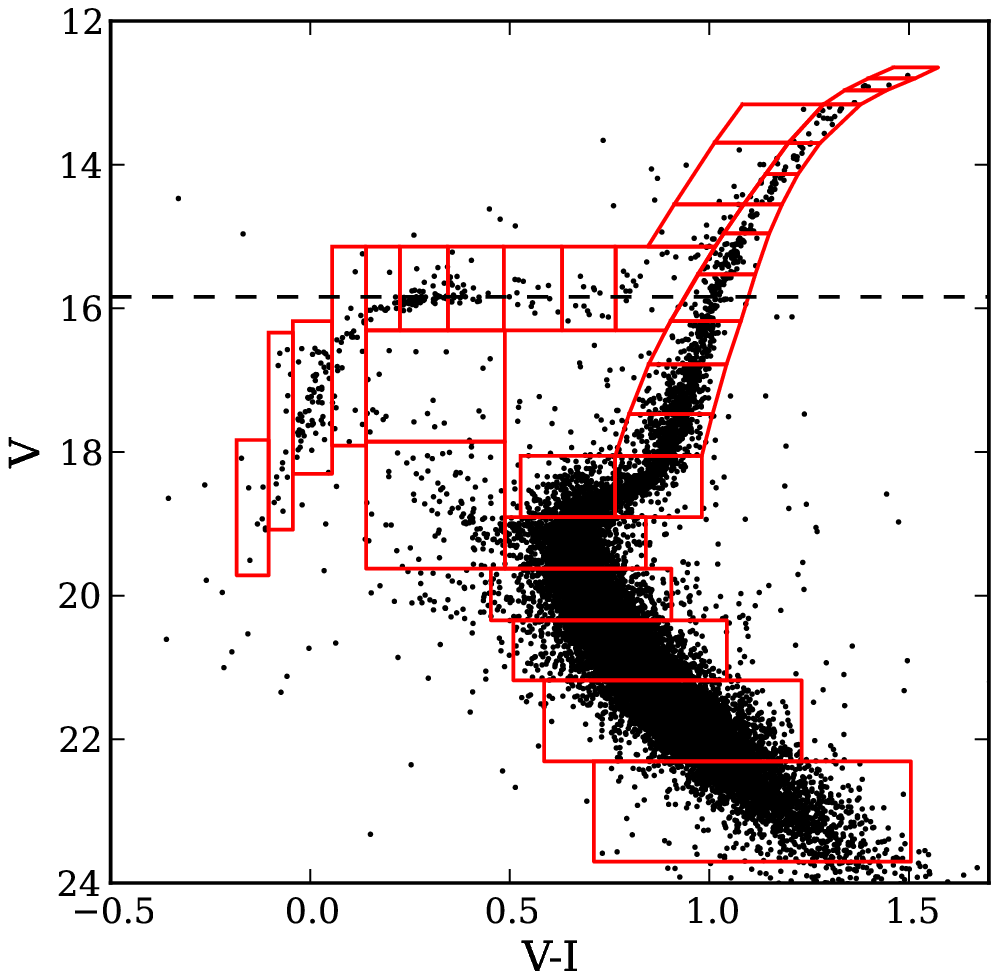}\label{fig:ComboCMD}}
\caption{Johnson's $V$, $I$ CMDs of the target clusters, from HST
\citep{Sarajedini2007,Anderson2008,Dotter2010,Dotter2011}.  The stars
have been selected based on their distances from the cluster center,
to reflect the populations sampled by the ILS.  Overlaid are the boxes
used to assign model atmospheres to the stars.  The 50\% V-band light
levels are shown by the horizontal dashed lines.  The brightest stars
that lay outside the boxes may be field stars---the possible effects
of these field stars will be investigated in a future paper, but are
not likely to have significant effect on the syntheses presented
here.\label{fig:CMDs}}
\end{center}
\end{figure*}

The CMD boxes for M3, M13, NGC 7006, and M15 are shown in Figure
\ref{fig:CMDs}, while Tables \ref{table:M3CMDBoxes} through
\ref{table:M15CMDBoxes} list the characteristics of these population
boxes (see MB08 for a figure and list of the 47~Tuc boxes).   The
stars are boxed in approximately the same way as 47~Tuc, i.e., the
number of red giant branch and main sequence boxes is approximately the
same, and the coarseness of the boxes increases down the RGB.  The
number of boxes on the HB differ between the clusters as a result of
the different HB morphologies.  A forthcoming paper (Sakari et al., in
prep.) shows that redefining the boxes has a negligible effect on the
output abundances.  The approximate evolutionary stages for each box
are also shown in Tables \ref{table:M3CMDBoxes} through
\ref{table:M15CMDBoxes}, while the 50\% $V$ magnitude light levels are
indicated by dashed lines in Figure \ref{fig:CMDs}.  This shows that
the brightest red giants, asymptotic giant branch stars, and HB stars
truly do dominate the $V$-band flux.

The number of stars drops slightly in the lower main sequence boxes.
This may be due to a combination of photometric incompleteness and
mass segregation.  Increasing the number of stars in the faintest two
boxes has an insignificant effect on the abundances, suggesting that
this effect can be neglected.

\begin{table*}
\centering
\begin{minipage}{130mm}
\caption{M3's CMD Boxes\label{table:M3CMDBoxes}}
  \begin{tabular}{@{}lcccccccc@{}}
  \hline
& $V_{0,\rm{avg}}$ & $(V-I)_{0,\rm{avg}}$ & $T_{\rm{eff}}$ (K) & $\log g$ &
$\xi$ (km/s) & $R$ (R$_{\sun}$) & $f(V)$ & $N$\\
\hline
RGB &  12.502 & 1.564 &  3995 & 0.333 & 1.87 &100.95 & 0.0141 &    1 \\
    &  12.675 & 1.450 &  4125 & 0.517 & 1.83 & 81.63 & 0.0361 &    3 \\
    &  12.754 & 1.377 &  4218 & 0.627 & 1.81 & 71.90 & 0.0224 &    2 \\
    &  12.981 & 1.334 &  4266 & 0.759 & 1.78 & 61.81 & 0.0544 &    6 \\
    &  13.332 & 1.218 &  4437 & 1.023 & 1.73 & 45.60 & 0.0394 &    6 \\
    &  13.683 & 1.164 &  4521 & 1.220 & 1.69 & 36.34 & 0.0332 &    7 \\
    &  14.270 & 1.081 &  4676 & 1.547 & 1.62 & 24.94 & 0.0761 &   28 \\
    &  14.748 & 1.014 &  4686 & 1.745 & 1.57 & 19.86 & 0.0284 &   16 \\
    &  15.398 & 0.952 &  4810 & 2.072 & 1.50 & 13.63 & 0.0712 &   74 \\
    &  16.446 & 0.879 &  4975 & 2.575 & 1.40 &  7.64 & 0.0695 &  197 \\
    &  17.470 & 0.824 &  5114 & 3.047 & 1.30 &  4.43 & 0.0419 &  296 \\
SGB &  18.185 & 0.769 &  5265 & 3.400 & 1.22 &  2.96 & 0.0252 &  338 \\
    &  18.512 & 0.619 &  6166 & 3.850 & 1.13 &  1.76 & 0.0530 &  960 \\
MS  &  18.959 & 0.568 &  6403 & 4.098 & 1.07 &  1.32 & 0.0630 & 1720 \\
    &  19.475 & 0.588 &  6336 & 4.283 & 1.03 &  1.07 & 0.0707 & 3120 \\
    &  20.147 & 0.651 &  6075 & 4.472 & 0.99 &  0.86 & 0.0552 & 4557 \\
    &  20.992 & 0.773 &  5586 & 4.647 & 0.96 &  0.70 & 0.0323 & 5928 \\
    &  22.113 & 0.999 &  4716 & 4.700 & 0.94 &  0.66 & 0.0067 & 3549 \\
AGB/HB&  13.697 & 1.052 &  4757 & 1.364 & 1.65 & 30.79 & 0.0363 &    8 \\
    &  14.471 & 0.948 &  4820 & 1.707 & 1.58 & 20.75 & 0.0362 &   16 \\
    &  15.084 & 0.861 &  5018 & 2.051 & 1.51 & 13.96 & 0.0129 &   10 \\
    &  15.468 & 0.733 &  5374 & 2.360 & 1.44 &  9.78 & 0.0170 &   19 \\
    &  15.615 & 0.599 &  5855 & 2.599 & 1.39 &  7.43 & 0.0144 &   18 \\
    &  15.718 & 0.521 &  6438 & 2.829 & 1.34 &  5.70 & 0.0080 &   11 \\
    &  15.654 & 0.436 &  6773 & 2.902 & 1.33 &  5.24 & 0.0069 &    9 \\
    &  15.610 & 0.292 &  7360 & 3.043 & 1.30 &  4.46 & 0.0097 &   12 \\
    &  15.481 & 0.179 &  7886 & 3.110 & 1.28 &  4.13 & 0.0267 &   30 \\
    &  15.529 & 0.103 &  8372 & 3.217 & 1.26 &  3.65 & 0.0111 &   13 \\
    &  15.834 & 0.028 &  9168 & 3.450 & 1.21 &  2.79 & 0.0163 &   25 \\
    &  15.953 &-0.021 & 10189 & 3.584 & 1.18 &  2.39 & 0.0045 &    8 \\
    &  17.501 &-0.144 & 15471 & 4.506 & 0.99 &  0.83 & 0.0001 &    1 \\
BS  &  18.145 & 0.355 &  7309 & 4.022 & 1.09 &  1.44 & 0.0042 &   57 \\
    &  17.356 & 0.146 &  8264 & 3.922 & 1.11 &  1.62 & 0.0030 &   20 \\
\hline
\end{tabular}
\end{minipage}\\
\medskip
\raggedright The average $V$ and $(V-I)$ colors of each
box are shown, along with the average effective temperature, surface
gravity, microturbulence, radius, fractional $V$-band flux, and
number of stars assigned to each box.  The different evolutionary
stages of the boxes are also shown, where RGB stands for Red Giant
Branch, SGB for Subgiant Branch, MS for Main Sequence, AGB for
Asymptotic Giant Branch, HB for Horizontal Branch, and BS for Blue
Stragglers.  Note that the 50\% light level for M3 is $V_{1/2} =
15.65$, which reaches the middle of the HB.
\end{table*}

\begin{table*}
\centering
\begin{minipage}{130mm}
\caption{M13's CMD Boxes\label{table:M13CMDBoxes}}
  \begin{tabular}{@{}lcccccccc@{}}
  \hline
& $V_{0,\rm{avg}}$ & $(V-I)_{0,\rm{avg}}$ & $T_{\rm{eff}}$ (K) & $\log g$ &
$\xi$ (km/s) & $R$ (R$_{\sun}$) & $f(V)$ & $N$\\
  \hline
RGB &  11.739 &  1.494 &  4083. & 0.400 & 1.86 &  93.39 & 0.0158 &    1 \\
    &  11.907 &  1.494 &  4070. & 0.458 & 1.85 &  87.35 & 0.0136 &    1 \\
    &  12.059 &  1.373 &  4221. & 0.649 & 1.81 &  70.16 & 0.0236 &    2 \\
    &  12.474 &  1.257 &  4377. & 0.933 & 1.75 &  50.57 & 0.0561 &    7 \\
    &  12.740 &  1.192 &  4477. & 1.110 & 1.71 &  41.23 & 0.0565 &    9 \\
    &  13.035 &  1.155 &  4541. & 1.269 & 1.67 &  34.36 & 0.0526 &   11 \\
    &  13.601 &  1.061 &  4717. & 1.600 & 1.60 &  23.46 & 0.0556 &   20 \\
    &  14.123 &  1.003 &  4707. & 1.803 & 1.56 &  18.58 & 0.0350 &   20 \\
    &  14.681 &  0.952 &  4809. & 2.081 & 1.50 &  13.48 & 0.0808 &   78 \\
    &  15.469 &  0.890 &  4948. & 2.467 & 1.42 &   8.65 & 0.0506 &  101 \\
    &  16.134 &  0.852 &  5040. & 2.776 & 1.35 &   6.06 & 0.0369 &  135 \\
    &  16.839 &  0.816 &  5133. & 3.100 & 1.29 &   4.17 & 0.0486 &  346 \\
SGB &  17.525 &  0.779 &  5247. & 3.424 & 1.22 &   2.87 & 0.0196 &  257 \\
    &  17.809 &  0.662 &  5987. & 3.808 & 1.13 &   1.85 & 0.0357 &  608 \\
MS  &  18.213 &  0.585 &  6329. & 4.074 & 1.08 &   1.36 & 0.0677 & 1676 \\
    &  18.738 &  0.580 &  6369. & 4.295 & 1.03 &   1.05 & 0.0805 & 3249 \\
    &  19.441 &  0.635 &  6146. & 4.507 & 0.99 &   0.83 & 0.0812 & 6352 \\
    &  20.367 &  0.762 &  5633. & 4.710 & 0.94 &   0.65 & 0.0478 & 8903 \\
    &  21.594 &  1.007 &  4699. & 4.778 & 0.93 &   0.60 & 0.0158 & 9561 \\
AGB &  13.473 &  0.969 &  4776. & 1.581 & 1.61 &  23.99 & 0.0223 &    7 \\
    &  14.022 &  0.841 &  5069. & 1.947 & 1.53 &  15.74 & 0.0134 &    7 \\
    &  14.646 &  0.817 &  5133. & 2.224 & 1.47 &  11.43 & 0.0044 &    4 \\
HB  &  14.851 &  0.194 &  7816. & 3.139 & 1.28 &   3.99 & 0.0153 &   17 \\
    &  14.879 &  0.054 &  8836. & 3.322 & 1.24 &   3.23 & 0.0261 &   32 \\
    &  15.389 & -0.013 &  9951. & 3.649 & 1.17 &   2.22 & 0.0247 &   46 \\
    &  16.169 & -0.105 & 13400. & 4.157 & 1.06 &   1.24 & 0.0122 &   53 \\
    &  16.745 & -0.214 & 20818. & 4.733 & 0.94 &   0.64 & 0.0013 &   10 \\
    &  18.333 & -0.247 & 24467. & 5.000 & 0.78 &   0.27 & 0.0024 &   67 \\
BS  &  16.465 &  0.193 &  7980. & 3.809 & 1.13 &   1.85 & 0.0019 &   10 \\
    &  17.656 &  0.341 &  7408. & 4.144 & 1.06 &   1.25 & 0.0022 &   34 \\
\hline
\end{tabular}
\end{minipage}\\
\medskip
\raggedright The 50\% light level for M13 is $V_{1/2} = 15.45$, which is
slightly below the reddest HB stars.
\end{table*}

\begin{table*}
\centering
\begin{minipage}{130mm}
\caption{NGC 7006's CMD Boxes\label{table:NGC7006CMDBoxes}}
  \begin{tabular}{@{}lcccccccc@{}}
  \hline
& $V_{0,\rm{avg}}$ & $(V-I)_{0,\rm{avg}}$ & $T_{\rm{eff}}$ (K) & $\log g$ &
$\xi$ (km/s) & $R$ (R$_{\sun}$) & $f(V)$ & $N$\\
\hline
RGB &  15.694 & 1.539 &  4022. &  0.371 &  1.87 &  96.61 & 0.0197 &    1 \\
    &  15.800 & 1.504 &  4058. &  0.445 &  1.85 &  88.69 & 0.0357 &    2 \\
    &  15.941 & 1.300 &  4333. &  0.726 &  1.79 &  64.21 & 0.0157 &    1 \\
    &  16.215 & 1.327 &  4271. &  0.793 &  1.78 &  59.40 & 0.0122 &    1 \\
    &  16.455 & 1.247 &  4389. &  0.975 &  1.74 &  48.20 & 0.0195 &    2 \\
    &  16.645 & 1.196 &  4469. &  1.107 &  1.71 &  41.40 & 0.0327 &    4 \\
    &  16.855 & 1.194 &  4465. &  1.188 &  1.69 &  37.69 & 0.0472 &    7 \\
    &  17.400 & 1.096 &  4644. &  1.517 &  1.62 &  25.82 & 0.0568 &   14 \\
    &  17.896 & 1.049 &  4735. &  1.768 &  1.57 &  19.33 & 0.0460 &   18 \\
    &  18.517 & 0.987 &  4738. &  2.018 &  1.52 &  14.50 & 0.0895 &   62 \\
    &  19.211 & 0.930 &  4856. &  2.357 &  1.44 &   9.81 & 0.0465 &   61 \\
    &  20.276 & 0.876 &  4981. &  2.845 &  1.34 &   5.59 & 0.0871 &  328 \\
SGB &  21.312 & 0.808 &  5156. &  3.339 &  1.23 &   3.17 & 0.0196 &  177 \\
    &  21.607 & 0.663 &  5978. &  3.765 &  1.14 &   1.94 & 0.0664 &  792 \\
MS  &  22.181 & 0.627 &  6151. &  4.048 &  1.08 &   1.40 & 0.0748 & 1516 \\
    &  22.774 & 0.660 &  6022. &  4.243 &  1.04 &   1.12 & 0.0491 & 1722 \\
    &  23.392 & 0.722 &  5774. &  4.409 &  1.01 &   0.92 & 0.0237 & 1467 \\
    &  23.998 & 0.812 &  5439. &  4.530 &  0.98 &   0.80 & 0.0070 &  759 \\
    &  24.733 & 0.955 &  4804. &  4.534 &  0.98 &   0.80 & 0.0021 &  454 \\
AGB &  16.429 & 1.101 &  4668. &  1.141 &  1.70 &  39.81 & 0.0299 &    3 \\
    &  17.233 & 1.008 &  4697. &  1.481 &  1.63 &  26.91 & 0.0281 &    6 \\
    &  17.694 & 0.941 &  4833. &  1.739 &  1.57 &  19.99 & 0.0372 &   12 \\
HB  &  18.496 & 0.823 &  5115. &  2.197 &  1.48 &  11.81 & 0.0265 &   18 \\
    &  18.718 & 0.663 &  5612. &  2.489 &  1.42 &   8.43 & 0.0266 &   22 \\
    &  18.747 & 0.493 &  6542. &  2.809 &  1.35 &   5.83 & 0.0244 &   21 \\
    &  18.638 & 0.244 &  7564. &  3.040 &  1.30 &   4.47 & 0.0363 &   28 \\
    &  18.713 & 0.093 &  8451. &  3.240 &  1.26 &   3.55 & 0.0308 &   26 \\
BS  &  21.184 & 0.407 &  7063. &  3.912 &  1.11 &   1.63 & 0.0039 &   33 \\
    &  20.583 & 0.209 &  7923. &  3.883 &  1.12 &   1.69 & 0.0053 &   26 \\
\hline
\end{tabular}
\end{minipage}\\
\medskip
\raggedright The 50\% light level is $V_{1/2} = 18.75$, i.e. in the
middle of the HB.
\end{table*}

\begin{table*}
\centering
\begin{minipage}{130mm}
\caption{M15's CMD Boxes\label{table:M15CMDBoxes}}
  \begin{tabular}{@{}lcccccccc@{}}
  \hline
& $V_{0,\rm{avg}}$ & $(V-I)_{0,\rm{avg}}$ & $T_{\rm{eff}}$ (K) & $\log g$ &
$\xi$ (km/s) & $R$ (R$_{\sun}$) & $f(V)$ & $N$\\
  \hline
RGB &  12.452 &   1.361 &  4349. & 0.446 & 1.85 &  88.58 & 0.0107 &       1 \\
    &  12.599 &   1.271 &  4464. & 0.594 & 1.82 &  74.73 & 0.0373 &       4 \\
    &  12.830 &   1.228 &  4513. & 0.724 & 1.79 &  64.33 & 0.0151 &       2 \\
    &  13.061 &   1.157 &  4617. & 0.884 & 1.76 &  53.51 & 0.0967 &      16 \\
    &  13.569 &   1.082 &  4735. & 1.160 & 1.70 &  38.93 & 0.0532 &      14 \\
    &  14.008 &   1.018 &  4710. & 1.323 & 1.66 &  32.29 & 0.0532 &      21 \\
    &  14.484 &   0.967 &  4805. & 1.566 & 1.61 &  24.41 & 0.0278 &      17 \\
    &  14.941 &   0.921 &  4898. & 1.796 & 1.56 &  18.73 & 0.0706 &      66 \\
    &  15.503 &   0.880 &  4987. & 2.064 & 1.51 &  13.75 & 0.0487 &      77 \\
    &  16.136 &   0.843 &  5073. & 2.356 & 1.44 &   9.83 & 0.0383 &     108 \\
    &  16.814 &   0.800 &  5180. & 2.673 & 1.38 &   6.82 & 0.0518 &     274 \\
    &  17.448 &   0.766 &  5270. & 2.963 & 1.31 &   4.88 & 0.0342 &     323 \\
SGB &  18.069 &   0.702 &  5460. & 3.285 & 1.25 &   3.37 & 0.0344 &     576 \\
    &  18.366 &   0.556 &  5985. & 3.588 & 1.18 &   2.38 & 0.0427 &     939 \\
MS  &  18.947 &   0.520 &  6612. & 4.009 & 1.09 &   1.46 & 0.0863 &    3261 \\
    &  19.649 &   0.554 &  6495. & 4.252 & 1.04 &   1.11 & 0.0621 &    4473 \\
    &  20.399 &   0.633 &  6164. & 4.451 & 1.00 &   0.88 & 0.0414 &    5994 \\
    &  21.285 &   0.770 &  5617. & 4.627 & 0.96 &   0.72 & 0.0183 &    6062 \\
    &  22.352 &   0.971 &  4797. & 4.715 & 0.94 &   0.65 & 0.0021 &    1886 \\
AGB/HB &  12.923 &   1.100 &  4721. & 0.892 & 1.75 &  53.02 & 0.0069 &       1 \\
    &  13.840 &   0.962 &  4814. & 1.312 & 1.67 &  32.69 & 0.0260 &       9 \\
    &  14.632 &   0.864 &  5022. & 1.733 & 1.58 &  20.14 & 0.0171 &      12 \\
    &  15.168 &   0.740 &  5345. & 2.087 & 1.50 &  13.39 & 0.0171 &      20 \\
    &  15.570 &   0.562 &  5959. & 2.474 & 1.42 &   8.58 & 0.0071 &      12 \\
    &  15.490 &   0.414 &  6832. & 2.716 & 1.37 &   6.49 & 0.0084 &      13 \\
    &  15.399 &   0.244 &  7537. & 2.864 & 1.34 &   5.47 & 0.0174 &      25 \\
    &  15.520 &   0.147 &  8051. & 3.021 & 1.30 &   4.57 & 0.0383 &      61 \\
    &  15.637 &   0.052 &  8853. & 3.195 & 1.26 &   3.74 & 0.0090 &      16 \\
    &  15.966 &  -0.032 & 10589. & 3.491 & 1.20 &   2.66 & 0.0101 &      28 \\
    &  16.613 &  -0.111 & 13883. & 3.958 & 1.10 &   1.55 & 0.0073 &      33 \\
    &  16.661 &  -0.199 & 19831. & 4.232 & 1.04 &   1.13 & 0.0016 &       8 \\
    &  18.304 &  -0.273 & 23306. & 4.983 & 0.88 &   0.48 & 0.0003 &       7 \\
BS  &  16.809 &   0.143 &  8198. & 3.557 & 1.19 &   2.47 & 0.0038 &      21 \\
    &  18.375 &   0.228 &  7923. & 4.115 & 1.07 &   1.30 & 0.0046 &     110 \\
\hline
\end{tabular}
\end{minipage}\\
\medskip
\raggedright The 50\% light level for M15 is $V_{1/2} = 15.84$, which runs
through the red HB stars.
\end{table*}

\subsubsection{Metallicities}\label{subsubsec:Metallicities}
The iron abundances were determined through an integrated light
equivalent width (EW) analysis, as described in MB08.  The EWs were
measured with a modified version of the automated program
DAOSPEC\footnote{DAOSPEC has been written by P.B. Stetson for the
Dominion Astrophysical Observatory of the Herzberg Institute of
Astrophysics, National Research Council, Canada.} \citep{DAOSPECref},
as described below.  The Fe lines used in the analysis, the atomic
data, and the measured EWs (in the Sun, Arcturus, and the target GCs)
are shown in Table  \ref{table:LineList}.  These lines are from the
ILS line lists in MB08 and \citet{Colucci2009}, with additional RGB
lines from \citet{Sakari2011} and \citet{Venn2012}.  Note that the
lines were restricted to the wavelength region covered by the HET
spectra---thus, 47~Tuc has fewer lines than in MB08.  Lines stronger
than 150 m\AA \hspace{0.025in} were not included in the Fe abundance
analysis (see \citealt{McWilliam1995b}).

\begin{table*}
\centering
\begin{minipage}{150mm}
\caption{The Fe Line List.$^{a}$\label{table:LineList}}
  \begin{tabular}{@{}lcccccccccc@{}}
  \hline
Wavelength & Element & E.P. & log gf & \multicolumn{7}{l}{Equivalent width (m\AA)}\\
(\AA) & & (eV) & & Sun & Arcturus & 47~Tuc & M3 & M13 & NGC~7006 &
M15\\
  \hline
5324.191 & \ion{Fe}{1} & 3.211 & -0.103 &  -    & -     & -     & 112.0 & 97.2  & 103.0 & 38.6  \\
5339.937 & \ion{Fe}{1} & 3.266 & -0.72  & 176.0 & -     & -     & 80.1  & 75.1  & 87.0  & -     \\
5367.476 & \ion{Fe}{1} & 4.415 &  0.443 & 177.0 & 142.3 & 114.0 & 70.0  & 68.0  & 78.0  & 22.9  \\
5369.974 & \ion{Fe}{1} & 4.371 &  0.536 &  -    & 144.0 & 147.0 & 72.5  & 73.2  & 75.1  & -     \\
5371.501 & \ion{Fe}{1} & 0.958 & -1.644 &  -    & -     & -     & 183.0 & 167.6 & 186.0 & 108.6 \\
\hline
\end{tabular}
\end{minipage}\\
\medskip
\raggedright Equivalent widths were measured in DAOSPEC; all
strong lines were checked and refined in \textit{splot}.  The lines
that were not measured in the Solar spectrum were those stronger than
EW $> 150$ m\AA.\\
$^{a}$Table \ref{table:LineList} is published in its entirety in the
electronic edition of \textit{MNRAS}. A portion is shown here for
guidance regarding its form and content.\\
\end{table*}

\paragraph{DAOSPEC and ILS}
DAOSPEC is intended for high resolution ($R>15000$), high SNR ($>30$)
spectra \citep{DAOSPECref}; to our knowledge, this is its first
application to ILS.  The program iteratively fits a continuum to the
input spectrum, detects the spectral lines in a defined region, and
fits Gaussian profiles to the lines, providing their EWs.  One major
advantage of DAOSPEC over hand-measurements with  IRAF's
\textit{splot} is the enforcement of a fixed full-width at half
maximum (FWHM) for all lines.\footnote{Though of course this fixed
FWHM, $\Delta \lambda$, is allowed to scale appropriately with
wavelength, given the constant resolving power of the spectrum,
$R~=~\lambda/\Delta\lambda$.} In individual stars, the FWHM is often
dominated by the instrumental broadening of the spectrograph; in ILS,
however, the velocity dispersion often dominates the line FWHM,
blending together many of the spectral lines.  To ensure that DAOSPEC
can detect these features, the program was modified to detect lines
that are separated by a least half of a FWHM (rather than the default
separation of one FWHM).

The success of DAOSPEC seems highly dependent on the choice of input
parameters, at least for ILS.  DAOSPEC has the option to determine its
own FWHM.  However, because ILS contain so many weak and/or blended
features, the FWHM was fixed to the input parameter.  This input FWHM
is the sum (in quadrature) of the broadening from the velocity
dispersion, the spectrograph, and the intrinsic broadening of Arcturus
(see Section \ref{subsec:DataReduction}), and is given in pixels.
This FWHM was then allowed to scale with wavelength.  Another
important input is the order of the polynomial-fit to the continuum.
For individual stars, \citet{DAOSPECref} found that higher-order
continuum fits produced the best results.  Again, however, ILS suffer
from contributions from a variety of stars, severe blends, and line
blanketing; high-order fits could therefore remove real features from
the spectra, leading to underestimates of the line strengths.  As
described in Section \ref{sec:Observations}, the HET spectra presented
here have been carefully and conservatively normalized---thus the
continuum was unchanged from the input.  Occasionally valuable lines were
mis-measured by DAOSPEC; these lines were remeasured with Gaussian
fits in IRAF's \textit{splot} task.

\begin{figure*}
\begin{center}
\centering
\subfigure[DAOSPEC vs. \textit{splot}]{\includegraphics[scale=0.46]{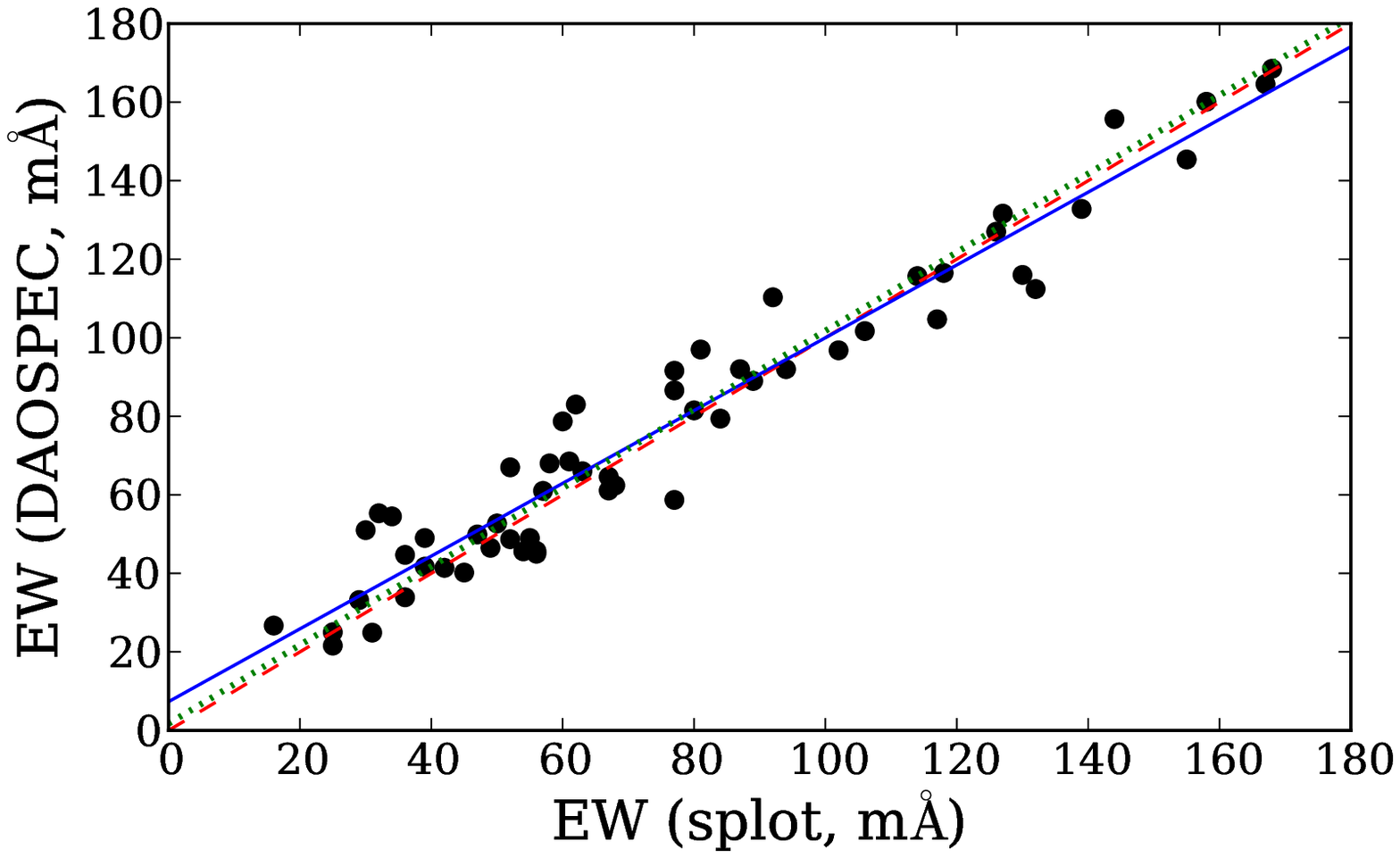}\label{fig:DAOSPECvsSplot}}
\subfigure[DAOSPEC vs. GETJOB]{\includegraphics[scale=0.46]{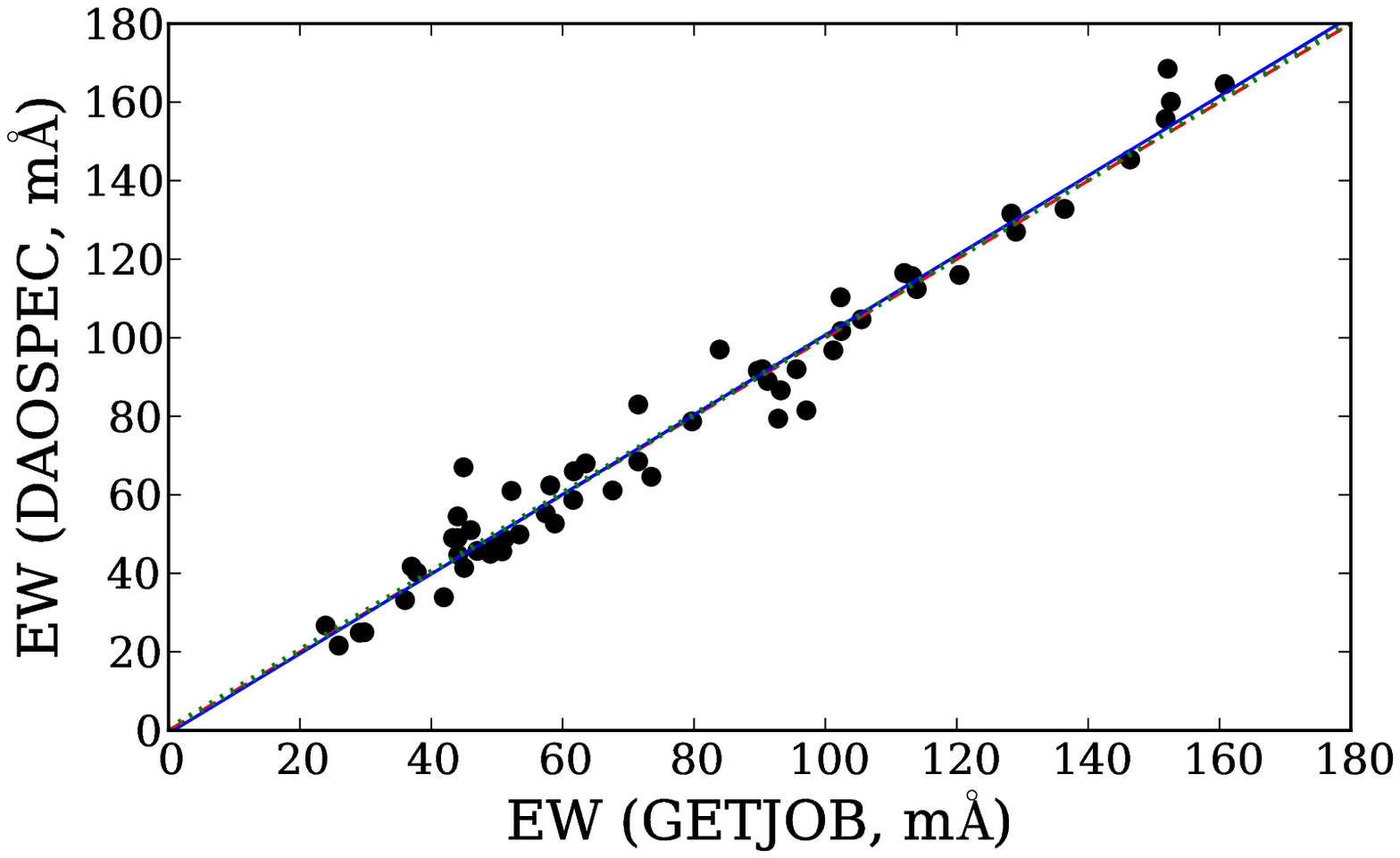}\label{fig:DAOSPECvsGETJOB}}
\caption{Comparisons of EW measurements on the ILS of 47 Tuc for the
Fe lines in common between the studies.  Figure
\ref{fig:DAOSPECvsSplot} compares the DAOSPEC measurements to those
from \textit{splot}, while Figure \ref{fig:DAOSPECvsGETJOB} compares
measurements from DAOSPEC and GETJOB (from MB08).  Each black point is
a separate spectral line.  The dashed red lines show perfect
agreement, while the dotted green lines show the average offset.  The
solid blue lines are linear least squares fits to the points.  The
agreement is excellent in both cases.\label{fig:DAOSPECvsThem}}
\end{center}
\end{figure*}

Previous tests have shown that with the correct input parameters
DAOSPEC is capable of reproducing \textit{splot}-measured EWs (in
IRAF) for individual stars \citep{DAOSPECref,Sakari2011}.  DAOSPEC
is also capable of reproducing \textit{splot} EWs in the ILS of
47 Tuc. Figure \ref{fig:DAOSPECvsSplot} shows that for Fe lines the
agreement between DAOSPEC and \textit{splot} EWs is excellent, with an
average offset of only 1.76 m\AA, though the scatter about the average
is significant ($\pm 9.93$ m\AA).  This scatter is likely caused by
blends in the ILS, which can be difficult to detect by eye.  DAOSPEC
also reproduces the EWs of MB08, which were measured with the program
GETJOB \citep{McWilliam1995a}.  The offset between the datasets remains
insignificant (0.37 m\AA, with the DAOSPEC EWs slightly higher), and
the scatter is smaller ($\pm 6.69$ m\AA).  Note that MB08 did not
combine their apertures; the spectral lines in the overlapping regions
were measured twice, and both measurements were included in their
analysis.  For this EW comparison the two EWs were averaged together
for all lines in the overlapping regions.  In individual stars,
DAOSPEC often underestimates the EWs of the strongest lines
(e.g. \citealt{Sakari2011})---thus, the EWs of these lines have been
verified or corrected in \textit{splot}.

DAOSPEC is therefore capable of accurately reproducing the EWs of the
lines in the 47 Tuc ILS.  Since 47 Tuc has a higher velocity
dispersion than most of the targets in this work (except for M15; see
Table \ref{table:Observations}), DAOSPEC should also be able to
accurately measure the EWs for all the target GCs.  However, care must
be taken with the strongest lines, to ensure they are properly
measured, and attention must be paid to lines in regions with
uncertain continuum normalization.

\paragraph{Fe abundances}
The \ion{Fe}{1} and \ion{Fe}{2} EWs for the target GCs were input into
ILABUNDS, along with the model atmospheres derived from the HST
photometry. The [Fe/H] values for each line were computed
differentially (i.e. relative to the derived Solar abundance
\textit{for that line}), in order to reduce uncertainties from, e.g.,
atomic data.  If the Solar lines were stronger than EW $\ga 200$
m\AA, a value of $\log \epsilon_{\rm{Fe},\sun} = 7.50$
\citep{Asplund2009} was instead adopted.  The average [\ion{Fe}{1}/H]
and [\ion{Fe}{2}/H] ratios for Arcturus and the target GCs are shown
in Table \ref{table:Fe}.  Also shown are the uncertainties in the
average Fe abundances, as determined from the line-to-line scatter,
$\sigma$, divided by $\sqrt{N}$, where $N$ is the number of lines.
Since only a single line is available for M15's \ion{Fe}{2} abundance,
the \citet{Cayrel} formula was used to determine the error in the
\ion{Fe}{2} line's EW, as described in \citet{Sakari2011}.  The
adopted uncertainty in abundance was then found by rerunning ILABUNDS
with the new EWs.

\begin{table*}
\centering
\begin{minipage}{200mm}
\caption{Fe Abundances.\label{table:Fe}}
  \begin{tabular}{@{}lcccccccccccc@{}}
  \hline
 & \multicolumn{2}{c}{Arcturus} & \multicolumn{2}{c}{47 Tuc} & \multicolumn{2}{c}{M3} & \multicolumn{2}{c}{M13} & \multicolumn{2}{c}{NGC 7006} & \multicolumn{2}{c}{M15} \\
 & [Fe/H] & $N$ & [Fe/H] & $N$ & [Fe/H] & $N$ & [Fe/H] & $N$ & [Fe/H] & $N$ & [Fe/H] & $N$ \\
$[$\ion{Fe}{1}/H$]$  & $-0.65 \pm 0.02$ & 91& $-0.81\pm0.02$& 68 & $-1.48\pm0.03$ & 100 & $-1.53\pm0.02$ & 72 & $-1.46\pm0.03$ & 76 & $-2.30\pm0.03$ & 32\\
$[$\ion{Fe}{2}/H$]$ & $-0.60 \pm 0.04$ & 5 & $-0.69\pm0.04$ & 4  & $-1.59\pm0.03$ &  5  & $-1.56\pm0.02$ &  3 & $-1.57\pm0.03$ & 6  & $-2.38\pm0.10$ & 1\\
& & & & & & & & & & & & \\
\textit{MB08} $[$\ion{Fe}{1}/H$]$& - & & $-0.75\pm0.03$ & & - & & - & & - & & - & \\
\textit{MB08} $[$\ion{Fe}{2}/H$]$& - & & $-0.72\pm0.06$ & & - & & - & & - & & - &\\
\textit{Literature} [Fe/H] & $-0.58\pm 0.03$ & & $-0.72$ & & $-1.50$ & & $-1.53$ & & $-1.52$ & &$-2.37$&\\
\hline
\end{tabular}
\end{minipage}\\
\medskip
\raggedright {\bf References: } Literature values are from
\citet{Yong} and \citet{Harris} for Arcturus and the target GCs,
respectively.  The MB08 values for 47~Tuc are also shown.\\
\end{table*}

As stated earlier, the Arcturus values agree well with \citet{Yong}.
The GC [Fe/H] values in Table \ref{table:Fe} agree quite well with the
literature values quoted in \citet{Harris}, which are also shown in
Table \ref{table:Fe}.  The 47~Tuc integrated light [\ion{Fe}{1}/H]
values are slightly different from the ILS values from MB08 (which are
also shown in Table \ref{table:Fe}), though the 47~Tuc integrated
light [\ion{Fe}{2}/H] ratios are in excellent agreement with MB08.
Given that the EWs (Figure \ref{fig:DAOSPECvsGETJOB}) and CMD boxes
are nearly identical, and since both [\ion{Fe}{1}/H] ratios have been
computed differentially, it is puzzling that the [\ion{Fe}{1}/H]
values are not in agreement.  The two $\log \epsilon_{\rm{Fe}}$ values
\textit{are} actually in agreement ($\log \epsilon_{\rm{Fe}} = 6.73$
compared to MB08's 6.77)---this suggests that there are differences in
how the differential ratios are computed.

For the remainder of this paper, all quoted [X/Fe] values
are computed with the integrated light Fe abundances derived here.

\subsubsection{Hot Stars}\label{subsubsec:HotStars}
Stars hotter than $\sim 8000$ K are expected to show chemical
abundance variations as a result of radiative levitation which may
bring the hottest stars up to Solar composition
(e.g. \citealt{Behr2000,Behr2003,Lovisi2012}).  To test this result,
syntheses of M13 were completed with Solar composition hot stars.
These tests are discussed in more detail in a forthcoming paper; here
it is simply noted that these abundance variations have a negligible
effect on the syntheses of the lines studied here.

Hot stars have also been observed to have large rotational velocities
(up to $\sim 60$ km/s; \citealt{Behr2003}).  Syntheses were also
performed on M13, assuming all stars hotter than 8000 K have Solar
composition \textit{and} rotational velocities of $v\sin i = 60$
km/s.  The effects are again negligible, since the increased
broadening weakens the line strengths in HB stars.  The changes in
composition and the high rotation of HB stars are therefore ignored in
this analysis.

\subsection{Input Line Lists}\label{subsec:LineList}
To demonstrate the necessity of having a complete, calibrated line
list, spectrum syntheses were performed with three different line
lists.
\begin{description}
\item[Minimal List: ] This list uses only lines found in standard RGB
equivalent width analyses.  The ILS-specific line lists from MB08 and
\citet{Colucci2009} were supplemented with additional RGB lines
(from the line lists assembled in \citealt{Sakari2011} and
\citealt{Venn2012}.)
\item[VALD List: ] This list consists of RGB lines from the Vienna
Atomic Line
Database\footnote{\url{http://www.astro.uu.se/~vald/php/vald.php}}
(VALD; \citealt{Kupka2000}).  These are lines that VALD has determined
would appear (to at least 2\%) in a tip of the RGB star at 47~Tuc's
metallicity.  None of the atomic data were changed from the VALD
values.
\item[Final List: ] The Final Line List consists of lines from the
Minimal List, with supplements from VALD for the coolest RGB stars,
warmer RGB stars, and hot stars, all at 47~Tuc's metallicity.  Lines
that should appear in the Solar spectrum were also included.  As shown
below, the atomic data from VALD are not capable of reproducing the
strengths or profiles of all the lines in the Solar and Arcturus
spectra.  Thus, atomic data (i.e. $\log gf$ values, damping
parameters, and wavelengths) were checked in both the
Kurucz\footnote{\url{http://kurucz.harvard.edu/linelists.html}} and
National Institute of Standards and Technology
(NIST)\footnote{\url{http://www.nist.gov/index.html}} databases.
These atomic data were then adjusted so that the Solar and Arcturus
spectra were accurately reproduced in the syntheses.  The Final Line
List also contains molecular lines from the Kurucz database, as
described in Section \ref{subsubsec:Molecules}.
\end{description}

\subsubsection{Hyperfine Structure}\label{subsubsec:HFS}
Hyperfine structure (HFS) occasionally affects lines in the
synthesized spectral regions. The HFS components are not included at
this time (except for the case of the \ion{Eu}{2} line in Section
\ref{subsec:Eu}) though their presence is noted.  All regions with
possible HFS components were ignored when determining continuum
levels.  In the future these components can also be incorporated into
the syntheses.

\subsubsection{Molecular Lines}\label{subsubsec:Molecules}
As discussed above, spectral lines from several important molecules
(e.g., CH, CN, and MgH) are also included in the Final List when the
lines were detected in the Arcturus Atlas \citep{Hinkle2003}.  The
MOOG 1997 default values for the molecular equilibrium calculations
were employed, with the exception of the MgH dissociation energy, for
which the MOOG 2010 value was adopted.

\begin{figure*}
\begin{center}
\centering
\subfigure[\ion{Mg}{1} 5528 \AA \hspace{0.025in} region]{\includegraphics[scale=0.46]{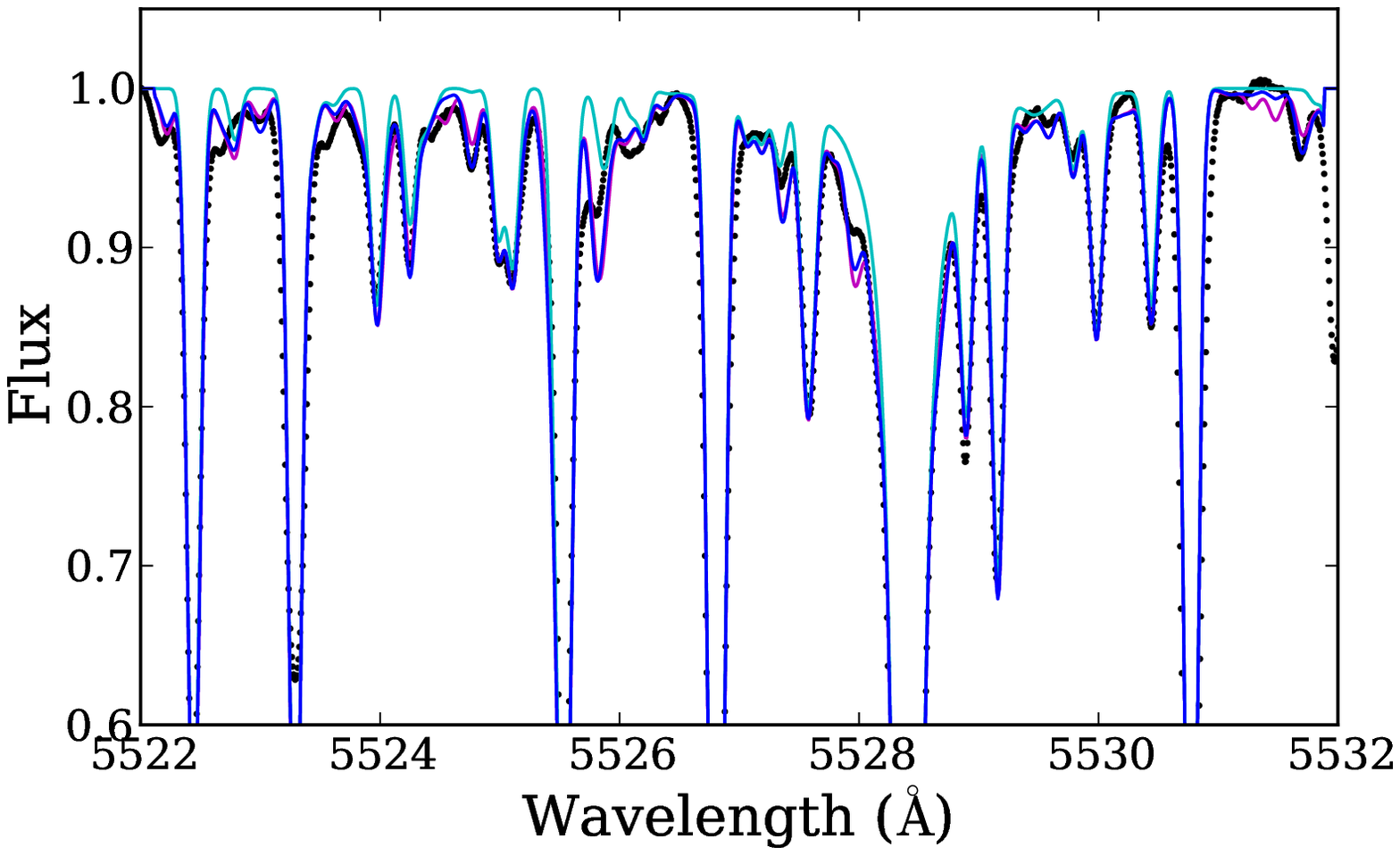}\label{fig:ArcturusMgIso}}
\subfigure[\ion{Eu}{2} 6645 \AA \hspace{0.025in} region]{\includegraphics[scale=0.46]{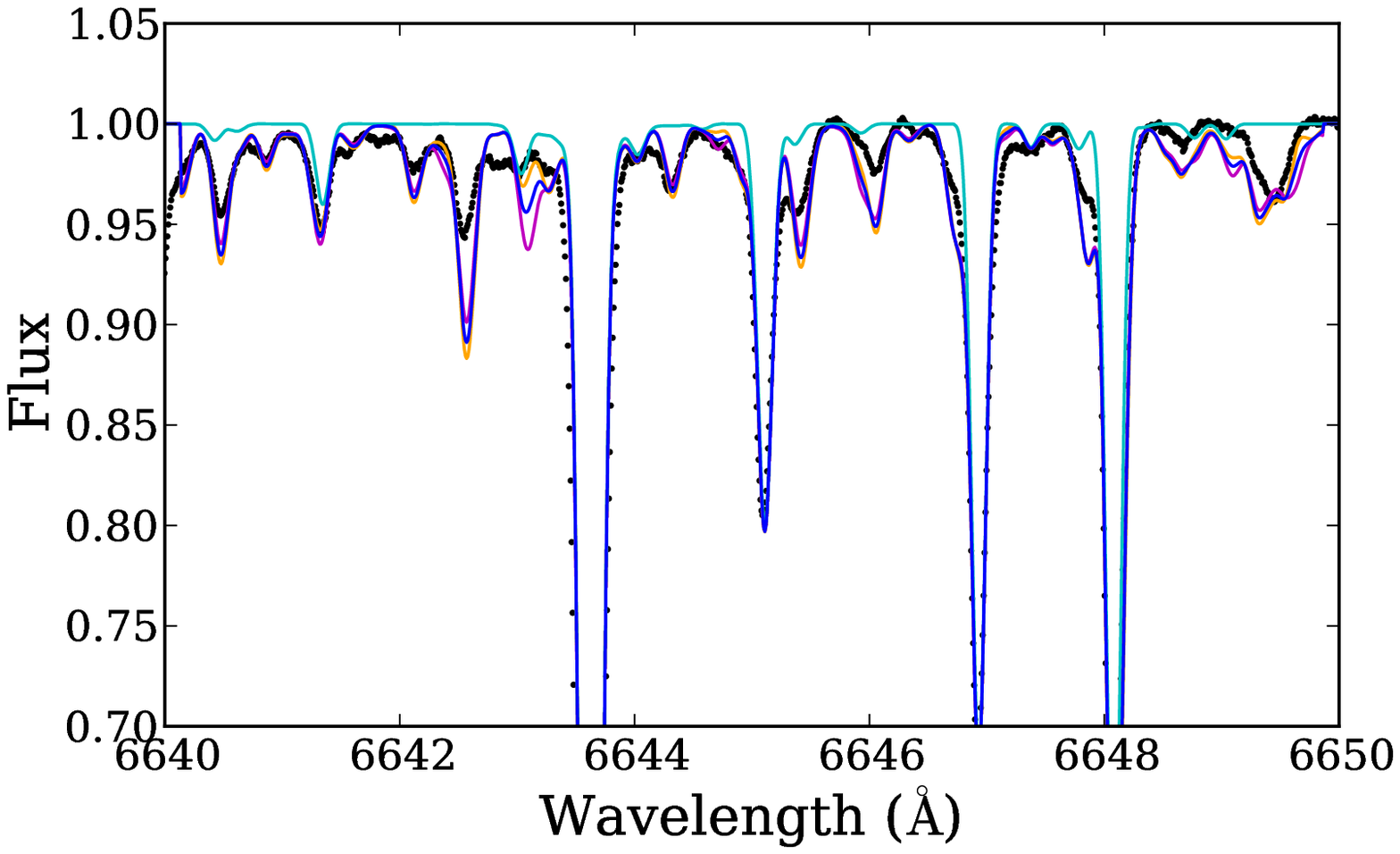}\label{fig:ArcturusEuIso}}
\caption{Syntheses of the Arcturus spectral regions around the 5528
\AA \hspace{0.025in} \ion{Mg}{1} (left) and 6645 \AA \hspace{0.025in}
\ion{Eu}{2} (right) lines.  The cyan lines show syntheses of only
atomic lines (i.e. no molecular lines were included), with isotopic
and HFS components included for the \ion{Eu}{2} line in Figure
\ref{fig:ArcturusEuIso}.  It is evident that atomic lines do not
account for all the lines in the regions. The \ion{Mg}{1} region
contains mainly MgH lines, while the \ion{Eu}{2} region contains CN
lines.  The blue lines in the lefthand image indicate Mg isotopic
ratios of $48:13:39$ for $^{24}$Mg:$^{25}$Mg:$^{26}$Mg
\citep{Yong2006}, while the magenta lines indicate Mg isotopic ratios
of $83:10:7$ \citep{Yong2003}.  The orange, blue, and magenta lines in
the righthand image show $^{12}$C/$^{13}$C ratios of 50, 9, and 4,
respectively.\label{fig:ArcturusIsotopes}}
\end{center}
\end{figure*}

Syntheses of these molecular features require input C and N
abundances, and $^{12}$C/$^{13}$C and $^{24}$Mg:$^{25}$Mg:$^{26}$Mg
ratios.  For each cluster, ``integrated'' C and N abundances are
derived to best fit the molecular lines---each integrated abundance is
adopted for the whole cluster, and star-to-star variations are not
considered.  The observed isotopic ratios from individual stars can
vary significantly, even within a single cluster.  The
$^{12}$C/$^{13}$C ratio has been observed to vary from $>50$ down to
$\sim 4$ \citep{LambertRies1981,GilroyBrown1991,Gratton2000}.  In
M~13, NGC~6752, and M~71, \citet{Yong2003,Yong2006} found
$^{24}$Mg:$^{25}$Mg:$^{26}$Mg ratios ranging from $48:13:39$ to
$83:10:7$.  Figure \ref{fig:ArcturusIsotopes} shows the 
effects of varying the isotopic ratios, as well as the effects of
neglecting all molecular lines, in the regions around the 5528 \AA
\hspace{0.025in} \ion{Mg}{1} and the 6645 \AA \hspace{0.025in}
\ion{Eu}{2} line (the regions with molecular lines, according to the
Arcturus Atlas).  Here $^{12}$C/$^{13}$C ratios of 4, 9, and 50 are
considered, as well as $^{24}$Mg:$^{25}$Mg:$^{26}$Mg ratios of
$48:13:39$ and $83:10:7$.  In these regions, the isotopic ratios do
not significantly affect the continuum levels, or the specific lines
that are being synthesized.  Hence, different isotopic ratios are not
investigated in the ILS syntheses.

It should also be noted that none of the molecular lines have been
calibrated to the Solar and Arcturus spectra (i.e. none of the atomic
data, etc. were changed from the Kurucz values).  Any regions with
mismatching/uncertain molecular features (whether from isotopic ratios
or atomic data) are identified in the syntheses with the Final Line
List.

\subsubsection{Damping}\label{subsubsec:Damping}
Damping (e.g., from pressure broadening) can affect the abundances
derived from strong lines.  For consistency, damping is included for
the strongest lines (in the Arcturus spectrum).  The damping is
implemented in ILABUNDS in a similar way as the 2010 version of MOOG,
i.e. the damping parameters from \citet{Barklem2000} and
\citet{Barklem2005} were converted to C6 parameters.  When damping
data were not available from the Barklem sources (or when they did not
provide satisfactory fits to the Solar/Arcturus spectra), values from
the VALD or Kurucz databases were included.

\section{Spectrum Synthesis Results}\label{sec:Syntheses}
Any rigorous spectrum synthesis method should yield abundances with
random errors that are less than or equal to the standard error from
an equivalent width analysis, whether for individual stars or an ILS
(though an equivalent width analysis may still be the preferable
choice when the line profiles are uncertain, or if the S/N is quite
low).  An equivalent width analysis of \ion{Fe}{1} lines from a $R
\approx 30,000$ and  S/N $\sim 100$ spectrum of an RGB star results in
a line to line scatter of $\sim \pm0.1$ dex
(e.g., \citealt{Sakari2011}). Since there are many \ion{Fe}{1} lines,
and each line is considered to be an independent measurement, the
average \ion{Fe}{1} abundance has a \textit{random} error of $\sim
0.1/\sqrt{N}$ dex (though systematic uncertainties may be larger, as
discussed by \citealt{McWilliam1995a}).   However, for a single line,
this random measurement error in the elemental abundance cannot be
determined directly.  Therefore $\pm 0.1$ dex is adopted as the
minimum uncertainty in the abundance of any single spectral line in an
$R=30,000$ and $\rm{S/N} = 100$ spectrum.  Since the GC ILS have these
same qualities, the goal of this analysis is to reduce the spectrum
synthesis abundance uncertainties to this minimum.

Sections \ref{subsec:Mg}, \ref{subsec:Na}, and \ref{subsec:Eu} show
the integrated light syntheses of Mg, Na, and Eu lines in the ILS of
the five Galactic GCs.  Each section begins with syntheses of the
47~Tuc ILS, using the three different line lists (see Section
\ref{subsec:LineList}).  For each case, the errors due to continuum
placement and profile fitting are discussed in detail, and it is shown
that neither the Minimal nor VALD Line Lists are sufficient for
reducing the errors to a satisfactory level. Each section then
presents syntheses on the other Galactic GCs, using only the Final
Line List.

\subsection{\ion{Mg}{1}: The 5528 and 5711 \AA \hspace{0.025in} lines}\label{subsec:Mg}
Magnesium is an important element in any abundance analysis. As an
$\alpha$-element, Mg can trace the chemical contributions from massive
stars, particularly Type II supernovae.  Additionally, its
nucleosynthetic history is simpler than many other elements that form
entirely through core burning in massive stars.  In GCs, however, Mg
has been observed to vary among stars in a single cluster
(e.g. \citealt{Shetrone1996,Gratton2004}), which means that an ILS
abundance may not reflect the ``primordial'' $\alpha$-abundance of the
cluster.  The effects of star-to-star chemical variations will be
investigated more in the future---the current work is limited to a
single ILS Mg abundance per cluster.

\subsubsection{Mg in 47~Tuc: The Minimal and VALD Line Lists}\label{subsubsec:MgRGB}
This section first presents spectrum syntheses of the 47~Tuc ILS
\ion{Mg}{1} lines using only the lines in the Minimal and VALD Line
Lists. This is to show that a complete line list that has been
calibrated to the Sun and Arcturus is necessary, and affects the
precision of the results.  The next section (Section
\ref{subsubsec:MgFull}) presents results with the best Final Line
List.

\paragraph{5528 \AA}
The 5528 \AA \hspace{0.025in} feature is typically strong in
metal-rich ($[\rm{Fe/H}] \ga -1.0$) stars.  In the 47~Tuc
spectrum, the equivalent width of this line is $\sim 230$ m\AA, making
the line too strong for an abundance analysis, since the
uncertainties in damping, stellar structure, and NLTE corrections
become too large.  However, this 5528 \AA \hspace{0.025in} feature is
not as strong in the more metal-poor clusters (such as M3, M13, NGC
7006, and M15), and the spectral region must therefore be calibrated.

\begin{figure*}
\begin{center}
\centering
\includegraphics[scale=0.9]{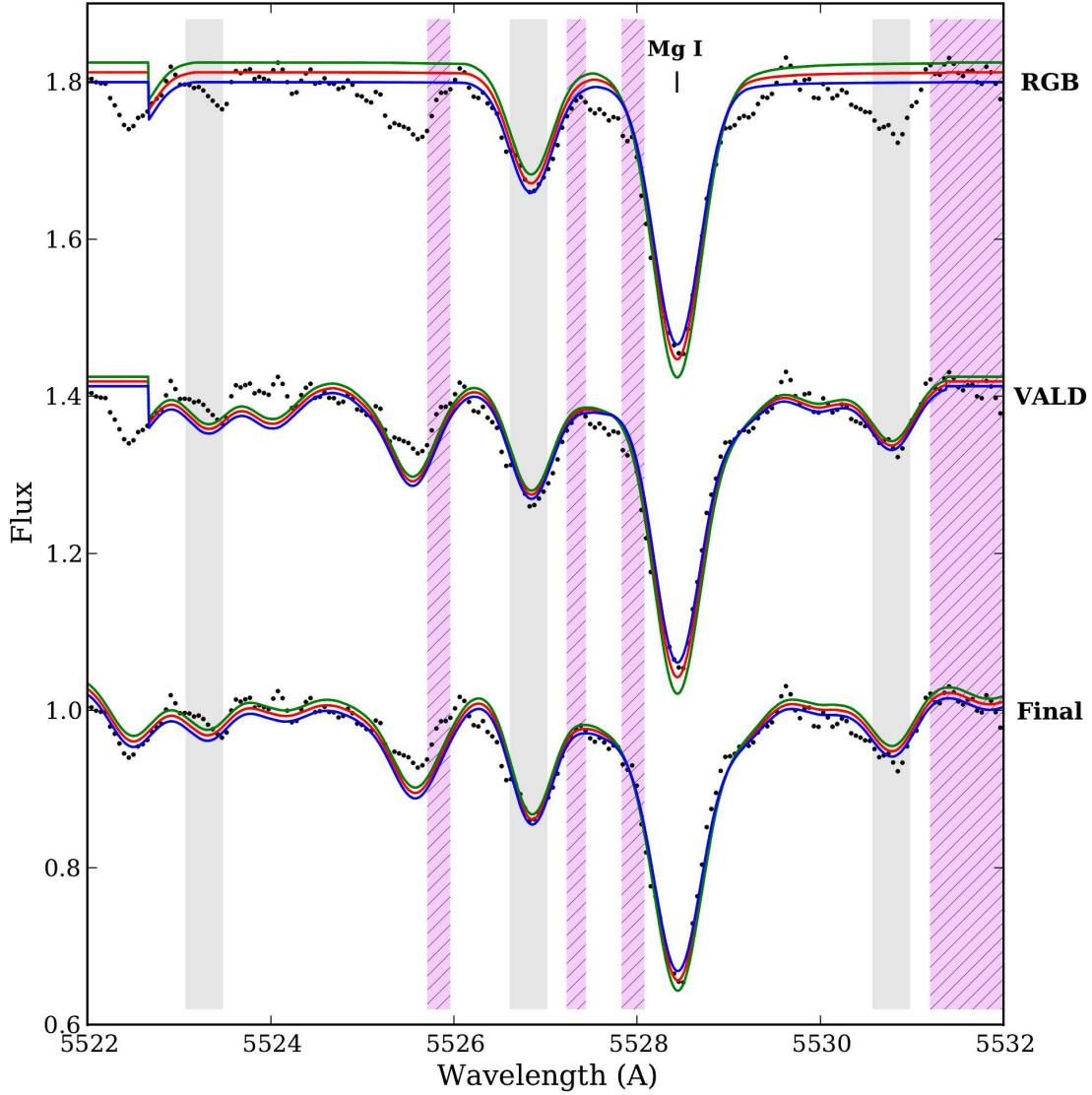}
\caption{Spectrum syntheses of the 5528 \AA \hspace{0.025in}
\ion{Mg}{1} line on the 47~Tuc ILS with the Minimal, VALD, and Final
Line Lists.  Uncertainties in continuum location and line profile
fitting are both considered.  The red lines show the average
abundance, while the green/blue lines show the $\pm 1 \sigma$
abundances, respectively.  The shaded grey regions indicate areas with
possible HFS components, while the hatched light purple regions
indicate uncertain molecular features. Both types of regions have been
ignored for continuum fits.\label{fig:Mg5528_RGB}}
\end{center}
\end{figure*}

Figure \ref{fig:Mg5528_RGB} shows syntheses of the 5528 \AA
\hspace{0.025in} line in 47~Tuc with the two basic line lists.
The top panel shows the syntheses with lines only from
the Minimal List, while the middle panel shows the
syntheses with the VALD Line List.  In addition to the \ion{Mg}{1}
line, only two other lines are in the Minimal List (an \ion{Fe}{1}
line at 5522.45 \AA, and a \ion{Sc}{2} line at 5526.82 \AA).  The
scarcity of lines makes it very difficult to distinguish weak lines
from noise.  This leads to large uncertainties in continuum placement,
as shown by the large vertical offsets. This uncertainty leads to Mg
abundance errors that are $\sim 0.20$ dex, which is insufficient for
distinguishing between a Mg-enhanced and non-Mg-enhanced cluster.

The VALD line list (which includes more lines) helps significantly
with continuum identification. Different continuum shifts are still
necessary in order to fit the different features, and it is still
difficult to distinguish weak features from noise.  In this case, the
best-fitting synthesis leads to Mg abundance errors that are $\pm 
0.15$ dex---this uncertainty is lower than before, but is still
large.

\begin{figure*}
\begin{center}
\centering
\includegraphics[scale=0.9]{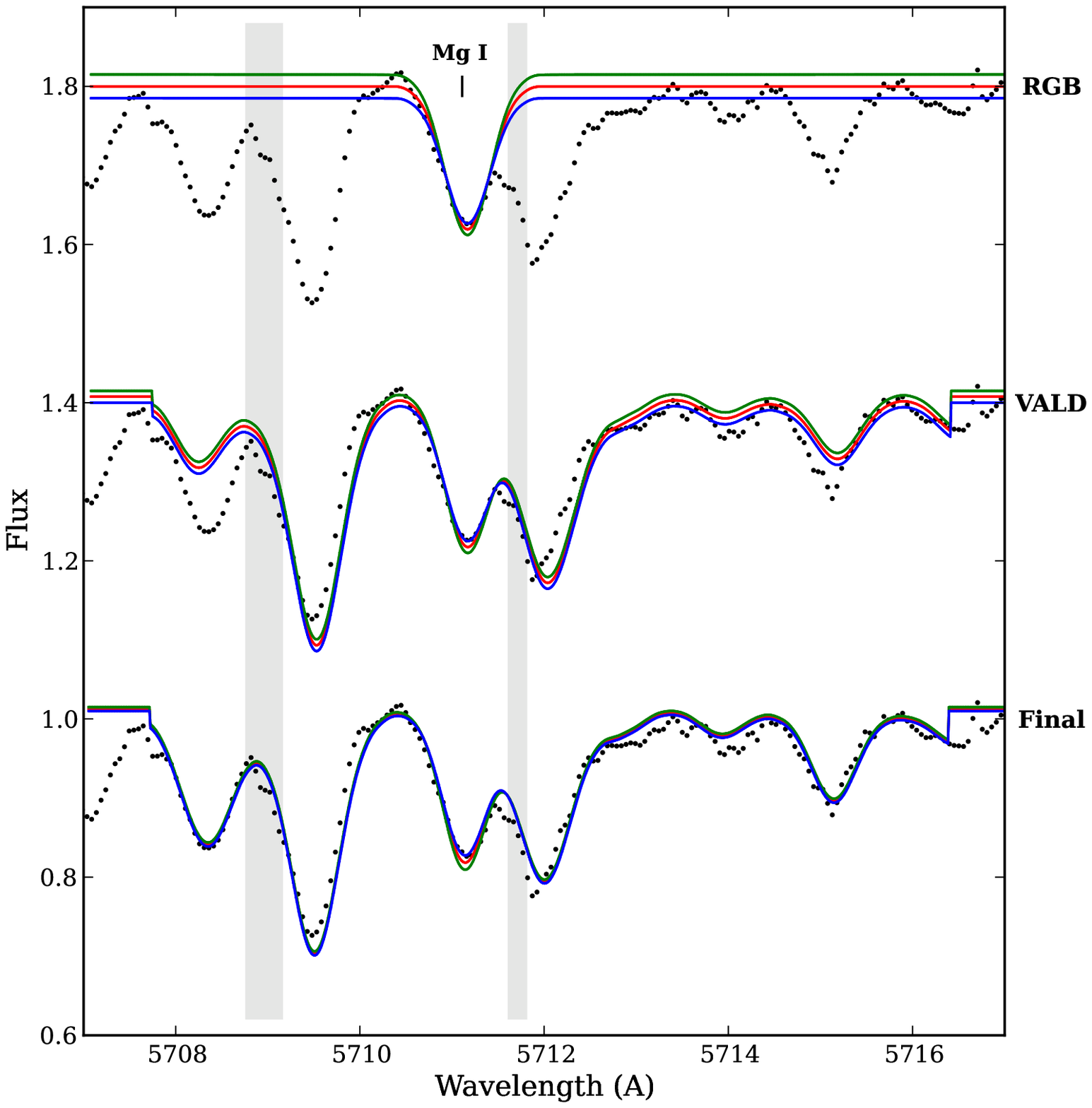}
\caption{Spectrum syntheses of the 5711 \AA \hspace{0.025in}
\ion{Mg}{1} line on the 47~Tuc ILS with the Minimal, VALD, and Final
Line Lists.  Lines are as in Figure
\ref{fig:Mg5528_RGB}.\label{fig:Mg5711_RGB}}
\end{center}
\end{figure*}

\paragraph{5711 \AA}
The only RGB line from the Minimal List that appears in this spectral
region is the 5711~\AA \hspace{0.025in} \ion{Mg}{1} line itself.
Without any additional lines in the region, it is difficult to locate
the continuum level in the region, as illustrated by the synthesis of
47~Tuc in Figure \ref{fig:Mg5711_RGB}.  In particular, it is unclear
whether the peak blueward of the 5711 \AA \hspace{0.025in} line is the
true continuum, noise, an improperly removed cosmic ray, etc.  It is
also unclear whether the width or depth of the line should be fit.
Considering all these factors, the uncertainty in the best-fitting
abundance ends up being $\pm 0.25$.

As before, the increased number of lines in the VALD Line List helps
to isolate the continuum level. However, many of the synthesized lines
in the region do not match the observed ones---some are stronger,
others are weaker, and some are missing altogether.  With the VALD
lines, the error in the best fitting abundance becomes $\pm 0.17$.
This level of uncertainty is better, but is still less than expected
for a spectrum synthesis on such a high S/N spectrum ($\sim 100$).

Thus, the Minimal and VALD line lists are insufficient for the ILS
syntheses of the regions around these two \ion{Mg}{1} lines, primarily
because the continuum level cannot be clearly identified.  These lists
may be missing spectral lines (which are blended together in the
ILS)---furthermore, the lines that \textit{are} in the lists do not
all fit the observed lines properly.  This suggests that the line
lists must be tested and \textit{calibrated} on well-studied stars,
such as the Sun and Arcturus.

\subsubsection{Mg in 47~Tuc: The Final Line List}\label{subsubsec:MgFull}
This section presents syntheses of the \ion{Mg}{1} lines in the Solar,
Arcturus, and 47~Tuc spectra with the complete, calibrated Final Line
Lists. The regions with possible HFS components are highlighted in
purple, while the regions with particularly uncertain molecular lines
are highlighted with grey, hatched rectangles. These regions (and any
with missing lines) were ignored in the final synthesis fits.

\begin{figure*}
\begin{center}
\centering
\includegraphics[scale=0.9]{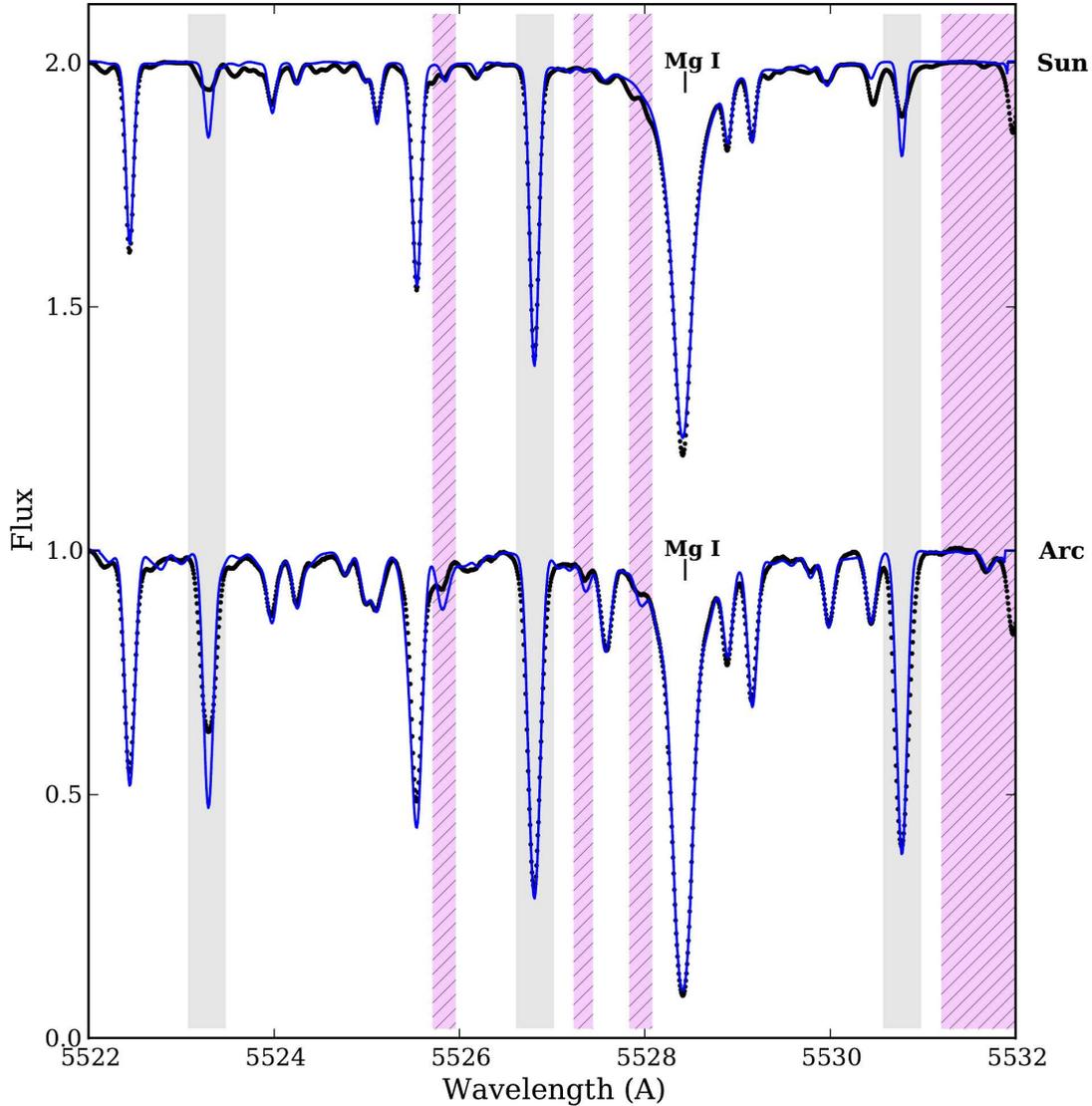}
\caption{Spectrum syntheses of the 5528 \AA \hspace{0.025in}
\ion{Mg}{1} line in the Solar (top) and Arcturus (bottom) spectra.
Lines are as in Figure \ref{fig:Mg5528_RGB}.  Note that only the best
fits are shown, as the differences in the $\pm 1\sigma$ syntheses are
generally too small to see.\label{fig:Mg5528}}
\end{center}
\end{figure*}

\paragraph{5528 \AA}
The syntheses to the 5528 \AA \hspace{0.025in} line are shown in
Figures \ref{fig:Mg5528_RGB} and \ref{fig:Mg5528}, while the final
abundances are tabulated in Table \ref{table:Abunds}.  The addition of
the MgH features in the Final Line List helps improve the continuum
identification, particularly for the blended features in 47~Tuc.
Despite the strength of the line, the synthesis of the 5528 \AA
\hspace{0.025in} feature fits the Solar spectrum quite well (see the
top panel of Figure \ref{fig:Mg5528}). The best-fitting value is
slightly higher than the \citet{Asplund2009} value, but is likely due
to the strength of the line and the uncertain atomic data. The
Arcturus abundance agrees well with the average literature value in
Table \ref{table:Abunds}.

In the 47~Tuc spectrum, the complete, calibrated line list leads to a
synthetic spectrum that is an excellent fit to the observed spectrum,
as shown in Figure \ref{fig:Mg5528}.  This is due to two reasons:
first, most of the lines in the region now fit better than before they
were calibrated, and second, the best continuum regions are evident in
the Solar and Arcturus spectra, and can be used to fit the 47~Tuc
continuum.  However, the 47~Tuc syntheses indicate a best-fitting
[Mg/Fe] ratio that is higher than the MB08 value of
$[\rm{Mg/Fe}]~=~0.22$ (which was determined from the EW of the 7387
\AA \hspace{0.025in} \ion{Mg}{1} line, and which was not calculated
differentially).  Adjusting the MB08 Solar abundance leads to a higher
[Mg/Fe] ratio in 47~Tuc, as shown below.

\paragraph{5711 \AA}

\begin{figure*}
\begin{center}
\centering
\includegraphics[scale=0.9]{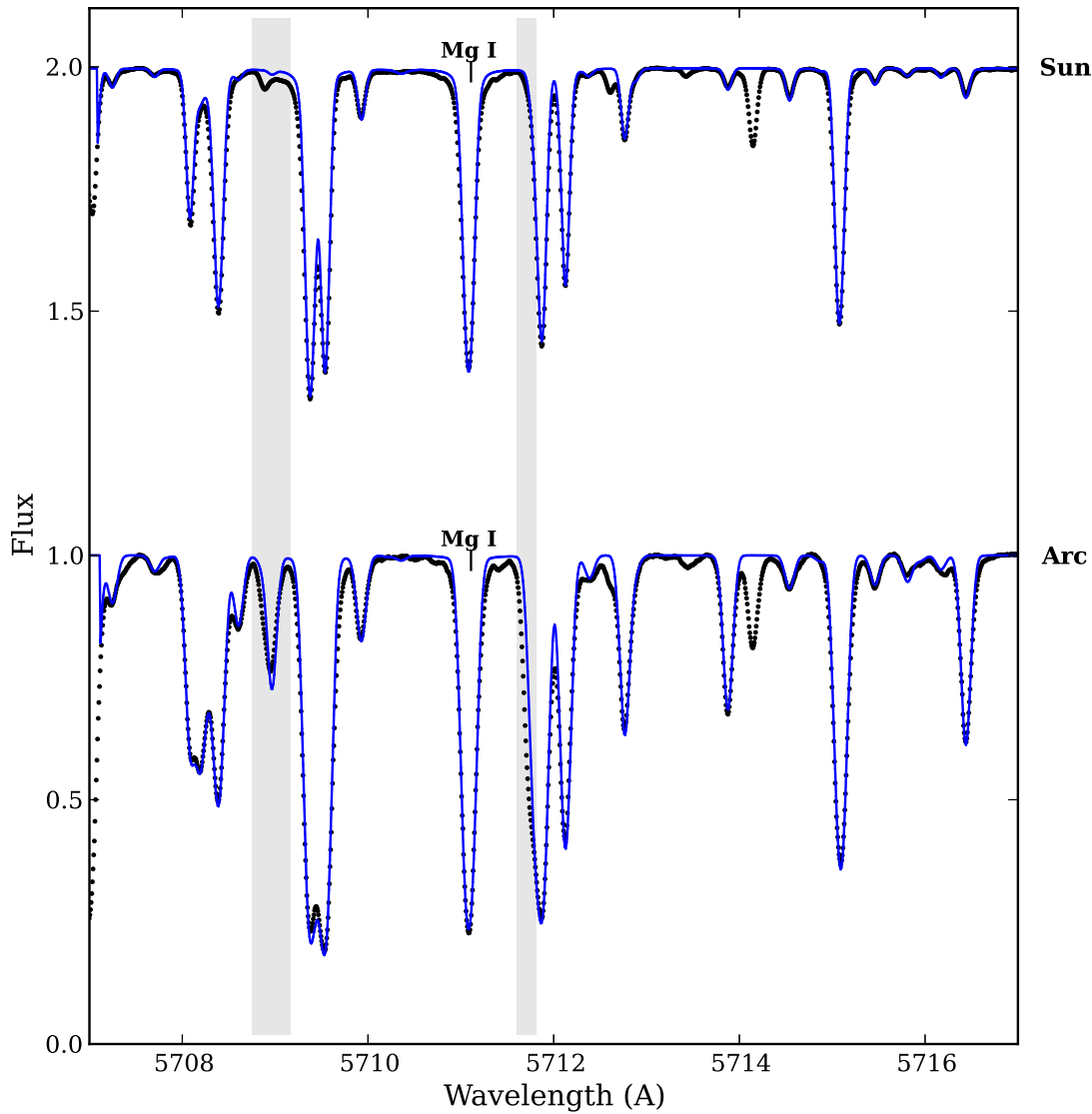}
\caption{Spectrum syntheses of the 5711 \AA \hspace{0.025in}
\ion{Mg}{1} line in the Sun and Arcturus.  Lines
are as in Figure \ref{fig:Mg5528}.\label{fig:Mg5711}}
\end{center}
\end{figure*}

The syntheses of the 5711 \AA \hspace{0.025in} \ion{Mg}{1} line are
shown in Figures \ref{fig:Mg5711_RGB} and \ref{fig:Mg5711}.  These
fits show that there are missing lines in the syntheses of this region
of the Solar and Arcturus spectra, even with a complete, calibrated
line list.  Several lines in the region also require HFS components,
e.g. the \ion{V}{1} and \ion{Sc}{1} lines.  Despite the missing
features, the strong lines are generally matched well in both the
individual and 47~Tuc spectra, with the exception of the feature at
5709 \AA.  This feature is a blend of \ion{Ti}{1}, \ion{Fe}{1},
\ion{Ni}{1}, and \ion{Ti}{2} features, where the \ion{Fe}{1} and
\ion{Ni}{2} features dominate the line strength.  It is unclear why
the lines match in the Solar spectrum, but not in the Arcturus or
47~Tuc spectra. Regardless, these features can be disregarded in the
analyses.

The derived abundances are again tabulated in Table
\ref{table:Abunds}, along with comparison literature abundances.   The
Solar abundance is in excellent agreement with the \citet{Asplund2009}
value.  The Arcturus [Mg/Fe] ratio is slightly higher than the average
literature value, but the values agree within the errors.  The 47~Tuc
[Mg/Fe] ratio is again considerably higher than the MB08 value
($[\rm{Mg/Fe}]~=~0.22$), but is in excellent agreement with the values
from the 5528~\AA \hspace{0.025in} line and with the recalculated MB08
value (see below).

Taken together, the two \ion{Mg}{1} lines provide a total
$[\rm{Mg\; I/Fe \; I}]~=~0.46~\pm~0.14$ for 47~Tuc.  This
\textit{qualitatively} agrees with the literature stellar
abundances assembled by \citet{Pritzl2005}, i.e. 47~Tuc is
Mg-enhanced, as expected for a Galactic GC at its metallicity, and the
ILS value is in good agreement with the literature average.
Ultimately, the calibrated Final Line List has reduced uncertainties
in the Mg abundance ratios from the Minimal and VALD Lists from $\sim
0.2$ and $\sim 0.18$ to $\sim 0.14$.  Thus, spectrum synthesis
techniques with the carefully calibrated Final Line List are able to
provide improved and precise abundances.

\begin{figure*}
\begin{center}
\centering
\includegraphics[scale=0.9]{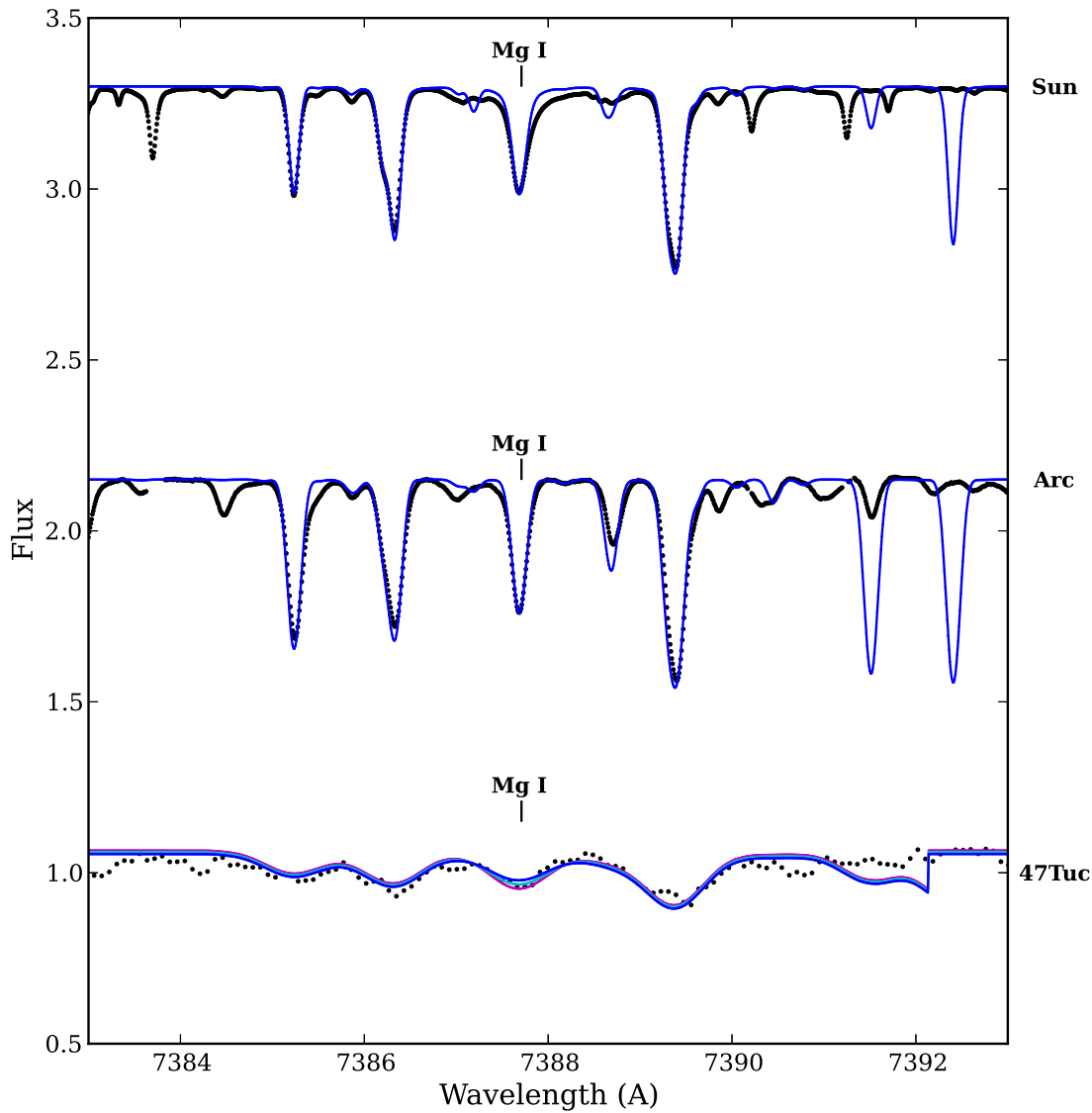}
\caption{Spectrum syntheses of the 7388 \AA \hspace{0.025in}
\ion{Mg}{1} line in the Solar (top), Arcturus (middle), and 47~Tuc
(bottom) spectra.  The HFS regions are not shown, though there are
several in this wavelength range.  Molecular lines are also not
included.  Lines are as in Figure
\ref{fig:Mg5528_RGB}.\label{fig:Mg7388}}
\end{center}
\end{figure*}

\paragraph{7387 \AA}
To compare the above \ion{Mg}{1} integrated light abundances with
MB08, the 7387 \AA \hspace{0.025in} line was synthesized in the Solar,
Arcturus, and 47~Tuc spectra using the VALD RGB Line List only (Figure
\ref{fig:Mg7388}).  This region was not calibrated because it falls
outside the observed spectral region of the HET clusters, and will not
be used in this analysis. Many of the lines in the region do not fit
well, precisely because they have not been calibrated.  Molecular
lines and HFS components were also not included, although the Arcturus
Atlas \citep{Hinkle2003} shows that there are CN lines in the region.

The Solar synthesis in Figure \ref{fig:Mg7388} shows that a Solar
abundance of $\log \epsilon_{\rm{Mg}}~=~7.30 \pm 0.02$ is required to
fit the observed feature; this value is significantly lower than the
\citet{Asplund2009} Mg abundance.  The line profile also cannot be fit
perfectly, as there seem to be extra components in the red wing of the
\ion{Mg}{1} line.  The Arcturus synthesis fit is better, but the Mg
abundance ($\log \epsilon_{\rm{Mg}}~=~7.15 \pm 0.03$) is lower than
from the other Mg lines; a differential comparison with the lower
Solar abundance leads to a normal $[\rm{Mg\; I/Fe \; I}]~=~0.50\pm
0.04$.  This suggests that the atomic data is systematically offset
for this line, and illustrates the importance of using differential
abundances and of checking all important lines in the Solar and
Arcturus spectra.

Syntheses of the 47~Tuc ILS yield $\log \epsilon_{\rm{Mg}}~=~7.21 \pm
0.20$, or $[\rm{Mg\; I/Fe \; I}]~=~0.72\pm 0.20$.  The large
uncertainty in the abundance reflects the uncertainty in the continuum
level, the uncertain line profile, and the low S/N at the line
center.  With the Solar Mg abundance from the 7387 \AA
\hspace{0.025in} line, the MB08 value (which comes from an EW analysis
of this line) can be recalculated \textit{differentially} to
$[\rm{Mg\; I/Fe \; I}]~=~0.56$---this value now agrees with the 7388
\AA, 5528 \AA, and 5711 \AA \hspace{0.025in} syntheses.  Thus, the
value quoted in MB08 is systematically lower than it should be, as a
result of the lower Solar abundance.

\subsubsection{Mg in the Other GCs}\label{subsubsec:MgOther}
\begin{figure*}
\begin{center}
\centering
\includegraphics[scale=0.9]{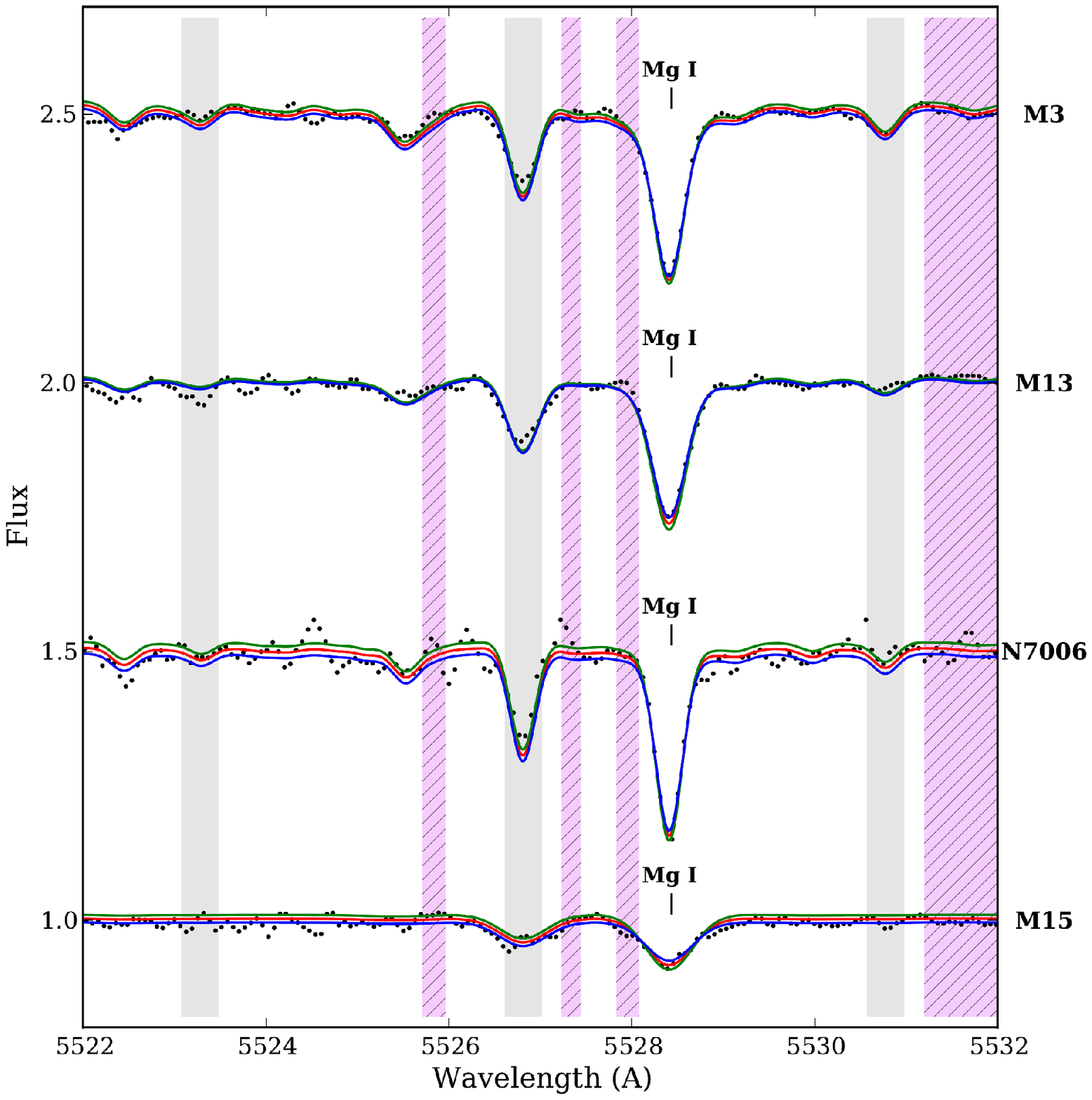}
\caption{Spectrum syntheses of the 5528 \AA \hspace{0.025in}
\ion{Mg}{1} line in M3, M13, NGC 7006, and M15. Lines
are as in Figure \ref{fig:Mg5528_RGB}.\label{fig:GCsMg5528}}
\end{center}
\end{figure*}

\begin{figure*}
\begin{center}
\centering
\includegraphics[scale=0.9]{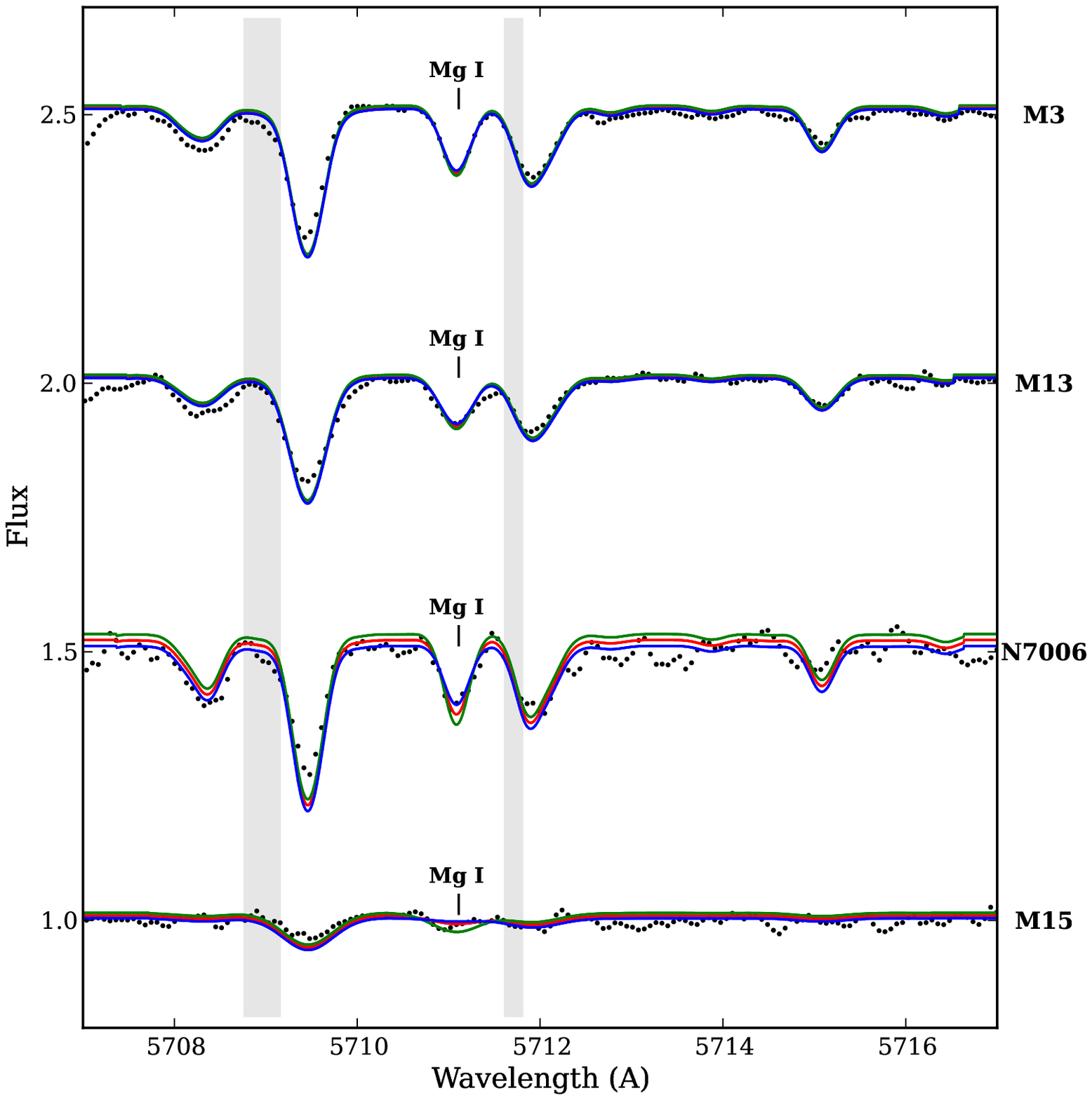}
\caption{Spectrum syntheses of the 5711 \AA \hspace{0.025in}
\ion{Mg}{1} line in M3, M13, NGC 7006, and M15.   Lines
are as in Figure \ref{fig:Mg5528_RGB}.\label{fig:GCsMg5711}}
\end{center}
\end{figure*}

The \ion{Mg}{1} abundances for M3, M13, NGC~7006, and M15, as
determined from the Final Line List, are shown in Table
\ref{table:Abunds} and Figures \ref{fig:GCsMg5528} and
\ref{fig:GCsMg5711}.

\begin{table*}
\centering
\begin{minipage}{160mm}
\caption{Solar, Arcturus, and Globular Cluster Abundances\label{table:Abunds}}
  \begin{tabular}{@{}lccccccc@{}}
  \hline
 & Sun & Arcturus & 47 Tuc & M3 & M13 & NGC 7006 & M15 \\
 & $\log \epsilon_{\sun}$ & [X/Fe] & [X/Fe] & [X/Fe] & [X/Fe] & [X/Fe] & [X/Fe]\\
 \hline
Mg 5528 \AA  & $7.75\pm0.05$ & $0.55\pm0.18$ & $0.39\pm0.13$ & $0.12\pm0.11$ & $0.13\pm0.11$ & $0.10\pm0.14$ & $-0.15\pm0.21$\\
Mg 5711 \AA  & $7.58\pm0.04$ & $0.59\pm0.06$ & $0.44\pm0.14$ & $0.20\pm0.07$ & $0.14\pm0.08$ & $0.13\pm0.20$ & $<0.55$ \\
Mg Total     & $-$           & $0.57\pm0.13$ & $0.42\pm0.14$ & $0.16\pm0.11$ & $0.14\pm0.10$ & $0.12\pm0.17$ & -  \\
{\it MB08}   & & & 0.22/0.56$^{a}$& & & & \\
{\it Literature Avg}& $7.60\pm0.04$ & $0.46\pm0.09^{b}$ & $0.40\pm0.03$ & $0.23\pm0.03$ & $0.11\pm0.03$ & $0.34\pm0.02$ & $0.35\pm0.05$ \\
{\it Lit. Range}    &  -            &  -                & $[0,+0.6]$    & $[-0.1,+0.6]$ & $[-0.2,+0.5]$ & $[+0.3,+0.4]$ & $[-0.4,+0.8]$ \\
 & & & & & & & \\
Na 6154 \AA  & $6.28\pm0.02$ & $0.20\pm0.03$ & $0.38\pm0.15$ & $0.27\pm0.13$ & $0.45\pm0.10$ & $0.66\pm0.14$ & $0.90\pm0.40$ \\
Na 6160 \AA  & $6.33\pm0.03$ & $0.20\pm0.04$ & $0.37\pm0.08$ & $0.17\pm0.13$ & $0.20\pm0.20$ & $0.26\pm0.13$ & $<1.05$ \\
Na Total     & $-$           & $0.20\pm0.04$ & $0.38\pm0.12$ & $0.22\pm0.13$ & $0.33\pm0.16$ & $0.41\pm0.14$ & $-$ \\
{\it MB08}   & & & $0.45\pm 0.10$ & & & & \\
{\it Literature Avg}& $6.24\pm0.04$ & $0.18\pm0.05$$^{b}$ & $0.45\pm0.01$ & $0.02\pm0.06$ & $0.27\pm0.06$ & $0.32\pm0.06$ & $0.39\pm0.06$ \\
{\it Lit. Range}    & -             &  -                  & $[-0.3,+1.0]$ & $[-0.3,+0.3]$ & $[-0.3,+0.6]$ & $[0,+0.4]$    & $[-0.6,+2.0]$\\
 & & & & & & & \\
Eu 6645 \AA  & $0.45\pm0.02$ & $0.28\pm0.05$ & $0.27\pm0.14$ & $0.75\pm0.11$ & $0.76\pm0.10$ & $0.72\pm0.15$ & $1.31\pm0.20$ \\
{\it MB08}   & & & $0.04$ & & & & \\
{\it Literature Avg}& $0.52\pm0.04$ & $0.26\pm0.04$ & $0.14\pm0.03$ & $0.51\pm0.02$ & $0.49\pm0.03$ & $0.36\pm0.02$ & $0.63\pm0.03$ \\
{\it Lit. Range}    & -             & -             & $[-0.4,+0.4]$ & $[+0.4,+0.8]$ & $[+0.3,+1.0]$ & $[+0.3,+0.4]$ & $[+0.2,+2.2]$\\
 & & & & & \\
\hline
\end{tabular}
\end{minipage}\\
\medskip
\raggedright The [X/Fe] ratios use \ion{Fe}{1} for neutral
species, and \ion{Fe}{2} for singly ionized species.  The mean
literature values are straight, un-weighted averages from all
available sources.   The quoted $\pm 1\sigma$ errors in the literature
mean do not reflect the observed range in the abundances.  The
literature ranges show the approximate extremes that have been
observed in the clusters. Flux-weighted averages may be more
appropriate for comparisons with integrated light abundances; however,
these averages require a reasonably complete sample that reflects the
scanned population. \\
{\bf References: } Literature Solar values are from \citet{Asplund2009}.
Arcturus literature values are an average of the values from
\citet{Yong}, \citet{Fulbright2007}, \citet{RamirezAllendePrieto2011},
and McWilliam et al. (2013, in prep; for Eu), after shifting to a
common $[\rm{Fe/H}] = -0.6$.  The GC literature data are from
observations of individual stars, and are from \citet{Pritzl2005},
with supplements from
\citet{BrownWallerstein1992,Sneden1997,Kraft1998,Sneden2004,Carretta2004,Jasniewicz2004,CohenMelendez2005,AlvesBrito2005,Preston2006,Wylie2006,KochMcWilliam2008,Carretta2009,Worley2009,Sobeck2011,Gratton2013}.
 The 47~Tuc values from MB08 are also shown.\\
$^{a}$The first value is the one quoted in MB08; the second is the
re-derived \textit{differential} value (see the text).\\
$^{b}$Values were adjusted to $[\rm{Fe/H}] = -0.6$.\\
\end{table*}

The fits to the \ion{Mg}{1} lines are quite good for M3 and M13, but
are much more uncertain for NGC 7006 (owing to its lower S/N) and M15
(as a result of its weaker lines).  The 5528 \AA \hspace{0.025in} line
is easily detectable in all spectra, and there are enough additional
lines to help isolate the continuum.  The 5711 \AA \hspace{0.025in}
region, however, is much more difficult for M15, as is evident in
Figure \ref{fig:GCsMg5711}; this line only provides an upper limit for
the Mg abundance in M15.  The final derived abundances are shown in
Table \ref{table:Abunds}.

The average \ion{Mg}{1} ILS abundance ratios for M3, M13, and NGC 7006
are in excellent agreement with each other.  With the exception of
M13, all the ILS values are slightly lower than the average literature
abundances, especially M15.  These differences between the ILS
abundances and the ``average'' literature abundances are due to the
known star-to-star Mg variations within the clusters.  For example,
the M15 Mg abundance quote in \citet{Pritzl2005} is an average of
eighteen bright RGB stars observed by \citet{Sneden1997}; the latter
analysis showed that M15 has strong star-to-star Mg variations,
ranging from $-0.4 \la [\rm{Mg/Fe}] \la +0.8$, with [Mg/Fe]
decreasing for stars higher up the RGB.  Since ILS are dominated by
the brightest stars, the M15 ILS is most likely dominated by the most
Mg-poor giants, which decreases the integrated light Mg abundance.
The derived ILS [Mg/Fe] ratio for M15 \textit{does} fall at the lower
end of the observed abundance range.  M3 and M13 also show signs of Mg
variations \citep{Sneden2004,CohenMelendez2005} which most likely
accounts for their slightly low average [Mg/Fe] abundances in
comparison to Milky Way field stars.  However, even though these
clusters also have star-to-star variations, their abundances
\textit{do} agree with the literature averages. This shows that
caution must be taken when comparing ILS abundances to ``average''
literature abundances from a limited sample of stars, as discussed in
Section \ref{subsec:Literature}.  Regardless, these comparisons with
the literature abundances \textit{do} show that spectrum syntheses of
the 5528 and 5711 \AA \hspace{0.025in} lines are capable of producing
Galactic GC \ion{Mg}{1} abundances that fall within the observed
ranges from individual stars, provided that the lines are sufficiently
strong, and that the S/N is sufficiently high.

\subsection{\ion{Na}{1} 6154/6160 \AA \hspace{0.025in} lines}\label{subsec:Na}
Sodium has always been an interesting element for GC studies
because, like magnesium, it varies strongly between stars in a single
cluster.  These variations are anticorrelated with O, and indicate
abundance changes up the RGB and/or the presence of multiple stellar
populations in GCs
(e.g. \citealt{Kraft1992,Gratton2004,Carretta2009}).  The Na/O
anticorrelation has been detected in all unambiguous Galactic GCs, and
seems to be a common signature of the GC formation process. Thus, Na
is another significant element in GC studies.

Sodium lines are known to suffer from strong non-LTE effects. To
minimize NLTE effects, \citet{Lind2011} suggest that
analyses at Solar metallicity should use the 6154/6160~\AA
\hspace{0.025in} doublet, while analyses of more metal-poor stars
($[\rm{Fe/H}] \la -1.0$) should use the 5682/5688~\AA
\hspace{0.025in} lines.  However, literature results use both sets of
lines, often without any NLTE corrections.  Here the results for the
6154/6160 doublet are presented for all the GCs.  The NLTE effects
make it difficult to fit the Solar 5682/5688 \AA \hspace{0.025in}
lines (in agreement with, e.g., \citealt{Baumueller1998}), which
prevents a differential analysis from being done. Without a
differential analysis the atomic data can lead to systematic offsets
between lines.  Further calibration work must therefore be done to
synthesize the 5682/5688~\AA \hspace{0.025in} doublet in ILS.

\citet{Mashonkina2000} also note that for all Na lines the strengths
of NLTE corrections depend on the temperature and surface gravity of
the stars, in addition to the stellar metallicity, and that Na
abundances of giants cannot be accurately determined to within 0.1 dex
without NLTE corrections. This could be particularly problematic for
ILS, whose spectral lines contain contributions from stars over a wide
range of $T_{\rm{eff}}$ and $\log g$.  Despite these concerns, NLTE
corrections are neglected in this analysis.  However, as NLTE
corrections are not generally applied to the comparison literature
data, differential comparisons should be reasonably robust.

\subsubsection{Na in 47~Tuc: The Minimal and VALD Line
Lists}\label{subsubsec:NaRGB} 
\begin{figure*}
\begin{center}
\centering
\includegraphics[scale=0.9]{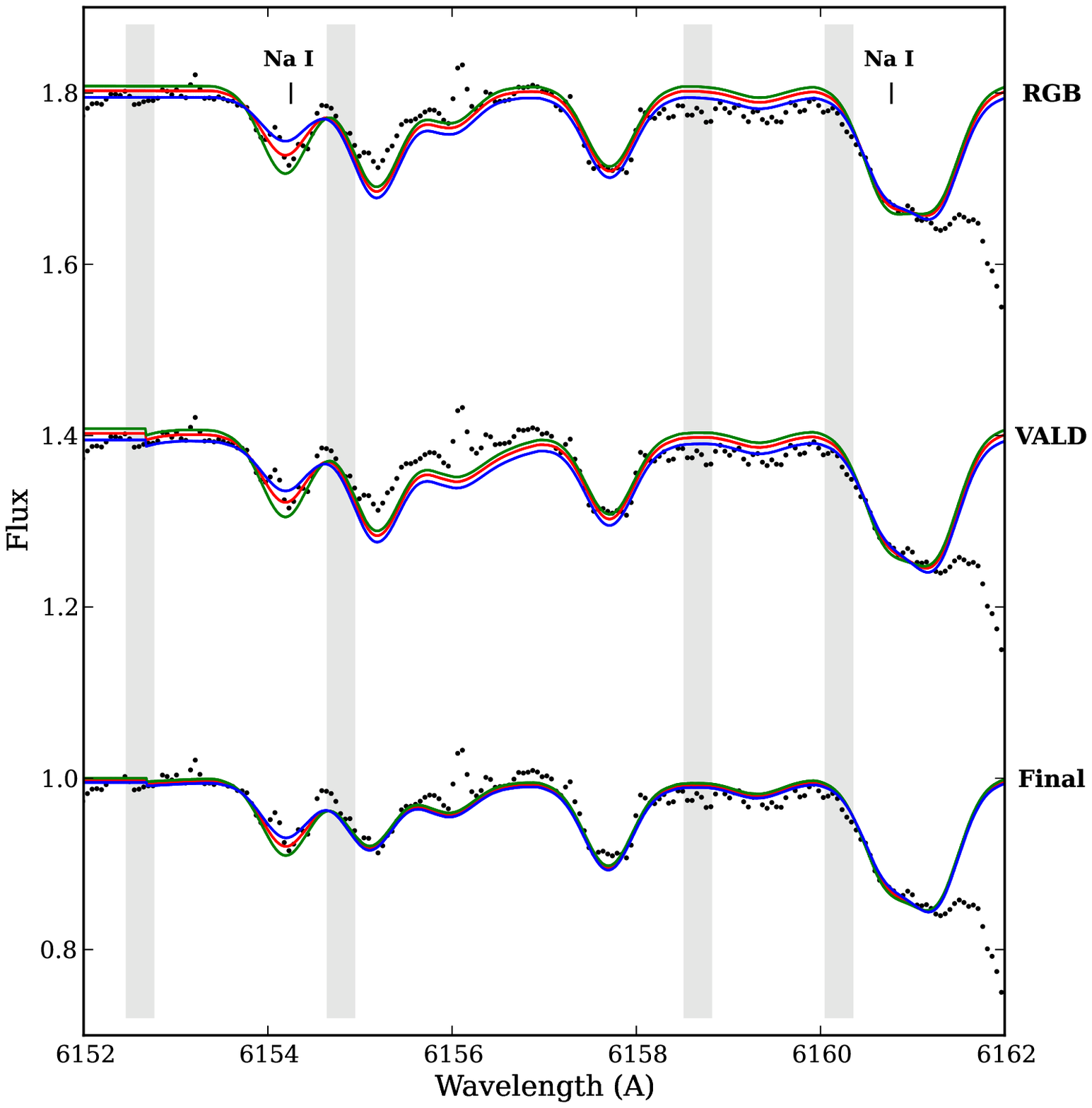}
\caption{Spectrum syntheses of the 6154/6160 \AA \hspace{0.025in}
\ion{Na}{1} lines in 47 Tuc, with the Minimal, VALD, and Final Line
Lists.   Lines are as in Figure
\ref{fig:Mg5528_RGB}.\label{fig:Na6154_RGB}}
\end{center}
\end{figure*}

Figure \ref{fig:Na6154_RGB} shows
spectrum syntheses of the 6154/6160 \AA \hspace{0.025in} doublet in 47
Tuc, using the Minimal and VALD Lists.  The synthesis with the Minimal
Line List (the top synthesis in Figure \ref{fig:Na6154_RGB}) only has
a few strong lines available for continuum identification.
Furthermore, the features are strongly blended in the 47~Tuc spectrum,
and there are few obvious continuum locations.    With only these
strong lines, the errors in the best-fitting abundances are $\pm0.30$
and $\pm0.13$ for 6154 and 6160 \AA, respectively. The VALD lines (the
middle spectrum in Figure \ref{fig:Na6154_RGB}) help slightly,
bringing the abundance errors to $\pm0.25$ and $\pm0.12$,
respectively.

With so many strong lines and blends in the region, it is easy to see
how fitting becomes very difficult in ILS.  In particular, it is
difficult to isolate continuum regions and to know which regions
\textit{should} match the synthetic spectra, and which could be
different as a result of improper atomic data, missing atomic or
molecular lines, or HFS.  Without Solar and Arcturus calibrations to
identify the best areas for continuum fitting, such a region is quite
difficult to fit in ILS.

\subsubsection{Na in 47~Tuc: The Final Line List}\label{subsubsec:NaFull}
\begin{figure*}
\begin{center}
\centering
\includegraphics[scale=0.9]{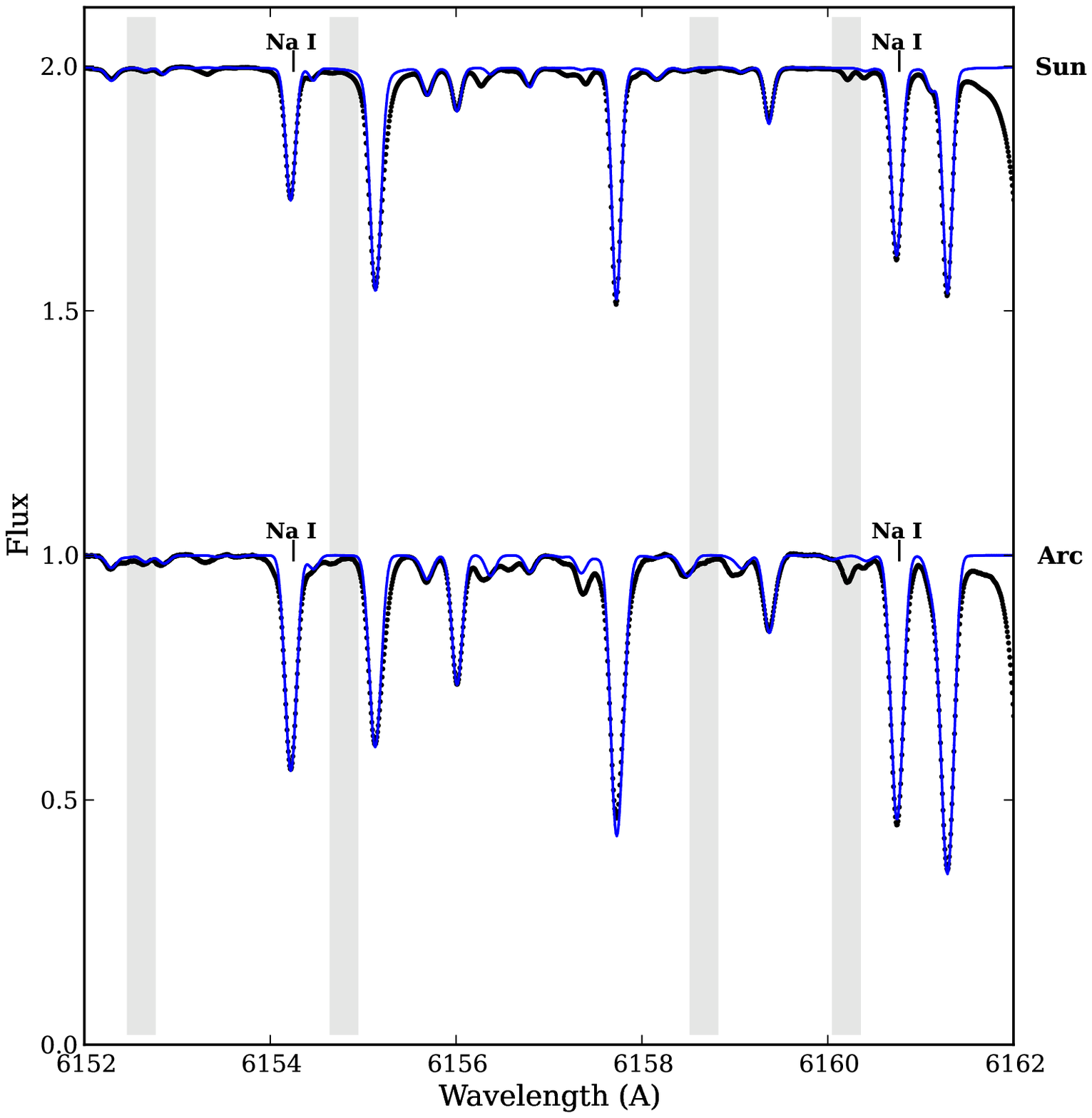}
\caption{Spectrum syntheses of the 6154/6160 \AA \hspace{0.025in}
\ion{Na}{1} lines on the Solar and Arcturus spectra.  Lines are as in
Figure \ref{fig:Mg5528_RGB}.\label{fig:Na6154}}
\end{center}
\end{figure*}

Figures \ref{fig:Na6154_RGB} and \ref{fig:Na6154} show syntheses of
the 6154 and 6160 \AA \hspace{0.025in} \ion{Na}{1} lines in the Solar,
Arcturus, and 47~Tuc spectra with the Final Line List.

HFS contamination seems to be minimal in this region. The Solar and
Arcturus syntheses show that there are clearly lines missing from the
line list, though they are mostly weak.  The Solar and Arcturus
spectra also show that the syntheses cannot perfectly reproduce the
shape of the 6154 \AA \hspace{0.025in} \ion{Na}{1} line---there seems
to be a missing component slightly redward of the line center.  Thus,
the syntheses to the 47~Tuc spectrum focused primarily on fitting the
depth of the line rather than the width.

The best fit abundances to the 6154/6160 \AA \hspace{0.025in}
\ion{Na}{1} lines are shown in Table \ref{table:Abunds}.  The Solar
values are in reasonable agreement with the \citet{Asplund2009} value.
The Arcturus value is also in excellent agreement with the average
literature value, after shifting to a common [Fe/H].  The 47~Tuc value
is in agreement with the MB08 value, which was derived from the same
lines.  The integrated light abundance also agrees well with the
average literature abundance in Table \ref{table:Abunds}.  As a
Galactic GC, 47~Tuc is one of the many Galactic GCs that have shown
star-to-star variations in Na, as mentioned earlier
(e.g. \citealt{Carretta2009}).  However, in this case the Na abundance
is not overly high, suggesting that the integrated light is not
dominated by Na-enhanced stars.

For the 6154 \AA \hspace{0.025in} sodium line, the Final Line List has
reduced the errors from $0.30$ (from the Minimal Line List) and $0.25$
(from the VALD Line List) to $0.15$.  In the case of the 6160 \AA
\hspace{0.025in} line the improvements are similarly excellent, with
decreases from $0.13$ and $0.13$ to $0.07$.  Again, the Final Line
List provides a significant improvement to the precision of the
derived ILS abundances.

\subsubsection{Na in the Other GCs}\label{subsubsec:NaOther}
\begin{figure*}
\begin{center}
\centering
\includegraphics[scale=0.9]{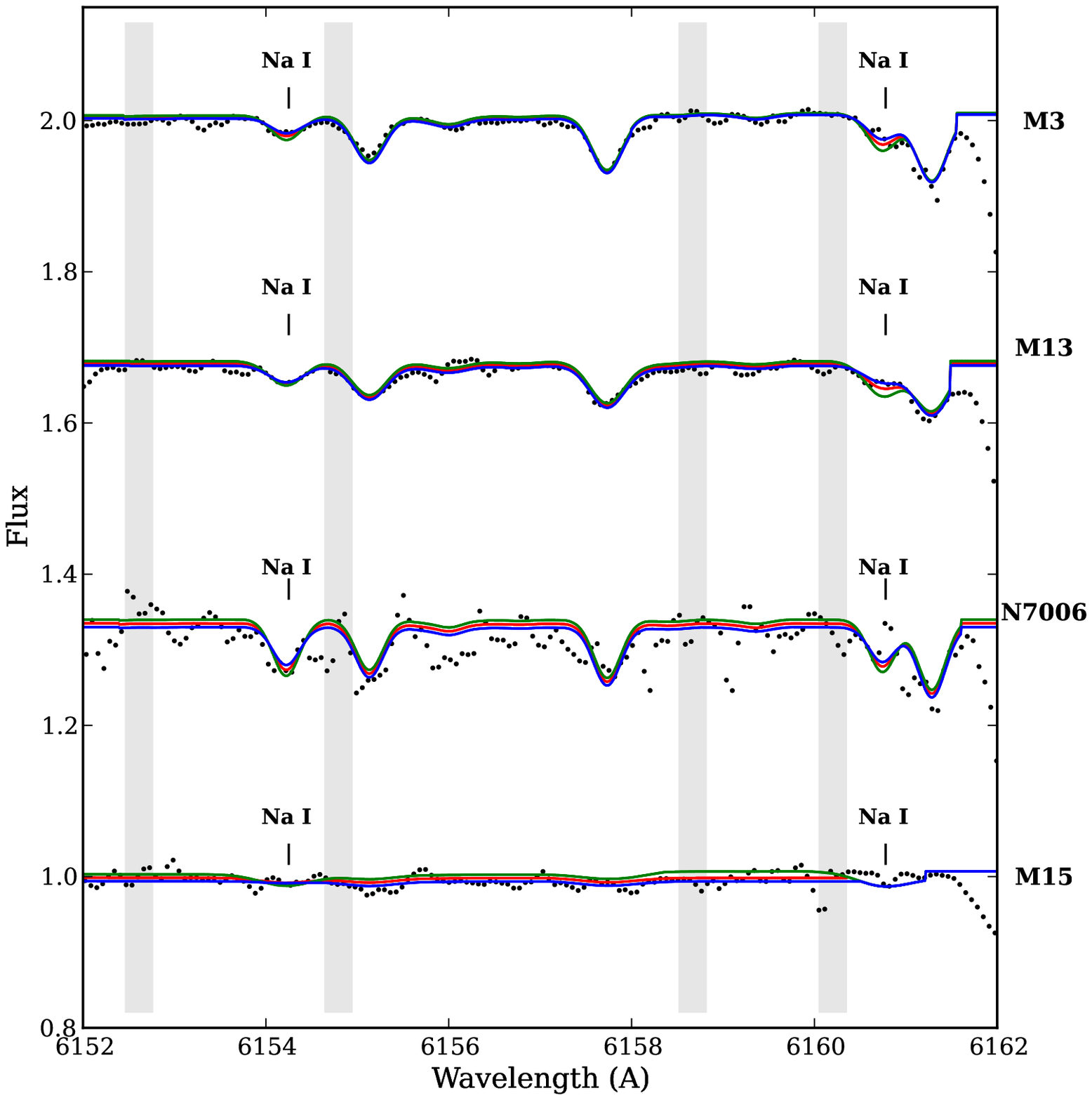}
\caption{Spectrum syntheses of the 6154/6160 \AA \hspace{0.025in}
\ion{Na}{1} lines on the M3, M13, NGC~7006, and M15 spectra.  Lines
are as in Figure \ref{fig:Mg5528_RGB}.  The 6160 \AA \hspace{0.025in}
line provides only an upper limit in M15.\label{fig:GCsNa6154}}
\end{center}
\end{figure*}

The Na 6154/6160 \AA \hspace{0.025in} syntheses on M3, M13, NGC~7006,
and M15 are shown in Figure \ref{fig:GCsNa6154}, while the abundances
are given in Table \ref{table:Abunds}.  The comparison literature
abundances are also shown in Table \ref{table:Abunds}.  These
literature abundances were determined with both the 6154/6160 and
5682/5688\AA \hspace{0.025in} lines (and occasionally also the Na D
lines).  In the case of M3, M13, and NGC~7006, the literature
abundances have \textit{not} had NLTE corrections, while some of the
M15 stars did have NLTE corrections.  The M3/M13 literature abundances
are from 36 and 60 stars from the base of the RGB, up to the tip of
the RGB, while the NGC~7006 literature abundances are only from six
tip of the RGB stars.  M15's literature abundances are mainly from
more evolved stars (e.g. RGB, HB, and AGB stars).

While M13 and NGC~7006's ILS abundances are in good agreement with the
literature averages, the IL Na abundance in M3 is quite a bit higher
than the average literature value.  The 6160 \AA \hspace{0.025in} line
is not well-resolved in the M15 ILS, and hence provides only an
upper limit.  The 6154 \AA \hspace{0.025in} line in M15, however,
provides a larger Na abundance than the literature average.  Again,
these discrepancies with the literature averages are results of the
intercluster Na variations, which vary from $-0.6 \la
[\rm{Na/Fe}] \la 2$ \citep{Sneden1997,Preston2006}.

\subsection{\ion{Eu}{2} 6645 \AA \hspace{0.025in}
line}\label{subsec:Eu}
The weak 6645 \AA \hspace{0.025in} \ion{Eu}{2} feature is commonly
synthesized in spectroscopic analyses, since it provides important
constraints on contributions from the r- (rapid) neutron capture
process.\footnote{At solar metallicity 97\% of Eu comes from the
r-process \citep{Burris2000}.}

\subsubsection{Eu in 47~Tuc: The Minimal and VALD Line Lists}\label{subsubsec:EuRGB}
\begin{figure*}
\begin{center}
\centering
\includegraphics[scale=0.9]{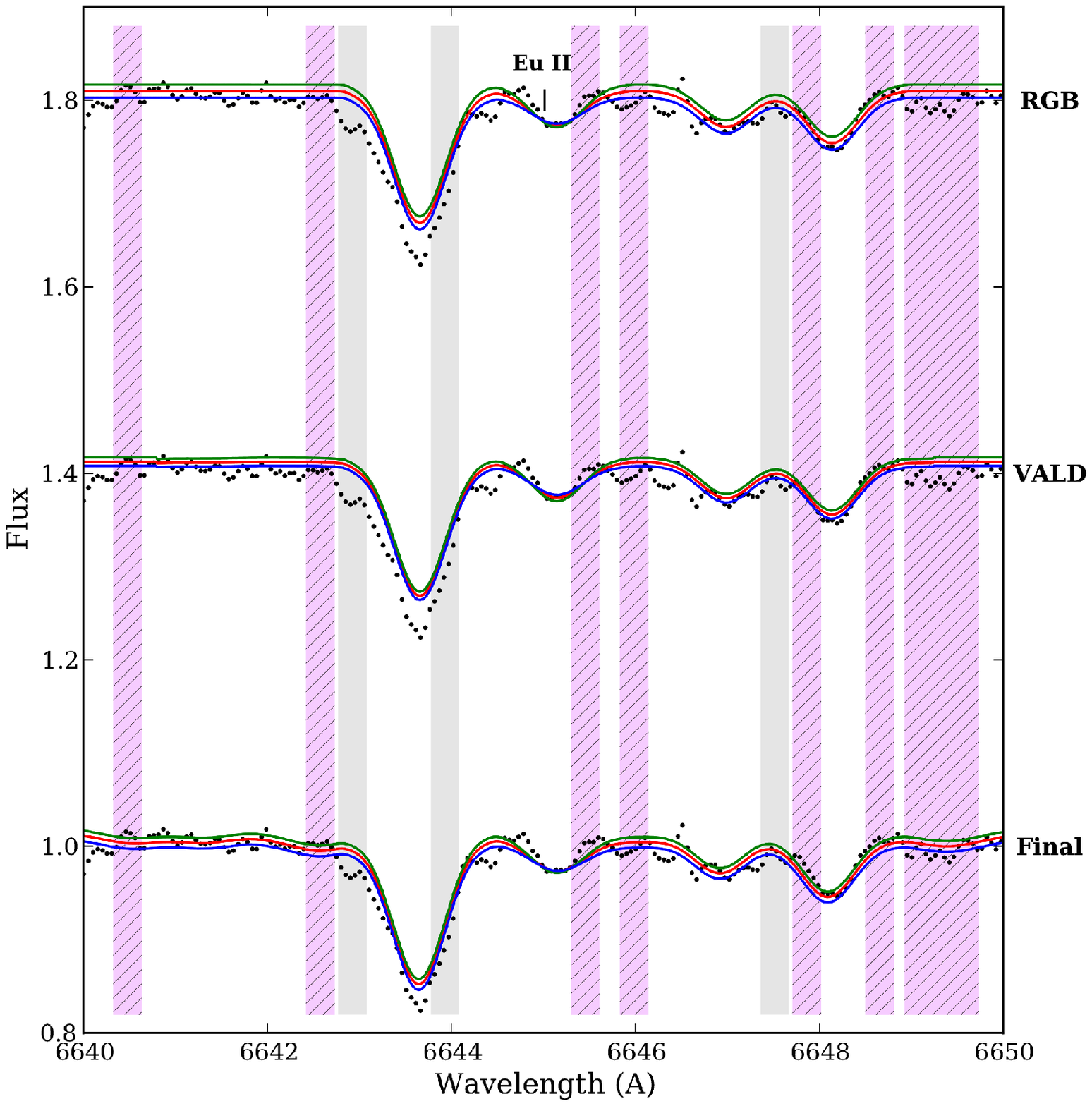}
\caption{Spectrum syntheses of the 6645 \AA \hspace{0.025in}
\ion{Eu}{2} line with the Minimal, VALD, and Final Line Lists. Lines
are as in Figure \ref{fig:Mg5528_RGB}.\label{fig:Eu6645_RGB}}
\end{center}
\end{figure*}

The 47~Tuc syntheses with the Minimal and VALD Lists are shown in
Figure \ref{fig:Eu6645_RGB}.  There appear to be fewer lines in this
region, making it easier to identify the continuum level.  Altogether,
the errors in the abundances are $\pm~0.17$ and $\pm~0.15$ for the
Minimal List and the VALD RGB List, respectively.

\subsubsection{Eu in 47~Tuc: The Final Line List}\label{subsubsec:EuFull}
With the Final Line List, syntheses of the \ion{Eu}{2} line
in the Solar, Arcturus, and 47~Tuc spectra are shown in Figures
\ref{fig:Eu6645_RGB} and \ref{fig:Eu6645}.  HFS and isotopic
components for the \ion{Eu}{2} line are included, using the data from
\citet{LawlerEu}---these corrections alter the shape of the line
slightly, but have a negligible effect on its equivalent width.  There
are several other lines with HFS components in the region of the
\ion{Eu}{2} line which are not included.

As discussed earlier, CN molecular lines are present throughout the
region around the 6645 \AA \hspace{0.025in} \ion{Eu}{2} line.  For the
Solar syntheses the C and N abundances from \citet{Asplund2009} are
used, while scaled-Solar C and N abundances are adopted for Arcturus.
For 47~Tuc it is less clear which C and N abundances to use.  The
effects of dredge-up in RGB stars (e.g. C depletion and N enhancement;
\citealt{LambertRies1981}) are likely to influence the integrated
light abundance.  In addition, all the stars in 47~Tuc show a
well-established CN bimodality, from the RGB
(e.g. \citealt{Briley1997}) down to the main sequence
(e.g. \citealt{Briley2004}).  To determine the extent to which the
input C and N abundances affect the ILS results, spectrum syntheses
were performed with the extreme values determined from individual
stars, i.e. $[\rm{C/Fe}] \approx -0.8$ and $[\rm{N/Fe}] \approx 0.0$
vs. $[\rm{C/Fe}] \approx +0.4$ and $[\rm{N/Fe}] \approx +2.0$
\citep{Briley2004}.

\begin{figure*}
\begin{center}
\centering
\includegraphics[scale=0.9]{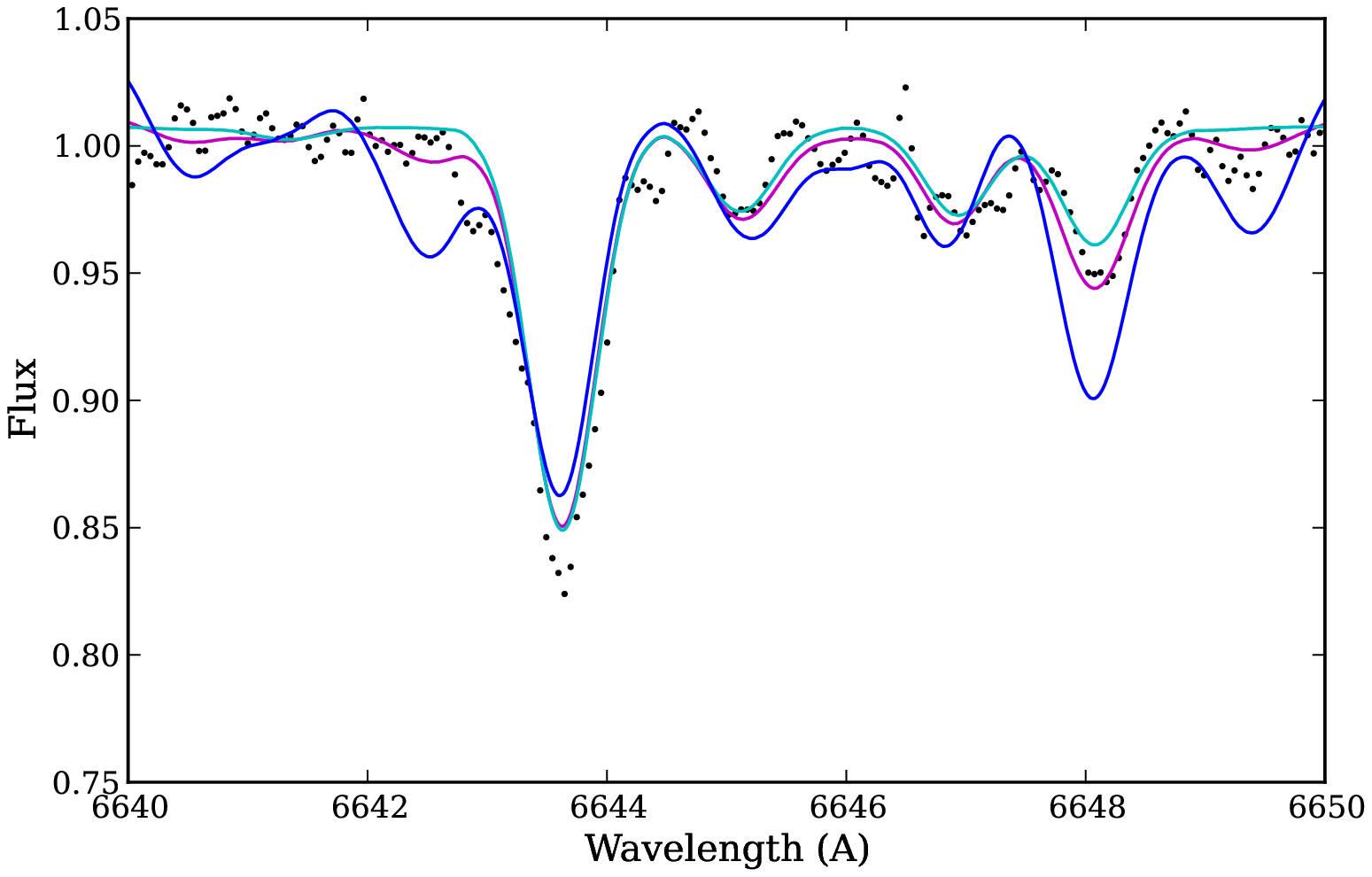}
\caption{Syntheses of the region around the \ion{Eu}{2} line in
47~Tuc with the CN molecules, and assuming a carbon isotopic ratio of
$^{12}$C/$^{13}$C = 9.  The cyan line assumes $[\rm{C/Fe}] = -0.8$ and
$[\rm{N/Fe}] = 0.0$, typical of the C-enhanced stars in 47~Tuc,
while the blue synthesis represents the C-deficient abundances,
$[\rm{C/Fe}] = 0.4$ and $[\rm{N/Fe}] = 2.0$ \citep{Briley2004}.  The
syntheses have been vertically shifted to fit the continuum, and the
Eu abundances were then altered to best fit the \ion{Eu}{2} 6645 \AA
\hspace{0.025in} line.  The Eu abundances differ by nearly 0.2 dex as
a result of the differing input carbon abundances.  The magenta line
shows the synthesis with the best-fitting carbon abundance, as
determined by fitting all the CN lines in the region; this case yields
a Eu abundance similar to the CN-strong
case. \label{fig:CNstrongVSweak}}
\end{center}
\end{figure*}

\begin{figure*}
\begin{center}
\centering
\includegraphics[scale=0.9]{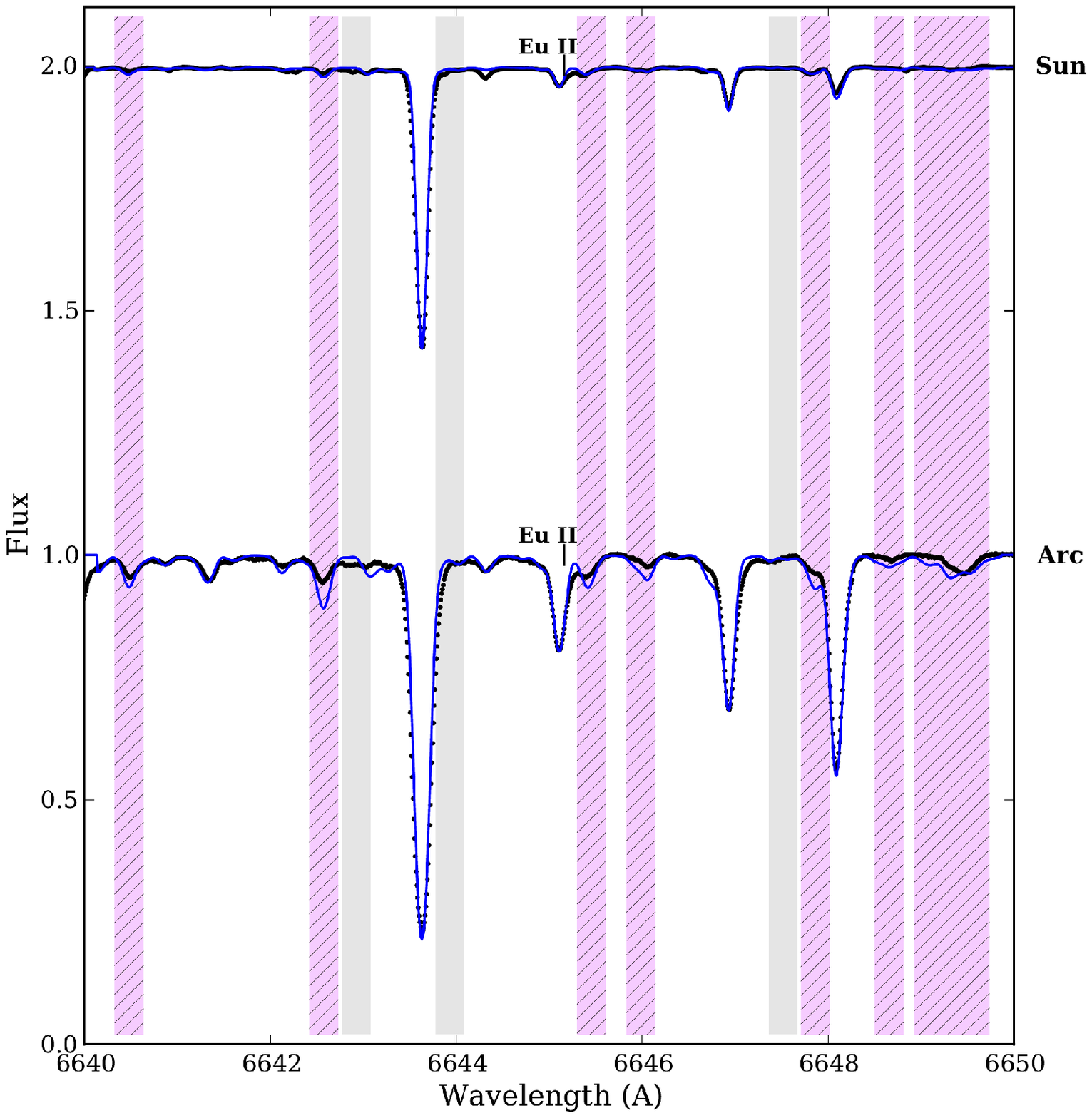}
\caption{Spectrum syntheses of the 6645 \AA \hspace{0.025in}
\ion{Eu}{2} line on the Solar and Arcturus spectra with the Final Line
List.  Lines are as in Figure \ref{fig:Mg5528}.\label{fig:Eu6645}}
\end{center}
\end{figure*}

The 47~Tuc spectrum syntheses with different carbon abundances are
presented in Figure \ref{fig:CNstrongVSweak}. All syntheses were
vertically shifted to fit the continuum regions, and the \ion{Eu}{2}
abundances were re-derived to best fit the line profile.  The
differences in the strengths of the CN lines are quite striking---in
general the CN-weak case (in blue, i.e. the higher carbon abundance)
has stronger spectral lines than the CN-strong case (in cyan).  The
unfortunate presence of a blended CN line redward of the \ion{Eu}{2}
line makes the Eu abundance particularly sensitive to the input C
abundance.  In order to force the CN-weak syntheses to best match the
observed line profile, a \ion{Eu}{2} abundance that is $\sim 0.2$ dex
lower than the CN-strong case must be adopted.  However, if the carbon
abundance is treated as a free parameter, and is determined by fitting
the CN lines in the region (the magenta line in Figure
\ref{fig:CNstrongVSweak}), the systematic errors in abundance can
be reduced.

The Final Line List spectrum synthesis abundances are shown in Table
\ref{table:Abunds}, along with the literature averages.  The Solar
value is slightly lower than the \citet{Asplund2009} value.  The
Arcturus abundance again agrees with the average literature value.
The 47~Tuc ILS [Eu/Fe] value is higher than MB08's $[\rm{Eu/Fe}] =
0.04$, which is based on an equivalent width analysis with the same
spectrum; measurements of this weak line may be more difficult than
originally realized.  The syntheses presented here show that the width
of the synthesized line is wider than the observed feature (see Figure
\ref{fig:Eu6645}), indicating that noise may have distorted the shape
of the \ion{Eu}{2} line.  In this case, the spectrum syntheses, with a
fixed FWHM broadening parameter, will do a better job of fitting the
true line profile.

The IL Eu abundance in 47~Tuc is slightly higher than the average
literature value, though the values do agree within the errors.  The
average literature [Eu/Fe] abundance in \citet{Pritzl2005} is based on
the abundances of 5 giants, 8 subgiants, and 3 turnoff stars, whose
abundances range from $-0.39 \la [\rm{Eu/Fe}] \la +0.44$.
The ILS abundance falls at the upper end of this observed range,
suggesting that the Eu-enhanced stars are dominating the ILS
\ion{Eu}{2} line strength.

Thus, the precision of the abundance from the weak \ion{Eu}{2} line
has significantly improved with the use of the Final Line List, with
the error decreasing to $0.10$ from $0.18$ and $0.15$ for the Minimal
and VALD Lists, respectively.  The spectrum synthesis technique has
also improved the accuracy of the derived abundance, as compared to
the EW analysis.

\subsubsection{Eu in the Other GCs}\label{subsubsec:EuOther}
\begin{figure*}
\begin{center}
\centering
\includegraphics[scale=0.9]{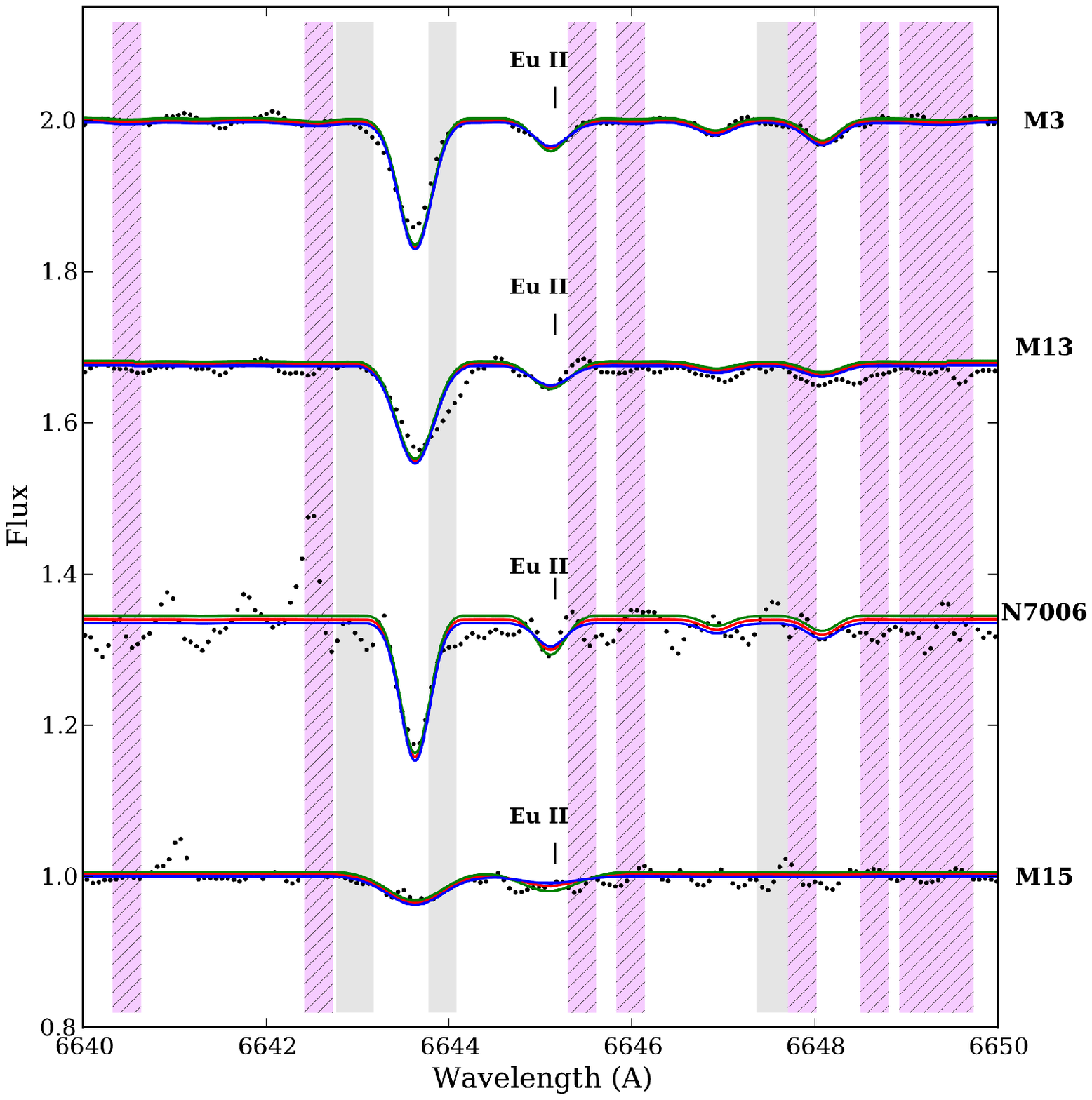}
\caption{Spectrum syntheses of the 6645 \AA \hspace{0.025in}
\ion{Eu}{2} line on the M3, M13, NGC~7006, and M15 ILS.  Lines are as
in Figure \ref{fig:Mg5528}.\label{fig:GCsEu6645}}
\end{center}
\end{figure*}

The syntheses of the \ion{Eu}{2} line in the ILS of the other GCs are shown in
Figure \ref{fig:GCsEu6645}, and the abundances are given in Table
\ref{table:Abunds}.  The input carbon abundances is less important for
these GCs, since the CN lines are weaker.  Table \ref{table:Abunds}
shows that these clusters are all enhanced in Eu, and that these
enhancements are considerably greater than the literature averages.
However, these literature averages do not reflect the Eu variations
that exist within the clusters.  \citet{Roederer2011} has shown that
large Eu dispersions are present in many Galactic GCs, including M3
($0.4 \la [\rm{Eu/Fe}] \la 0.8$), M13 ($0.2 \la
[\rm{Eu/Fe}] \la 1.0$), and M15 ($0.2 \la [\rm{Eu/Fe}]
\la 2.2$).  NGC~7006 does not show a significant dispersion in Eu
($0.30 \la [\rm{Eu/Fe}] \la +0.44$) but these abundances are
based on observations of only six giants \citep{Kraft1998}.  All of
the integrated light synthesis-based abundances fall at the upper end
of the literature ranges, again suggesting that the ILS-derived
abundance is dominated by the most Eu-rich stars in the GCs.

\section{Discussion}\label{sec:Discussion}
Section \ref{sec:Syntheses} presented spectrum syntheses of
high-resolution ILS of the five Galactic GCs 47~Tuc, M3, M13,
NGC~7006, and M15, clusters which cover a wide range in metallicity
(from $[\rm{Fe/H}] \approx -0.7$ to $-2.4$) and HB morphology.  This
Section presents a discussion of the important findings from these
syntheses.

\subsection{The Nature of Integrated Light Spectrum
Syntheses}\label{subsec:Nature}
\begin{figure*}
\begin{center}
\centering
\includegraphics[scale=0.9]{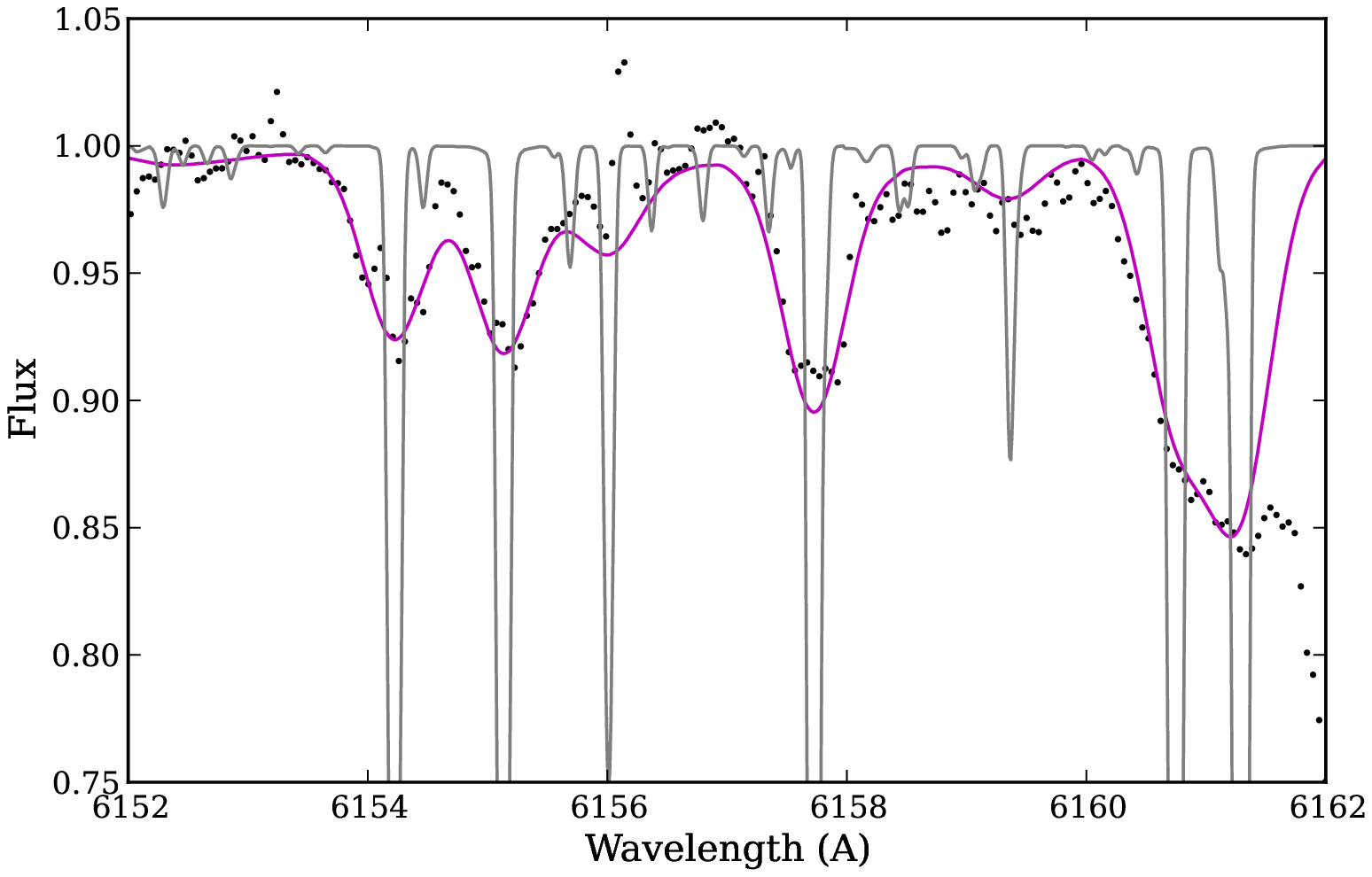}
\caption{Syntheses of the \ion{Na}{1} 6154/6160 \AA \hspace{0.025in}
doublet in the 47~Tuc spectrum.  The magenta line shows the
best-fitting abundance from Section \ref{subsubsec:NaFull}.  The grey
line shows the same synthesis, \textit{before it has been broadened by
the velocity dispersion and instrumental broadening}.  It is clear
that the magenta line never fully reaches the continuum level of
the grey synthesis, due to blending of the weak
features.\label{fig:47TucBlendVSNoBlend}}
\end{center}
\end{figure*}

The results from Section \ref{sec:Syntheses} clearly show that
determining abundances from spectrum syntheses of ILS is more
difficult than with individual, resolved stars.  The severe blends in
an ILS make fitting individual lines and identifying the continuum
level difficult.  Comparisons between the individual Solar/Arcturus
spectra and the 47~Tuc ILS clearly show that 47~Tuc does not have
``traditional'' continuum regions (i.e. regions that are free of any
spectral lines)---even the most featureless regions in the 47~Tuc
spectrum are blends of continuum \textit{and} weak absorption
features.  This is particularly evident in Figure
\ref{fig:47TucBlendVSNoBlend}, which shows syntheses of the 6154/6160
\AA \hspace{0.025in} \ion{Na}{1} lines in the 47~Tuc ILS; both the
unbroadened and broadened (by the velocity dispersion and instrumental
broadening) spectra are shown. In this crowded region, the broadened
spectrum never reaches the continuum level of the unbroadened
spectrum.  Thus, even the basic step of fitting the continuum level
requires some \textit{a priori} knowledge of the weak lines that are
involved in the blending, and the abundances of the elements that
cause those spectral lines.  This once again illustrates the
importance of having a complete and calibrated input line list.

Despite this concern, the uncertainties in continuum placement
introduced by blended weak features may not be significant in all
cases.  The target GCs are all more metal-poor than the Sun and
Arcturus, suggesting that fewer spectral lines are present throughout
the GC spectra.  The cluster velocity dispersion, which blends
the lines together, also makes the weak lines weaker.  Even if the
lines are present and are strong enough to influence the continuum, the
lines in the Final Line List have been calibrated to the Sun and
Arcturus, meaning that once the integrated light abundances of the
elements with the most lines (e.g., Fe, Ca, Ti, and Ni) have been
determined (e.g., through an EW analysis), the synthesized lines
should match the observed ones.  Thus, the well-calibrated strong
lines from these well-known elements may also be used to fit the
continuum. Ultimately,  however, this problem is unavoidable in
integrated light spectrum syntheses, and continuum level uncertainty
should be considered as an important source of uncertainty in the
integrated light abundances.

\subsection{Comparisons with Literature Abundances}\label{subsec:Literature}
As is clear from the abundances presented in Section
\ref{sec:Syntheses}, caution must be taken when comparing ILS
abundances to average literature values. \textit{The ILS abundances do
not represent average cluster abundances.}  The stellar
contributions to an ILS are flux-weighted, meaning that the brightest
RGB, AGB, and post-AGB stars contribute the most light to the observed
spectrum (and therefore have more influence on a line's shape and
strength). Furthermore, the contributions to individual \textit{line
strengths} can also depend on line properties (e.g. ionization state,
excitation potential, etc.; see Figure 9 in MB08). For elements that
are not expected to vary significantly between stars in a given
cluster (e.g. Fe), these effects will be unimportant (assuming the
population is properly modeled).  For elements that \textit{are}
expected to vary (such as Mg, Na, and Eu), the observed line strengths
will also depend on the abundance spreads in the stars that dominated
the light.  The literature averages will also depend on the number and
types of stars observed.  It is important to consider these effects
when comparing with literature abundances.  Flux-weighted literature
averages may help with comparisons with ILS abundances, but this
requires a reasonably complete sample of stars that are selected only
from the observed regions.

For elements with star-to-star variations within a GC, such as Mg, Na,
and Eu, it is therefore more instructive to consider the observed
literature \textit{ranges}, especially those observed among the
brightest evolved stars that dominate the integrated light.  These
ranges will reflect any star-to-star abundance variations, whether due
to effects that may occur up the RGB (e.g. mixing, \citealt{Korn2007})
or as a result of separate populations throughout the cluster.  With
this caveat in mind, the ILABUNDS spectrum synthesis method presented
here provides accurate abundances that fall within the ranges of
literature abundances from individual stars.

\subsection{The Chemical Signatures of Multiple Populations?}\label{subsec:MultiplePops}
As standard Galactic GCs, 47~Tuc, M3, M13, NGC~7006, and M15 all show
signs of multiple populations (under the definition of multiple
populations as ``synonymous with `multiple generations of stars' that
can be distinguished either from their spectra or from multiple
sequences in the colour magnitude diagram;''
\citealt{Gratton2012}). These multiple populations are often
distinguished through star-to-star C, N, O, Na, Mg, and Al abundance
variations (e.g. \citealt{Carretta2009}).  These abundance variations
are known to affect certain features in lower resolution ILS
(e.g. \citealt{Coelho2011} find that the Ca 4227, G4300, CN$_{1}$,
CN$_{2}$, and NaD Lick indices are all affected by the second
generation abundance differences).  The high-resolution integrated
light Na and Mg abundances presented here are therefore likely
affected by these multiple populations within the GCs.

The abundance trends from the syntheses are generally qualitatively
similar among the GCs. With the exception of M15, the GCs are all
found to be Mg-enhanced in their integrated light, as compared to the
Sun. Similarly, all GCs were found to be Na-enhanced.  The magnitudes
of these enhancements vary between the clusters.  For example, 47~Tuc
has the highest integrated light Mg enhancement and a moderate
integrated light Na enhancement, while M15 shows Mg depletion and a
much stronger Na enhancement.  These findings qualitatively agree with
the Na and Mg star-to-star variations in Galactic GCs, and suggest
that in the observed regions, \textit{stars from different populations
may be dominating the  integrated light in these clusters}
(e.g. ``normal'' stars may dominate the 47~Tuc light, while more
``second generation'' Na-rich/Mg-poor stars may dominate the M15
integrated light).  This is likely a stochastic effect, and the
integrated light abundances will likely depend on the area observed in
the ILS.  More theoretical work with multiple populations must be done
in order to understand exactly how these effects will alter the
integrated light abundances.

\subsection{The Effects of Horizontal Branch
Morphology}\label{subsec:HBMorphology}
Metallicity variations alone cannot explain differences in HB
morphologies in Galactic GCs (e.g. \citealt{SandageWildey1967}). Thus,
there must be at least a \textit{second parameter} governing HB
morphology, such as age, although other factors are expected
to contribute to differences in HB morphology
\citep{Dotter2010,Gratton2010}. Because the physical causes behind HB
morphology are not well understood, HB stars cannot be accurately
modeled for all cluster types \textit{a priori}.  This means that for
a given GC, the temperatures and surface gravities of the synthesized
HB stars may not be accurate.  In addition, uncertain physical
processes (e.g. radiative levitation; \citealt{Michaud2008}) can lead
to abundance anomalies in hot HB stars (e.g. Fe;
\citealt{Lovisi2012}), which complicates their contributions to the
ILS.\footnote{However, these chemical variations in the hottest stars
are unlikely to have any effect on the syntheses of the \ion{Mg}{1},
\ion{Na}{1}, and \ion{Eu}{2} lines studied here, as discussed in
Section \ref{subsubsec:HotStars}.}  Although HB stars do not
contribute much of the optical light in Galactic GCs
\citep{Schiavon2002} and although HB morphology does not have a
drastic effect on the integrated optical colors of GCs
\citep{SmithStrader2007}, certain spectral lines may be affected
(i.e. those affected by hot stars, e.g. the Balmer lines,
\citealt{SchiavonHB}; or partially ionized lines,
MB08). \citet{Colucci2009} found that manually adding in blue HB stars
did not significantly change the age and Fe estimates for M31 GCs, but
the effects on spectrum syntheses have not yet been investigated.

The ILS and HST observations of the second parameter triad M3, M13,
and NGC~7006 provide a unique opportunity to test the effects of
HB morphology on synthetic spectra.  If these Galactic clusters were
unresolved, their HB morphologies would not be known \textit{a
priori}---the adopted models might synthesize HBs that are bluer
or redder than the real HB.  With resolved photometry, the HB
morphology of one of these second parameter clusters, M3, is
deliberately mis-modeled (using the HB photometry from the other
second parameter clusters).  The effects of an improperly modeled HB
on the spectrum syntheses of Mg, Na, and Eu can then be tested
directly.  The effects of a purely red HB can also be investigated, by
putting all the HB stars into the reddest HB box.  In all cases the
total number of HB stars is preserved.

\begin{figure*}
\begin{center}
\centering
\includegraphics[scale=0.9]{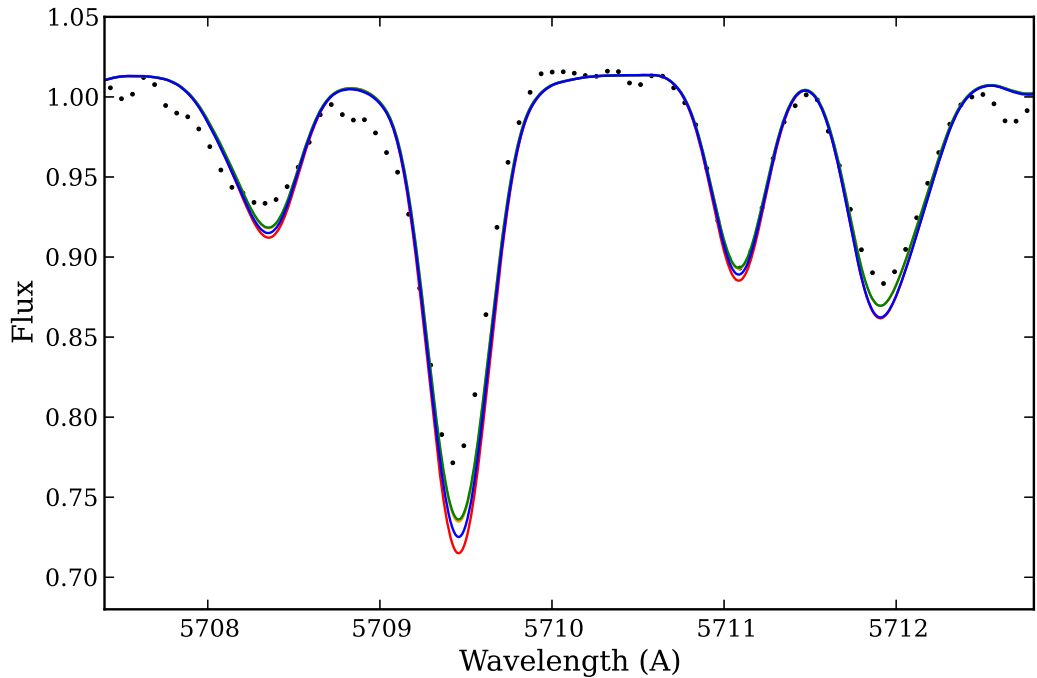}
\caption{Spectrum syntheses of the 5711 \AA \hspace{0.025in}
\ion{Mg}{1} line in the M3 ILS, with different HB morphologies.  The
green line shows M3's original intermediate HB morphology.  The blue
line shows M3's synthesis with M13's blue HB, the orange line shows
M3's synthesis with NGC~7006's redder HB, and the red line shows M3
with a purely red HB.\label{fig:Mg5711_hb}}
\end{center}
\end{figure*}

The resulting abundance changes are shown in Table \ref{table:HB}.
The abundance differences are quite small, in all cases; in
particular, the abundance differences between the syntheses with M3's
HB and NGC~7006's HB are insignificant.  The largest abundance
differences occur for the synthesis of the 5528 \AA \hspace{0.025in}
\ion{Mg}{1} line with the purely red HB, though the other lines show
much smaller abundance differences with the reddest HB.  In general,
the differences with a HB that is too blue are all insignificant (with
$\Delta \log \epsilon \la 0.03$).  Though the resulting abundance
changes are small, the effects are not isolated to the lines of
interest.  A different HB morphology affects lines throughout the
spectrum, as shown in Figure \ref{fig:Mg5711_hb}.  All lines in the
region around the 5711 \AA \hspace{0.025in} line, including the Fe
lines, are affected by the HB morphology, which can affect the
continuum placement.  In this spectral region, the continuum changes
are $<1$\%, which causes a negligible error in abundance.  However,
this test shows that some spectral lines and regions are more
sensitive to the HB morphology than others, even at red wavelengths,
since the blue HB stars add continuum flux.  At bluer wavelengths the
blue HB stars are likely to have an even stronger effect.

\begin{table}
\centering
\begin{minipage}{60mm}
\caption{Abundance Changes with Various HB Morphologies\label{table:HB}}
  \begin{tabular}{@{}lccc@{}}
  \hline
 & M13's HB & NGC~7006's HB & Purely Red HB \\
 & $\Delta \log \epsilon$ & $\Delta \log \epsilon$ & $\Delta \log
\epsilon$\\
\hline
Mg 5528 & -0.03 & 0.0 & -0.06 \\
Mg 5711 & -0.02 & 0.0 & -0.02 \\
Na 6154 & -0.03 & 0.0 & -0.02 \\
Eu 6645 & -0.03 & 0.0 & -0.12 \\
 & & &  \\
\hline
\end{tabular}
\end{minipage}\\
\medskip
\raggedright The differences in abundance are with respect to the
original synthesis, with M3's true HB stars.  Three cases are
considered: M3 with M13's HB, M3 with NGC~7006's HB, and M3 with a
purely red HB. Lines throughout the region are affect by the HB
morphology, which could influence continuum placement and line profile
fitting.\\
\end{table}

\section{Conclusions}\label{sec:Summary}
This work has introduced a high resolution integrated light spectrum
synthesis technique, and has applied the method to five Galactic
globular clusters over a wide range of metallicities and HB
morphologies.  Previous high-resolution ILS analyses were either only
equivalent width analyses (e.g. MB08) or adopted different spectrum
syntheses methods (e.g. \citealt{Colucci2009}).  The spectrum
synthesis method introduced here is fully consistent with the
MB08 equivalent width technique, and is capable of reproducing the
integrated abundances of 47~Tuc, as derived from an equivalent width
analysis.

In particular, this work has produced the following key
findings:

\begin{itemize}
\item  Spectrum syntheses of GC ILS can yield abundances with $\sim
0.1$ dex precision, comparable to the accuracy obtained with high
quality spectral analyses of individual stars.    To achieve this
level of precision, attention must be given to the completeness of the
input line list, which needs to be calibrated, e.g., to the Solar and
Arcturus spectra.   Molecular features do affect spectral lines of
interest and the continuum determinations, and need to be included in
the syntheses.

\item The abundances determined here from the ILS of 47~Tuc, M3, M13,
NGC~7006, and M15 fall within the abundance ranges in the literature
from individual cluster member stars. 

\item The integrated light abundances may not represent the average
cluster abundances in the literature, due to star-to-star abundance
variations within each GC.  The signatures of star-to-star abundance
variations in Mg, Na, and Eu seem to be evident in the integrated
light abundances of all the target GCs.

\item HB morphology has only a negligible effect on the final Mg, Na,
and Eu abundances from syntheses of the four spectral regions
investigated here.  Composition changes in the hottest stars, e.g.,
from radiative levitation, and rotational variations have a negligible
effect on the abundances from these lines.
\end{itemize}

This work shows that high resolution ILS analyses can be used to
determine precise elemental abundances in GCs, at least for certain
lines of Mg, Na, and Eu, when the spectral line list is carefully
considered.   This method works over the observed range of
metallicities and HB morphologies found in the target Galactic GCs.

\section*{Acknowledgments}
The authors thank the referee for helpful comments and suggestions.
The authors thank R. Bernstein for the use of her 47~Tuc spectrum.
CMS acknowledges funding from the Natural Sciences \& Engineering
Research Council (NSERC), Canada, via the Vanier CGS program.  KAV
acknowledges funding through the NSERC Discovery Grants program.  The
Hobby-Eberly Telescope (HET) is a joint project of the University of
Texas at Austin, the Pennsylvania State University, Stanford
University, Ludwig-Maximilians-Universität München, and
Georg-August-Universität Göttingen. The HET is named in honor of its
principal benefactors, William P. Hobby and Robert E. Eberly. The
authors wish to thank the night operations staff of the HET for their
assistance and expertise with these unusual observations.

\end{document}